\documentclass[a4paper,fleqn]{cas-sc}

\usepackage[numbers,sort&compress]{natbib}

\usepackage{amsmath}
\usepackage{amssymb}
\usepackage{empheq}
\usepackage{booktabs}
\usepackage{graphicx}
\usepackage[percent]{overpic}
\usepackage{float}
\usepackage{subcaption}

\usepackage{tikz}
\usetikzlibrary{arrows.meta,decorations.pathmorphing,patterns.meta,calc,tikzmark,positioning}
\usepackage{pgfplots}
\usepackage{pgfplotstable}
\usepgfplotslibrary{fillbetween,groupplots}
\pgfplotsset{compat=1.18}

\usepackage{helvet}
\providecommand{\sansmath}{}
\usepackage{etoolbox}
\usepackage{fix-cm}

\tikzset{
  every picture/.style={
    font=\sffamily\sansmath
  }
}

\pgfplotsset{
  every axis/.append style={
    axis line style={black, line width=0.55pt},
    tick style={black, line width=0.45pt},
    tick align=inside,
    tick pos=both,
    major tick length=2.0pt,
    minor tick length=1.2pt,
    xlabel style={
      font=\fontsize{10}{10}\selectfont\sffamily\sansmath
    },
    ylabel style={
      font=\fontsize{10}{10}\selectfont\sffamily\sansmath
    },
    tick label style={
      font=\fontsize{8.8}{8.8}\selectfont\sffamily\sansmath
    },
    title style={
      font=\fontsize{10}{10}\selectfont\sffamily\sansmath
    },
    legend style={
      font=\fontsize{8.2}{8.2}\selectfont\sffamily\sansmath,
      draw=none,
      fill=none,
      cells={anchor=west},
      inner sep=1.5pt,
      row sep=1pt
    },
    legend image post style={line width=0.8pt},
  },
  every axis plot/.append style={
    line width=0.85pt,
    mark size=1.7pt
  },
}

\pgfsetplotmarksize{1.7pt}
\newcommand{\figaxislabel}[1]{%
    {\fontsize{10}{8}\selectfont\sffamily\sansmath $#1$}%
}

\newlength\figurewidth
\newlength\figureheight

\begin{document}

% CAS front matter
\let\WriteBookmarks\relax
\def\floatpagepagefraction{1}
\def\textpagefraction{.001}

\shorttitle{A coupled lattice Boltzmann model for pore-scale mushy-layer dynamics with application to sea ice}
\shortauthors{N. Trippella and D. Maggiolo}

\title[mode=title]{A coupled lattice Boltzmann model for pore-scale mushy-layer dynamics with application to sea ice}

\author{Nicola Trippella}[orcid=0009-0009-7267-7911]
\cormark[1]
\ead{nicolat@chalmers.se}

\author{Dario Maggiolo}

\affiliation{%
  organization={Division of Fluid Dynamics, Chalmers University of Technology},
  postcode={41296},
  city={Gothenburg},
  country={Sweden}
}

\cortext[cor1]{Corresponding author}

\begin{abstract}
A pore-scale lattice Boltzmann model is developed for reactive transport in phase-changing porous media, coupling solute transport in the liquid with conjugate heat transfer across the solid and liquid phases. Two methodological developments are introduced: a novel link-based Volume-of-Pixel formulation for direction-dependent interface motion, and a generalized Robin boundary treatment for solute redistribution at moving interfaces. The phase-change formulation is validated against thermal benchmarks, while the Robin treatment is assessed through literature test cases, exhibiting convergence rates between first and second order. The coupled model is then applied to a local pore-scale segment of a sea ice brine channel in a purely diffusive regime. The simulations recover the expected ideal mushy-layer behaviour under appropriate scale separation. At sea-ice-relevant concentration ratios, thermal perturbations produce relatively large variations in brine salinity, causing freezing rather than melting to reshape the pore geometry in a non-uniform way. The resulting framework provides a pore-scale numerical tool for resolving coupled temperature, solute redistribution, and phase-boundary dynamics in mushy layers.
\end{abstract}

% Graphical abstract and highlights can be added later if required by the submission system.
% \begin{graphicalabstract}
% \includegraphics{...}
% \end{graphicalabstract}
%
% \begin{highlights}
% \item ...
% \item ...
% \item ...
% \end{highlights}


\maketitle

% Main manuscript
\section{Introduction}
Mushy layers are reactive porous media in which a solid matrix is permeated by a multicomponent liquid maintained close to local thermodynamic equilibrium by coupled heat and mass exchange~\cite{Huppert_Worster_85, Anderson_Guba_20}. Mushy layers occur in a broad range of natural and industrial systems, including sea ice, magma chambers, the inner core--outer core boundary, and metal castings~\cite{amiri24}. Among these systems, sea ice is the focus of the present work. Sea ice is a two-phase, two-component reactive porous medium consisting of fresh ice crystals and saline brine~\cite{feltham06, worster97}. External temperature variations generate thermal gradients within the sea ice matrix, driving heat diffusion through both ice and interstitial brine. Via the local thermodynamic coupling between temperature and salinity, these gradients promote freezing or melting. Freezing rejects salt into the surrounding brine, whereas melting releases fresh water and locally dilutes it~\cite{notz2005, worsterMushy92, Worster2015}. The resulting redistribution of heat and salt continuously modifies the local equilibrium state and reshapes the pore geometry. Resolving these pore-scale processes is essential for linking brine dynamics to the macroscopic salt fluxes that regulate ocean stratification, deep-water formation, large-scale circulation, and nutrient redistribution in or beneath sea ice~\cite{Hunke2011,untersteiner68,Thomas10}. 

Early experimental studies established the importance of brine drainage in mushy layers~\cite{campbell86,huppert88a,woods91,kerr94,wettlaufer97}. Continuum mushy-layer theories subsequently provided the basis for modeling coupled heat, solute, and fluid transport related to chimney convection~\cite{worster00,worster13,Cox1988,Vancoppenolle2009,Wells2011}. These approaches do not directly resolve local transport processes or microstructural evolution, motivating the development of pore-scale numerical models for mushy-layer dynamics. Lattice Boltzmann methods (LBMs) have been widely applied to heat, mass and momentum transport, including multiphase, multicomponent and reactive flows~\cite{PonceDawson1993,He2000,Shan1995,Tian2014}. As a mesoscopic approach, LBM evolves particle distribution functions, from which macroscopic transport quantities are recovered. Its algorithmic simplicity, intrinsic parallelism and ability to handle complex geometries make it particularly attractive for pore-scale problems~\cite{Maggiolo2026}. In the LBM framework, a key challenge is the treatment of the moving solid--liquid interface. Several approaches have been developed, ranging from interface-capturing to sharp-interface methods. Diffuse-interface approaches, including Level Set, Volume-of-Fluid, and phase-field methods, represent the interface through an auxiliary or continuous field distributed over a finite region~\cite{Huang2005,Li2008,Luo2012,Hossain2026,Dinh2023}. In contrast, sharp-interface methods describe the interface as a discontinuity separating the two phases~\cite{Kang2006,Yu2010}. The Volume-of-Pixel (VOP) method is a front-tracking approach in which the computational domain is discretized into pixels whose phase fraction evolves according to the local interface motion. Its simple numerical formulation makes it attractive for problems involving heterogeneous surface reactions~\cite{Kang2004,Kang2002}. Kang \emph{et al.}~\cite{Kang2003} coupled VOP-based geometry evolution with fluid flow, solute transport, and heterogeneous reactions in multicomponent systems~\cite{Kang2006,Kang2010}, while Chen \emph{et al.}~\cite{chen13} later applied the method to thermally driven brine-inclusion migration in salt crystals. Transport-induced phase change in mushy layers poses an additional challenge. Heat and solute fluxes may differ from one direction to another as thermal and solutal perturbations reach the interface through distinct local pathways. In addition, the complex pore geometry introduces interfaces with different orientations and irregularities. Their combined effect can produce direction-dependent departures from equilibrium. Standard nodal VOP formulations do not explicitly resolve this directional variability, since the local phase-change response is represented by a single contribution at each node. The present work therefore evaluates the phase-change rate separately on each interfacial link through a link-based VOP formulation, allowing the solid--liquid boundary to respond directionally to the local transport conditions.

% Second novelty (extension of Huang)
Another numerical challenge in the LBM framework concerns the treatment of solute transport at that evolving boundary. As the interface moves along a lattice link, the associated solute redistribution must remain consistent with the local interface velocity and position. He \emph{et al.}~\cite{He2000} introduced a lattice Boltzmann Robin treatment for heterogeneous surface reactions, later extended by Kang \emph{et al.}~\cite{Kang2002} to dissolution problems at fixed interfaces. Walsh and Saar~\cite{Walsh2010} subsequently generalized Robin boundary conditions to arbitrarily positioned moving interfaces using interpolation, although the required stencil reconstruction can become cumbersome in evolving complex geometries. More recent formulations~\cite{Ju2020,analytical} treat dissolution- and precipitation-driven reactive transport, but do not couple interface displacement with the local solute balance. Among the available approaches, the Robin boundary scheme proposed by Huang \emph{et al.}~\cite{huang15} is particularly attractive because it tracks the interface locally without requiring multi-node interpolation. Its applicability, however, is limited to a restricted range of interface positions along each lattice link. In the present work, this range is extended to cover the full lattice-cell span crossed by the interface during melting and solidification, and the original two-dimensional scheme is generalized to three dimensions on a D3Q19 lattice. The extended treatment achieves second-order accuracy over part of the admissible range and between first- and second-order accuracy elsewhere. Its performance is assessed through a hierarchy of benchmark problems.

This paper is organized as follows. Section~\ref{sec:numerical_methodology} presents the numerical methodology. It first introduces the governing equations for heat transport, solute redistribution, and phase-boundary motion, Sec.~\ref{subsec:gov_eqt}, and then describes the lattice Boltzmann framework adopted in this work, Sec.~\ref{subsec:lbdm_overview}. Section~\ref{sec:results} presents the validation and application of the model. Thermal phase-change simulations are first discussed in Sec.~\ref{subsec:thermal_overview_sim}, followed by solute-transport tests in Sec.~\ref{subsec:solut_sim}. Finally, Sec.~\ref{subsec:coupled_sim} applies the complete coupled formulation to idealized sea ice mushy-layer configurations. Section~\ref{sec:conclusions} summarizes the main conclusions of this work.
\begin{figure}[pos=htpb]
    \centering
    \begin{tikzpicture}
\begin{axis}[
width=0.35\textwidth,
height=0.23\textwidth,
    scale only axis,
    xmin=0,
    xmax=100,
    ymin=-40,
    ymax=40,
    xlabel={$w_{\mathrm{NaCl}}\;[\mathrm{wt.\%}]$},
    ylabel={$T\;[^\circ\mathrm{C}]$},
    axis lines=box,
    tick align=inside,
    tick pos=both,
    xtick={0,20,40,60,80,100},
    ytick={-40,-30,-20,-10,0,10,20,30,40},
    minor x tick num=3,
    minor y tick num=4,
    major tick length=4pt,
    minor tick length=2pt,
    tick label style={font=\fontsize{11}{13}\selectfont\sffamily},
    label style={font=\fontsize{13}{15}\selectfont\sffamily},
    xlabel style={yshift=1pt},
    ylabel style={yshift=-2pt},
    clip=false
]

\addplot[
    red,
    line width=1.35pt,
    smooth
]
coordinates {
    (0,0)
    (3,-1.4)
    (6,-3.0)
    (9,-5.0)
    (12,-7.5)
    (15,-10.7)
    (18,-14.6)
    (21,-19.1)
    (23.3,-21.1)
};

\addplot[
    black,
    line width=1.15pt,
    smooth
]
coordinates {
    (23.3,-21.1)
    (24.5,-15)
    (26,-9)
    (28,-3)
    (29.2,0)
    (30,5)
    (32,15)
    (34,25)
    (36,35)
    (37,40)
};

\addplot[
    black,
    line width=1.15pt
]
coordinates {
    (0,-21.1)
    (37.5,-21.1)
};

\addplot[
    black,
    line width=1.15pt
]
coordinates {
    (29.2,0)
    (100,0)
};

\addplot[
    black,
    line width=1.15pt
]
coordinates {
    (37.5,-40)
    (37.5,0)
};

\node[
    align=center,
    font=\fontsize{8}{10}\selectfont\sffamily
]
at (axis cs:8.5,-13)
{ice +\\brine};

\node[
    font=\fontsize{8}{10}\selectfont\sffamily
]
at (axis cs:15,15)
{Salt water};

\end{axis}
\end{tikzpicture}
    \caption{Schematic phase diagram of the binary water--NaCl system, adapted from~\cite{mousavi2022}.}
    \label{fig:nacl_phase_diagram}
\end{figure}
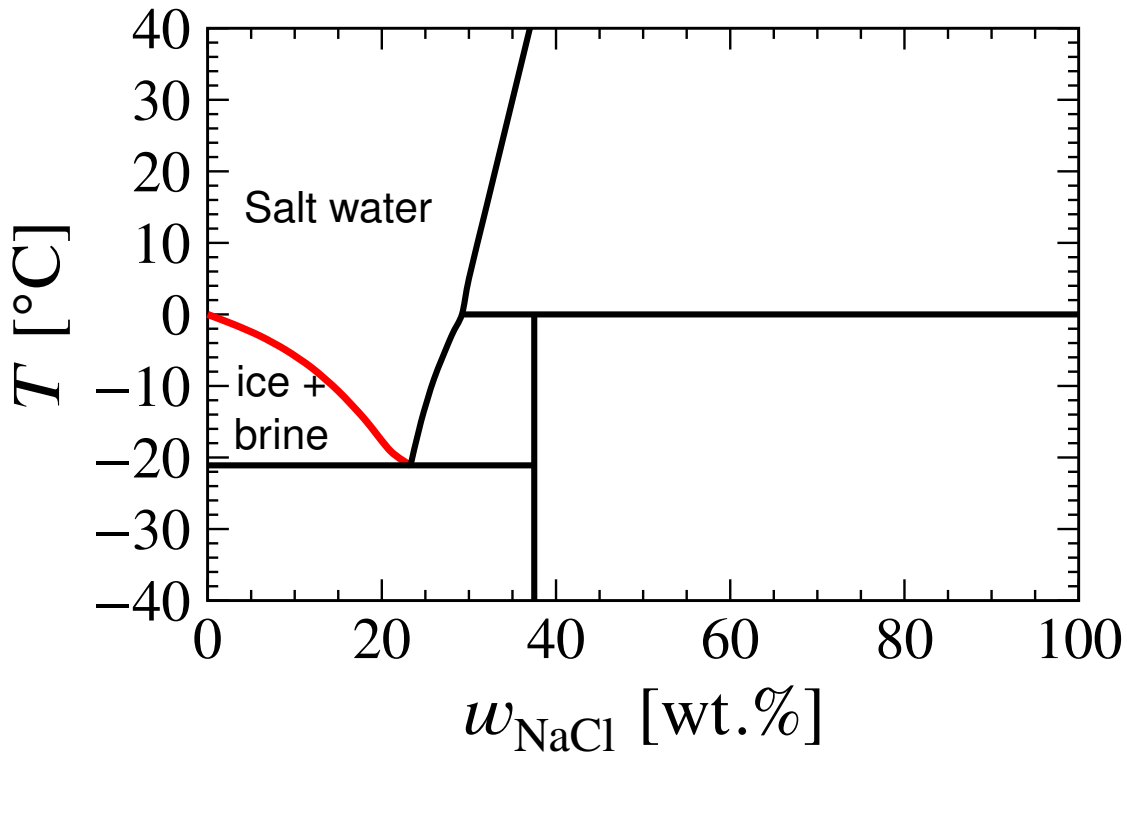
\section{Numerical methodology}
\label{sec:numerical_methodology}
The present study focuses on a purely diffusive regime, neglecting fluid motion and the associated advective transport of heat and solute. This approximation is relevant to diffusion-dominated stages of sea ice evolution, including slow mushy-ice desalination~\cite{du2025sea} and conditions preceding the onset of buoyancy-driven convection~\cite{NotzWorster2009,GriewankNotz2013}. The assumption is adopted here to isolate the newly developed formulations under controlled diffusive conditions and does not imply that advection is generally negligible in sea ice. The model further relies on the following simplifying assumptions:
\begin{itemize}
    \item Thermophysical properties are constant within each phase.
    \item Density variations between the solid and liquid phases are neglected, so that $\rho_s=\rho_l=\rho$.
    \item A sub-eutectic binary solution is considered, consistent with the phase behaviour of sea ice~\cite{wells19}; the corresponding liquidus branch is highlighted in red in Fig.~\ref{fig:nacl_phase_diagram}.
\end{itemize}

\subsection{Governing equations}
\label{subsec:gov_eqt}

Conservation of energy in a medium undergoing solid--liquid phase change can be formulated by accounting for heat transport within both phases and energy exchange associated with the motion of the phase boundary~\cite{worsterMushy92,worster00}. The thermal problem is therefore governed by Eqs.~\eqref{eq:heat_liquid}--\eqref{eq:jump_condition}. 
\par\vspace{-1em}
\noindent
\begin{minipage}[t]{0.44\textwidth}
\begin{empheq}[left=\empheqlbrace]{align}
&\frac{\partial T_l}{\partial t}
= \kappa_l \nabla^2 T_l,
\qquad \text{in } \Omega_l,
\label{eq:heat_liquid}
\\[4pt]
&\frac{\partial T_s}{\partial t}
= \kappa_s \nabla^2 T_s,
\qquad \text{in } \Omega_s,
\label{eq:heat_solid}
\\[4pt]
&T_s(\mathbf{x},0)
= T_w,
\qquad \text{in } \Omega_s,
\label{eq:initial_temperature_solid}
\\[4pt]
&T_l(\mathbf{x},0)
= T_i,
\qquad \text{in } \Omega_l.
\label{eq:initial_temperature_liquid}
\end{empheq}
\end{minipage}
\hfill
\begin{minipage}[t]{0.54\textwidth}
\begin{empheq}[left=\empheqlbrace]{align}
&\left.
\frac{\partial T}{\partial n}
\right|_{\Gamma_N}
= 0,
\qquad \text{on } \Gamma_N,
\label{eq:neumann}
\\[4pt]
&T\big|_{\Gamma_D}
= T_D,
\qquad \text{on } \Gamma_D,
\label{eq:dirichlet}
\\[4pt]
&-L\rho\frac{d\widetilde{R}}{dt}
=
-k_s\left.
\frac{\partial T_s}{\partial n}
\right|_{\Gamma_{sl}^{+}}
+k_l\left.
\frac{\partial T_l}{\partial n}
\right|_{\Gamma_{sl}^{-}},
\qquad \text{on } \Gamma_{sl}.
\label{eq:jump_condition}
\end{empheq}
\end{minipage}
\par\vspace{0.7em}
Here, $T_l$ and $T_s$ denote the temperature fields in the liquid domain $\Omega_l$ and in the solid domain $\Omega_s$. The thermal diffusivities are defined as $\kappa_l=k_l/(\rho c_{p,l})$ and $\kappa_s=k_s/(\rho c_{p,s})$, where $k_l$ and $k_s$ are the thermal conductivities, and $c_{p,l}$ and $c_{p,s}$ are the specific heat capacities of the liquid and solid phases, respectively. Equations~\eqref{eq:heat_liquid} and~\eqref{eq:heat_solid} describe heat diffusion within $\Omega_l$ and $\Omega_s$, while Equations~\eqref{eq:initial_temperature_solid} and~\eqref{eq:initial_temperature_liquid} specify the initial thermal conditions. Eq.~\eqref{eq:neumann} represents a homogeneous Neumann boundary condition imposed on $\Gamma_N$. Here, $n$ is the outward unit normal to the boundary. Equation~\eqref{eq:dirichlet} defines a Dirichlet boundary condition with prescribed temperature $T_D$. Depending on the tested configuration, either Dirichlet or Neumann conditions are applied at the external boundaries. At the solid--liquid interface $\Gamma_{sl}$, Eq.~\eqref{eq:jump_condition} expresses the Stefan jump condition. Here, $L$ denotes the latent heat per unit mass and $\mathrm {d}\widetilde{R} / \mathrm{d}t$ represents the normal velocity of the solid--liquid interface. According to the convention adopted in the numerical solution, freezing corresponds to $\mathrm{d}\widetilde{R}/\mathrm{d}t<0$, whereas melting corresponds to $\mathrm{d}\widetilde{R}/\mathrm{d}t>0$. The superscripts $+$ and $-$ indicate the solid and liquid sides of the interface, according to the adopted normal convention.
\par\vspace{-1em}
\noindent
\begin{minipage}[t]{0.44\textwidth}
\begin{empheq}[left=\empheqlbrace]{align}
&\frac{\partial C}{\partial t}
= D \nabla^2 C,
\qquad \text{in } \Omega_l,
\label{eq:solute_liquid}
\\[4pt]
&C(\mathbf{x},0)
= C_i,
\qquad \text{in } \Omega_l.
\label{eq:initial_concentration}
\end{empheq}
\end{minipage}
\hfill
\begin{minipage}[t]{0.54\textwidth}
\begin{empheq}[left=\empheqlbrace]{align}
&\left.
\frac{\partial C}{\partial n}
\right|_{\Gamma_N^C}
= 0,
\qquad \text{on } \Gamma_N^C,
\label{eq:solute_neumann}
\\[4pt]
&C\big|_{\Gamma_D^C}
= C_D,
\qquad \text{on } \Gamma_D^C,
\label{eq:solute_dirichlet}
\\[4pt]
&\rho D
\left.
\frac{\partial C}{\partial n}
\right|_{\Gamma_{sl}}
= \rho 
\left[
C_s-C\big|_{\Gamma_{sl}}
\right]
\frac{d\widetilde{R}}{dt},
\qquad \text{on } \Gamma_{sl}.
\label{eq:solute_robin}
\end{empheq}
\end{minipage}
\par\vspace{0.7em}
For the solute field, transport is considered only in the liquid phase~\cite{Oerter1992} and is governed by the diffusion problem in Eqs.~\eqref{eq:solute_liquid}--\eqref{eq:solute_robin}. Here, $C$ denotes the solute concentration in the liquid phase, $D$ is the solutal diffusivity and the salt-free solid phase has concentration $C_s=0$. Equation~\eqref{eq:solute_liquid} governs solute diffusion in $\Omega_l$, while Eq.~\eqref{eq:initial_concentration} prescribes a uniform initial concentration $C_i$ throughout the liquid phase. Equations~\eqref{eq:solute_neumann} and~\eqref{eq:solute_dirichlet} define the external boundary conditions, imposing respectively a zero-flux condition on $\Gamma_N^C$ and a prescribed concentration $C_D$ on $\Gamma_D^C$. As for the thermal problem, the choice between Dirichlet and Neumann conditions at the external boundaries depends on the specific configuration considered. The Robin boundary condition~\eqref{eq:solute_robin} expresses the interfacial solute mass balance, governing solute redistribution at the moving solid--liquid interface.

In a mushy-layer system, temperature and solute concentration are intrinsically coupled through the liquidus relation, which defines the local thermodynamic equilibrium between the solid and liquid phases. The red curve in Fig.~\ref{fig:nacl_phase_diagram} shows the liquidus curve considered in this study, which can be approximated as linear~\cite{wells19,notz2005,mousavi2022}, such that $
T_{\mathrm{eq}}(C) = T_m + mC,$
where $T_m$ is the melting temperature of the pure solvent, $m<0$ is the slope and $T_{eq}$ is the equilibrium temperature. Any local departure of the interfacial temperature from the equilibrium one drives melting or solidification. The corresponding interface velocity, $\mathrm{d}\widetilde{R}/\mathrm{d}t$, enters both the solute balance in Eq.~\eqref{eq:solute_robin} and the thermal interfacial balance in Eq.~\eqref{eq:jump_condition}. To close the coupled system, an additional kinetic relation is introduced as follows~\cite{kerr90}
\begin{equation}
\frac{\mathrm{d}\widetilde{R}}{\mathrm{d}t}
=
B
\left[
T(\widetilde{R})
-
T_{\mathrm{eq}}\!\left(C(\widetilde{R})\right)
\right].
\label{eq:interface_kinetics}
\end{equation}
The kinetic coefficient $B$ controls the interface response to departures from thermodynamic equilibrium, with $T(\widetilde{R})$ and $C(\widetilde{R})$ evaluated at the solid--liquid interface. Rather than imposing instantaneous liquidus equilibrium, Eq.~\eqref{eq:interface_kinetics} uses the departure $T-T_{\mathrm{eq}}$ to drive the local melting or solidification rate. 

\subsection{Pore-scale simulations overview}
\label{subsec:lbdm_overview}
In this work, the \texttt{lbdm} code is employed~\cite{lbdm}, building on its previous application to solute transport and reactive processes in porous-media systems~\cite{Maggiolo2023}. The lattice Boltzmann equation is formulated using the Bhatnagar--Gross--Krook (BGK) approximation. D2Q5 lattices, characterized by five discrete directions, are mainly employed for two-dimensional validation tests, whereas D3Q19 lattices, comprising nineteen directions, are used for selected two- and three-dimensional simulations. The quantities $N_Q$ and $N_O$ denote the total number of lattice directions and the number of orthogonal directions, respectively. For the D2Q5 stencil, all non-rest moving directions are orthogonal, whereas the D3Q19 stencil also contains diagonal directions. The discrete lattice vectors are denoted by $\mathbf{e}_i$. The lattice spacing is $\Delta x=1$, while the lattice speed of sound is $c_s=1/\sqrt{3}$. A double-distribution-function (DDF) approach is adopted for temperature and concentration~\cite{li16}. Only temperature and concentration distribution functions are evolved with a time step $\delta_t=1$, while density is held constant and fluid velocity is set to zero.

\subsection{Thermal LB method}
The temperature-based DDF LB model from~\cite{shan97} gives the following equation for the temperature distribution function $t$ along the $i$-th lattice direction:
\begin{flalign}
&\hspace{1cm}
t_i\left(\mathbf{x}+\mathbf{e}_i\delta_t,t+\delta_t\right)
-
t_i(\mathbf{x},t)
%\nonumber\\[-0.2em]
%&\hspace{1cm}
=
-\frac{1}{\tau_{T,j}}
\left[
t_i(\mathbf{x},t)-t_i^{\mathrm{eq}}(\mathbf{x},t)
\right]
+
\phi_{i,j}\delta_t,
\qquad
i=1,\ldots,N_Q.
&&
\label{eq:thermal_LBM}
\end{flalign}
To ensure conservation of conductive heat fluxes across the solid--liquid interface, a conjugate source contribution $\Phi^{\mathrm{conj}}_j$ is introduced for each $j$-th node~\cite{karani15}. This term can be written along the Cartesian direction $\alpha$ as
\begin{equation}
\Phi^{\mathrm{conj}}_{\alpha, j}
=
0.5\,
\nabla_{\alpha} \left(1 / c_p\right)_j \,
\left(-k_j \nabla_{\alpha} T\right)
=
0.5\,
\nabla_{\alpha} \left(1 / c_p\right)_j
\,
q^{*}_{\alpha,j},
\qquad
\alpha\in\{x,y,z\}.
\qquad
j\in\Omega_s,\Omega_l.
\label{eq:thermal_conj_source}
\end{equation}
The prefactor $0.5$ in Eq.~\eqref{eq:thermal_conj_source} distributes the source contribution equally between the neighbouring fluid and solid boundary nodes. The Cartesian components of the nodal conductive heat flux are obtained as~\cite{imani12,chai13}
\begin{equation}
q^{*}_{\alpha,j}
=
-k_j \frac{\partial T}{\partial x_\alpha}
=
\left(1-\frac{1}{2\tau_{T,j}}\right)
\sum_{i=1}^{N_Q}
\left(t_i-t_i^{eq}\right)e_{i\alpha},
\label{eq:local_heat_flux}
\end{equation}
where $e_{i\alpha}$ is the $\alpha$-component of the discrete lattice velocity $\mathbf{e}_i$. The relaxation time $\tau_{T,j}$ can assume two values, depending on the two phases' thermal diffusivities, through $
\kappa_j
=
c_s^2\left(\tau_{T,j}-0.5\right)$.
The equilibrium temperature distribution function $t_i^{eq}$ is evaluated according to Eq.~\eqref{eq:eql_t_eval}. The heat-capacity gradient in~\eqref{eq:thermal_conj_source} is discretized by a one-sided finite-difference, assuming that the interface is located at the mid-link~\cite{Chen2008,Mohamad2009,Astaneh2022}:
\begin{equation}
\frac{\partial}{\partial x_{\alpha}}
\left(
\frac{1}{c_p}
\right)_j
\simeq
\frac{
\displaystyle
1 / c^\mathrm{int}_{p}
-
1 / c_{p,j}
}{
\Delta x/2
},
\qquad
c^\mathrm{int}_{p}
=
\frac{
c_{p,j}
+
c_{p,j+1}
}{2}.
\label{eq:cp_gradient_fd}
\end{equation}
Here, $j$ denotes a lattice node in $\mathbf{x}$, while $j+1$ denotes the neighbouring node located at $\mathbf{x}+\Delta x\,\mathbf{e}_i$, with $\mathbf{e}_i$ aligned with the $\alpha$-direction. Accordingly, $c_{p,j}$ and $c_{p,j+1}$ can take either the liquid or solid heat-capacity value, $c_{p,l}$ or $c_{p,s}$. As a consequence, $\Phi^{\mathrm{conj}}_{\alpha,j}$ is non-zero only at interfacial nodes, where $\partial_{\alpha}(1/c_p)_j$ differs from zero. The final nodal conjugate source term is obtained by summing over the Cartesian contributions, $\Phi^{\mathrm{conj}}_j=\sum_{\alpha}\Phi^{\mathrm{conj}}_{\alpha,j}$.

A second source term accounts for the latent heat released or absorbed during phase change, $\Phi^{\mathrm{lat}}=-\rho\,(L/c_p)\,\mathrm{d}\widetilde{R}/\mathrm{d}t$ (from Eq.~\eqref{eq:jump_condition}). In the present numerical implementation, $\Phi^{\mathrm{lat}}$ is not-null only at the liquid boundary nodes. In addition, each orthogonal interfacial link contributes according to its directional rate
$\mathrm{d}\widetilde{R}_i / \mathrm{d}t$. The $j$-th nodal source is obtained by summing over the orthogonal directions as per Eq.~\eqref{eq:thermal_sources_LH_directional}.
\begin{equation}
\Phi^{\mathrm{lat}}_{i,j}
=
-
\rho\frac{L}{c_{p,j}}
\frac{\mathrm{d} \widetilde{R}_i}{\mathrm{d}t},
\qquad
\Phi^{\mathrm{lat}}_{j}
=
\sum_{i=1}^{N_O}
\Phi^{\mathrm{lat}}_{i,j},
\qquad
j\in\Omega_l.
\label{eq:thermal_sources_LH_directional}
\end{equation}
The equilibrium temperature distribution is given by Eq.~\eqref{eq:eql_t_eval}, with $\Phi_j = \Phi_j^{\mathrm{lat}} + \Phi_j^{\mathrm{conj}}$~\cite{wang2022}:
\begin{equation}
t_i^{\mathrm{eq}}(\mathbf{x},t)=w_i T(\mathbf{x},t),
\qquad
\text{with }
T(\mathbf{x},t)=
\sum_{i=1}^{N_Q} t_i(\mathbf{x},t)
+
\frac{\delta_t}{2}\Phi_j
.
\label{eq:eql_t_eval}
\end{equation}
The source term $\phi_{i,j}$ in Eq.~\eqref{eq:thermal_LBM} can finally be expressed in terms of $\Phi_j$ as $\phi_{i,j}
=
w_i
\left(
1- (1 / 2\tau_{T,j})
\right)
\Phi_j$~\cite{wang2022}.
As shown in~\cite{karani15,wang2022}, a Chapman--Enskog expansion of Eq.~\eqref{eq:thermal_LBM} recovers the conservative form of the energy equation,
$\partial(\rho T)/\partial t
=
\mathbf{\nabla}\cdot\left(\kappa_j\,\mathbf{\nabla}(\rho T)\right)
+
\Phi_j$,
where $\kappa_j$ denotes the local phase thermal diffusivity. The initial thermal conditions are $t_i(\mathbf{x},0)=w_i T_w$ in the solid
domain and $t_i(\mathbf{x},0)= w_i T_i$ in the liquid. The Dirichlet and Neumann boundary conditions on temperature are implemented using the same LBM treatment described in Sec.~\ref{subsec:baseline_robin}, with $C$ replaced by $T$ and $C_D$ by $T_D$ in Eq.~\eqref{eq:huang_dirichlet_halfway}.

\subsection{Concentration LB method}
\label{subsec:conc_LBM}
In the LBM framework, Eq.~\eqref{eq:solute_liquid} is modeled through a DDF concentration-based equation. The concentration distribution function satisfies the following evolution equation:
\begin{equation}
c_i\left(\mathbf{x}+\mathbf{e}_i\delta_t,t+\delta_t\right)
-
c_i(\mathbf{x},t)
=
-\frac{1}{\tau_C}
\left[
c_i(\mathbf{x},t)-c_i^{\mathrm{eq}}(\mathbf{x},t)
\right],
\qquad
i=1,\ldots,N_Q.
\label{eq:concentration_LBM}
\end{equation}

where $c_i^{\mathrm{eq}}(\mathbf{x},t)$ is the equilibrium distribution function which for $\mathbf{u}=0$ reduces to
$c_i^{\mathrm{eq}}=w_i C$, with
$C=\sum_{i=1}^{N_Q}c_i$~\cite{Succi2001}.
The concentration relaxation time $\tau_C$ is related to solute diffusion through
$D=c_s^2(\tau_C-0.5)$. The asymptotic analysis developed in~\cite{Junk03,Junk05,Yoshida10} is recalled to establish the consistency of the lattice Boltzmann formulation with the macroscopic equation~\eqref{eq:solute_liquid}. In this analysis, the diffusive scaling $\delta_t=a\Delta x^2$ is introduced, where $a$ is an adjustable scaling parameter. The distribution functions and the concentration are then expanded as
\begin{align}
c_i(\mathbf{x},t)
\sim
c_i^{(0)}(\mathbf{x},t)
+
\Delta x\,c_i^{(1)}(\mathbf{x},t)
+
\Delta x^2 c_i^{(2)}(\mathbf{x},t)
+
\Delta x^3 c_i^{(3)}(\mathbf{x},t)
+
\mathcal{O}(\Delta x^4),
\label{eq:asymptotic_ci}
\\[0.5em]
C(\mathbf{x},t)
\sim
C^{(0)}(\mathbf{x},t)
+
\Delta x\,C^{(1)}(\mathbf{x},t)
+
\Delta x^2 C^{(2)}(\mathbf{x},t)
+
\Delta x^3 C^{(3)}(\mathbf{x},t)
+
\mathcal{O}(\Delta x^4).
\label{eq:asymptotic_C}
\end{align}
Substituting these expansions into the lattice Boltzmann equation, Taylor-expanding the streaming operator, imposing $\mathbf{u}=0$, and collecting terms of equal order in $\Delta x$ gives
\begin{equation}
\begin{aligned}
c_i^{(0)}
=
w_i C^{(0)},
\quad
c_i^{(1)}
=
w_i C^{(1)}
-
\tau_C w_i
\left(\mathbf{e}_i\cdot\mathbf{\nabla}\right)C^{(0)},
\quad
c_i^{(2)}
=
w_i C^{(2)}
-
\tau_C w_i A,
\end{aligned}
\label{eq:asymptotic_coefficients}
\end{equation}
where
$A=D_{i,2}c_i^{(0)}+D_{i,1}c_i^{(1)}$
(see~\cite{huang15} for the definition of $D_{i,1}$ and $D_{i,2}$).
Summing the second-order expression over all lattice directions and using the symmetry properties of the discrete velocity set yields $
\partial C^{(0)} / \partial t
+
\mathbf{\nabla}\cdot
\left(
\mathbf{u}C^{(0)}
\right)
=
D\,\mathbf{\nabla}^2 C^{(0)}.$ The leading-order macroscopic quantity $C^{(0)}$ represents the concentration recovered in the continuum limit. The initial condition for the solute field is imposed as $c_i(\mathbf{x},0)
=
w_i C(\mathbf{x},0)
=
w_i C_i.$ The Dirichlet and homogeneous Neumann boundary-condition formulations are derived in Eq.~\eqref{eq:huang_dirichlet_halfway}.

\subsubsection{Baseline Robin boundary formulation}
\label{subsec:baseline_robin}
At the moving solid--liquid interface, the concentration field requires a dedicated treatment. Since the present lattice discretization represents curved boundaries using a stair-step approximation, each solid--fluid link can locally be treated as a straight boundary. Therefore, the straight-wall formulation proposed by Huang~\cite{huang15} is adopted. The interfacial solute Robin boundary condition in Eq.~\eqref{eq:solute_robin} can be rearranged in the general Robin form as
\begin{equation}
\frac{\mathrm{d}\widetilde{R}_i}{\mathrm{d}t}\,C
+
D\,\frac{\partial C}{\partial n}
=
\frac{\mathrm{d}\widetilde{R}_i}{\mathrm{d}t}\,C_s
\quad \Longrightarrow \quad
A_2 C
+
A_1\,\frac{\partial C}{\partial n}
=
A_3,
\quad \Longrightarrow \quad
A_2 C^{(0)}
-
A_1
\left(
\mathbf{e}_i\cdot\mathbf{\nabla}
\right)
C^{(0)}
=
A_3,
\label{eq:robin_general_A}
\end{equation}
where $A_1=D$, $A_2=\mathrm{d}\widetilde{R}_i/\mathrm{d}t$, $A_3=A_2C_s=0$ and the outward normal satisfies $\mathbf{n}=-\mathbf{e}_i$. 

\begin{figure}[pos=t]
\centering
\begin{subfigure}[t]{0.40\textwidth}
    \centering
    \setlength\figurewidth{0.86\linewidth}
    \setlength\figureheight{0.54\linewidth}
    \begin{tikzpicture}[font=\sffamily\sansmath]

\begin{axis}[
width=\figurewidth,
height=\figureheight,
scale only axis,
xmin=0.00,
xmax=1.00,
ymin=-1.00,
ymax=4.00,
xlabel={$\gamma$},
ylabel={$p_1$},
axis line style={black},
tick align=inside,
tick pos=both,
grid=none,
tick label style={
    font=\fontsize{11}{13}\selectfont\sffamily\sansmath
},
label style={
    font=\fontsize{13}{15}\selectfont\sffamily\sansmath
},
xlabel style={
    yshift=1pt
},
ylabel style={
    yshift=-2pt
}
]

% Original Huang p1 coefficient
\addplot[
black,
solid,
line width=1.15pt,
unbounded coords=jump
]
table[
x=gamma,
y=p1
]{figures/p1_huang/p1_original.dat};

\end{axis}

\end{tikzpicture}
    \label{fig:p1_huang}
\end{subfigure}
\hfill
\begin{subfigure}[t]{0.40\textwidth}
    \centering
    \setlength\figurewidth{0.86\linewidth}
    \setlength\figureheight{0.54\linewidth}
    \begin{tikzpicture}[font=\sffamily\sansmath]

\begin{axis}[
width=\figurewidth,
height=\figureheight,
scale only axis,
xmin=0.00,
xmax=1.00,
ymin=-1.50,
ymax=1.50,
xlabel={$\gamma$},
ylabel={$p_2$},
axis line style={black},
tick align=inside,
tick pos=both,
grid=none,
tick label style={
    font=\fontsize{11}{13}\selectfont\sffamily\sansmath
},
label style={
    font=\fontsize{13}{15}\selectfont\sffamily\sansmath
},
xlabel style={
    yshift=1pt
},
ylabel style={
    yshift=-2pt
}
]

% Original Huang p2 coefficient
\addplot[
black,
solid,
line width=1.15pt,
unbounded coords=jump
]
table[
x=gamma,
y=p2
]{figures/p2_huang/p2_original.dat};

\end{axis}

\end{tikzpicture}
    \label{fig:p2_huang}
\end{subfigure}
\vspace{-1.05em}
\caption{Huang coefficients over $\gamma\in[0,1]$ for
$\tau_C=0.53$, $A_2=0.005$, $A_1=0.01$, and $\omega_i=1/18$.}
\label{fig:huang_coefficients}
\end{figure}
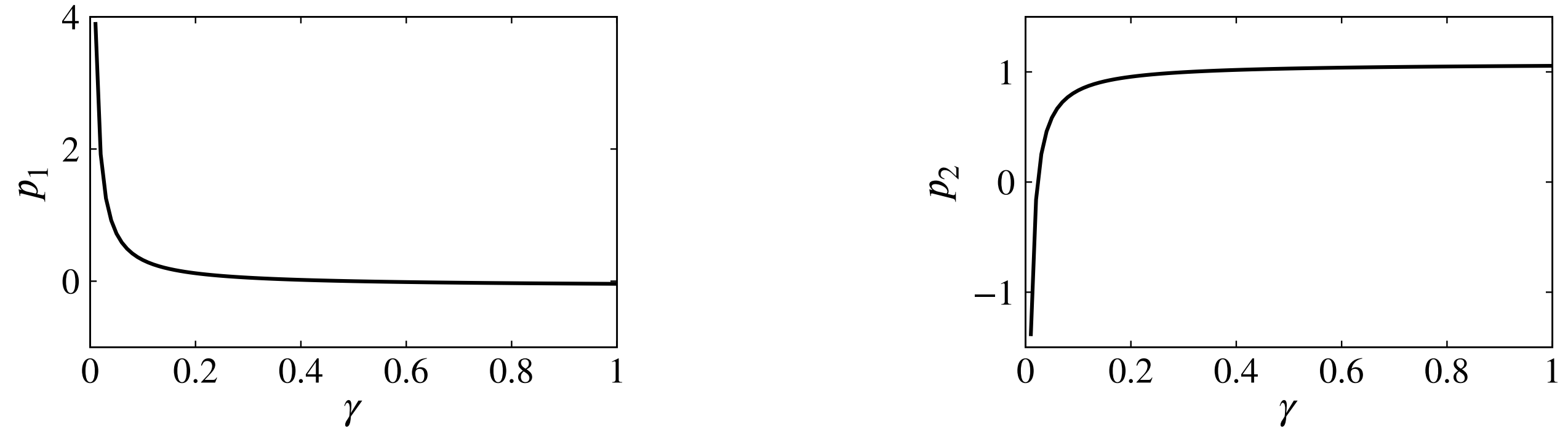

The unknown distribution at the near-boundary fluid node $\mathbf{x}$ is $c_i(\mathbf{x},t+\delta_t)$, where $\mathbf{e}_i$ points from the solid node towards the fluid node. This population is reconstructed from the post-collision ones $\widetilde{c}_i$ and $\widetilde{c}_{\bar{i}}$, with $\mathbf{e}_{\bar{i}}=-\mathbf{e}_i$. The $\gamma$ parameter, defining the normalized interface position relative to the near-boundary fluid node, is computed as
$\gamma=\left\|\mathbf{x}-\mathbf{x}_b\right\|/\Delta x$,
where $\mathbf{x}_b$ is the physical interface position. When $\mathbf{u}=0$ and $A_3=0$, the Huang scheme reduces to
\begin{equation}
c_i(\mathbf{x},t+\delta_t)
=
p_2\,\widetilde{c}_{\bar{i}}(\mathbf{x},t)
+
p_1\,\widetilde{c}_{i}(\mathbf{x},t),
\label{eq:huang_boundary_scheme}
\end{equation}
with
\begin{equation}
p_1
=
\frac{
\Delta x\,p(2\gamma-1)
+
2w_i A_1
\left(\gamma-\frac{1}{2}\right)
(2\tau_C-1)
}{
2\Delta x\,\gamma p
+
4 w_i A_1\gamma(\tau_C-1)
},
\quad
p_2
=
-
\frac{
\Delta x\,p
+
2 w_i A_1
\left(\gamma-\tau_C+\frac{1}{2}\right)
}{
2\Delta x\,\gamma p
+
4 w_i A_1\gamma(\tau_C-1)
}.
\label{eq:huang_p2}
\end{equation}
These two coefficients depend on
$p=A_2\omega_i(\tau_C-1)(2\tau_C-1)$. Figure~\ref{fig:huang_coefficients} shows $p_1$ and $p_2$ over the $\gamma$-range in which they are defined.
For the D3Q19 lattice, the evaluation of $A_2$ requires an additional distinction between orthogonal and diagonal links. Along the $N_O$ orthogonal lattice links aligned with the Cartesian directions, $A_2=\mathrm{d}\widetilde{R}_i/\mathrm{d}t$. Regarding the diagonal links, the interface velocity is obtained by projecting the orthogonal propagation rates onto the diagonal lattice direction, such that
$A_2=(\mathbf{e}_i/\lVert\mathbf{e}_i\rVert)\cdot(\mathrm{d}\mathbf{R}_i/\mathrm{d}t)$. Here, $\mathrm{d}\mathbf{R}_i/\mathrm{d}t$ collects the orthogonal interface propagation rates associated with the $i$-th diagonal link, while $\lVert\mathbf{e}_i\rVert=\sqrt{2}$. Consistently with the boundary-value reconstruction discussed by Li \emph{et al.}~\cite{LiMeiKlausner2013}, the concentration at the actual interface location, $C_b=C(\mathbf{x}_b)$, is first evaluated from the distribution functions. Once $C_b$ is known, the corresponding normal concentration gradient $\left.\partial C/\partial n\right|_b$ follows directly from the Robin condition in Eq.~\eqref{eq:robin_general_A}. These two quantities are therefore computed as
\begin{equation}
C_b
=
\left[
2\gamma\,c_i(\mathbf{x},t+\delta_t)
-
(2\gamma-1)\widetilde{c}_i(\mathbf{x},t)
+
\widetilde{c}_{\bar{i}}(\mathbf{x},t)
\right] /
2 w_i,\quad
\left.
\partial C / \partial n
\right|_b
=
-A_2 C_b / A_1.
\label{eq:huang_cb_gradient}
\end{equation}
The general Robin boundary formulation also recovers Dirichlet- and Neumann-type boundary conditions, which reduce to, respectively~\cite{Yoshida10,zhang12}
\begin{equation}
c_i(\mathbf{x},t+\delta_t)
=
-\widetilde{c}_{\bar{i}}(\mathbf{x},t)
+
2\omega_i C_D, \quad
c_i(\mathbf{x},t+\delta_t)
=
\widetilde{c}_{\bar{i}}(\mathbf{x},t).
\label{eq:huang_dirichlet_halfway}
\end{equation}
\subsubsection{Extended Robin boundary formulation}
\label{subsec:extended_robin}
\begin{figure}[pos=htbp]
\centering

\begin{subfigure}[t]{0.45\textwidth}
    \centering
    \resizebox{!}{4.7cm}{%
        \begin{tikzpicture}[font=\sffamily\sansmath]

% =========================================================
% Parameters
% =========================================================
\def\dx{1.8}   % horizontal lattice spacing
\def\dy{1.6}   % vertical lattice spacing

% Coordinates:
% solid/open nodes: x = 0
% fluid nodes: x = 1.8, 3.6, 5.4
% fictitious cell boundaries: x = -0.9, 0.9, 2.7, 4.5
% y = 0, 1.6, 3.2, 4.8

% =========================================================
% COLOURED REGIONS
% =========================================================

% Extended range: beyond the solid node
% from the left fictitious-cell boundary to the solid node
\fill[cyan!35, opacity=0.18]
    (-0.9,-0.6) rectangle (0,5.4);

% Original Huang range:
% between solid node and first fluid node
\fill[violet!35, opacity=0.18]
    (0,-0.6) rectangle (1.8,5.4);

% Extended range: beyond the first fluid node
% from the fluid node to the next fictitious-cell boundary
\fill[cyan!35, opacity=0.18]
    (1.8,-0.6) rectangle (2.7,5.4);

% =========================================================
% Dashed fictitious cell boundaries
% =========================================================

% Horizontal dashed lines
\foreach \y in {0.8,2.4,4.0}{
    \draw[dashed, line width=0.7pt]
    (-0.9,\y) -- (6.0,\y);
}

% Vertical dashed lines
\foreach \x in {-0.9,0.9,2.7,4.5}{
    \draw[dashed, line width=0.7pt]
    (\x,-0.6) -- (\x,5.4);
}

% =========================================================
% Solid lattice lines
% =========================================================

% Horizontal lattice lines
\foreach \y in {0,1.6,3.2,4.8}{
    \draw[black, line width=0.8pt]
    (-0.9,\y) -- (6.0,\y);
}

% Vertical lattice lines
\foreach \x in {0,1.8,3.6,5.4}{
    \draw[black, line width=0.8pt]
    (\x,-0.6) -- (\x,5.4);
}

% =========================================================
% First possible wall position
% inside the original Huang range:
% between dashed line x=0.9 and fluid node x=1.8
% =========================================================

\def\xwallA{1.25}

\draw[gray, line width=1.1pt]
    (\xwallA,-0.6) -- (\xwallA,5.4);

% Square markers on wall A
\foreach \y in {0,1.6,3.2,4.8}{
    \fill[gray]
    ({\xwallA-0.07},{\y-0.07})
    rectangle
    ({\xwallA+0.07},{\y+0.07});
}

% =========================================================
% Second possible wall position
% inside the extended range:
% beyond the first fluid node, but before x=2.7
% =========================================================

\def\xwallB{2.30}

\draw[gray, line width=1.1pt]
    (\xwallB,-0.6) -- (\xwallB,5.4);

% Square markers on wall B
\foreach \y in {0,1.6,3.2,4.8}{
    \fill[gray]
    ({\xwallB-0.07},{\y-0.07})
    rectangle
    ({\xwallB+0.07},{\y+0.07});
}

% =========================================================
% Solid nodes: open circles
% =========================================================

\foreach \y in {0,1.6,3.2,4.8}{
    \draw[
        fill=white,
        draw=black,
        line width=0.9pt
    ]
    (0,\y) circle (4pt);
}

% =========================================================
% Fluid nodes: filled circles
% =========================================================

\foreach \x in {1.8,3.6,5.4}{
    \foreach \y in {0,1.6,3.2,4.8}{
        \fill[black] (\x,\y) circle (4pt);
    }
}

% =========================================================
% Gamma-position labels
% Same typography as draw_huang2.tikz
% =========================================================

\tikzset{
    gamma label/.style={
        above,
        inner sep=1pt,
        font=\fontsize{15}{17}\selectfont\sffamily\sansmath
    }
}

\node[gamma label] at (-0.9,5.4) {$\mathsf{1.5}$};
\node[gamma label] at (0,5.4)    {$\mathsf{1}$};
\node[gamma label] at (0.9,5.4)  {$\mathsf{0.5}$};
\node[gamma label] at (1.8,5.4)  {$\mathsf{0}$};
\node[gamma label] at (2.7,5.4)  {$\mathsf{-0.5}$};

\end{tikzpicture}

    %
    }
    \caption{Lattice representation with the cells associated to each node shown by dashed lines. The numbers indicate representative $\gamma$ values along the lattice links.}
    \label{fig:huang_scheme}
\end{subfigure}
\hfill
\begin{subfigure}[t]{0.45\textwidth}
    \centering
    \resizebox{!}{4.7cm}{%
        \begin{tikzpicture}[
    remember picture,
    font=\sffamily\sansmath,
    >=Latex,
    line cap=round,
    line join=round
]

% =========================================================
% Colours
% =========================================================
\definecolor{flowblue}{RGB}{35,70,245}
\definecolor{heatred}{RGB}{205,35,35}
\definecolor{interfacegray}{RGB}{120,120,120}
\definecolor{curveorange}{RGB}{220,110,35}
% =========================================================
% Geometry
% =========================================================
\def\xsolid{0.00}
\def\xfluid{2.05}
\def\xwall{3.65}
\def\noder{0.20}

\def\ytop{3.00}
\def\ybot{0.75}

% =========================================================
% Lattice links
% =========================================================
\foreach \y in {\ytop,\ybot}{
    \draw[black, line width=0.85pt]
    (-0.65,\y) -- (4.25,\y);
}

% =========================================================
% Interface wall and boundary point
% =========================================================
\foreach \y in {\ytop,\ybot}{
    \draw[interfacegray, line width=1.25pt]
    (\xwall,\y-0.42) -- (\xwall,\y+0.42);

    \fill[interfacegray]
    ({\xwall-0.075},{\y-0.075})
    rectangle
    ({\xwall+0.075},{\y+0.075});
}

% =========================================================
% Solid and fluid nodes
% =========================================================
\foreach \y in {\ytop,\ybot}{
    \draw[
        fill=white,
        draw=black,
        line width=0.9pt
    ]
    (\xsolid,\y) circle (5pt);

    \fill[black] (\xfluid,\y) circle (5pt);
}

% =========================================================
% Time labels
% =========================================================
\node[
    anchor=west,
    heatred,
    font=\fontsize{11}{13}\selectfont\sffamily\sansmath
] at (4.45,\ytop)
{$t+\delta_t$};

\node[
    anchor=west,
    font=\fontsize{11}{13}\selectfont\sffamily\sansmath
] at (4.45,\ybot)
{$t$};

% =========================================================
% Top populations at t + delta_t
% =========================================================

% c_bar_i(t + dt), left-going population
\draw[
    heatred,
    line width=1.05pt,
    -{Latex[length=2.5mm,width=1.8mm]}
]
({\xfluid-\noder},\ytop) -- (0.82,\ytop);

\node[
    heatred,
    anchor=south,
    font=\fontsize{10.5}{12}\selectfont\sffamily\sansmath
] at (0.92,\ytop+0.30)
{$c_{\bar{i}}(t+\delta_t)$};

% c_i(t + dt), right-going population
\draw[
    heatred,
    line width=1.05pt,
    -{Latex[length=2.5mm,width=1.8mm]}
]
({\xfluid+\noder},\ytop) -- ({\xwall-0.34},\ytop);

\node[
    heatred,
    anchor=south,
    font=\fontsize{10.5}{12}\selectfont\sffamily\sansmath
] at (2.74,\ytop+0.30)
{$c_i(t+\delta_t)$};

% Interface concentration
\node[
    anchor=north,
    font=\fontsize{11}{13}\selectfont\sffamily\sansmath
] (CbLab) at ({\xwall-0.03},{\ytop-0.38})
{$C_b$};

% =========================================================
% Bottom post-collision populations at t
% =========================================================

% c_bar_i tilde(t)
\draw[
    black,
    line width=1.00pt,
    -{Latex[length=2.5mm,width=1.8mm]}
]
({\xfluid-\noder},\ybot) -- (0.82,\ybot);

\node[
    below,
    yshift=-7pt,
    font=\fontsize{10.5}{12}\selectfont\sffamily\sansmath
] at (0.98,\ybot)
{$\widetilde{c}_{\bar{i}}(t)$};

% c_i tilde(t)
\draw[
    black,
    line width=1.00pt,
    -{Latex[length=2.5mm,width=1.8mm]}
]
({\xfluid+\noder},\ybot) -- ({\xwall-0.34},\ybot);

\node[
    below,
    yshift=-7pt,
    font=\fontsize{10.5}{12}\selectfont\sffamily\sansmath
] at (2.62,\ybot)
{$\widetilde{c}_{i}(t)$};

% =========================================================
% Blue reconstruction arrows
% =========================================================

% Contribution from c_bar_i(t) to c_bar_i(t+dt)
\draw[
    flowblue,
    line width=0.85pt,
    -{Latex[length=2.4mm,width=1.7mm]}
]
(0.98,\ybot+0.08)
.. controls (0.90,1.55) and (0.90,2.36) ..
(0.98,\ytop-0.10);

% Contribution from c_i(t) to c_bar_i(t+dt)
% Endpoint moved left, so it points to the red left-going population,
% not to the black node.
\draw[
    flowblue,
    line width=0.85pt,
    -{Latex[length=2.4mm,width=1.7mm]}
]
(2.62,\ybot+0.08)
.. controls (2.42,1.55) and (2.05,2.35) ..
(1.72,\ytop-0.10);

% =========================================================
% Red dependency arrows
% =========================================================

% c_bar_i(t+dt) -> c_i(t+dt)
\draw[
    curveorange,
    line width=0.95pt,
    -{Latex[length=2.6mm,width=1.9mm]}
]
(1.36,\ytop+0.85)
.. controls (1.72,\ytop+1.20) and (2.30,\ytop+1.13) ..
(2.58,\ytop+0.82);

% C_b -> c_i(t+dt)
% Shorter and lower, pointing to the right-going red population
% without overlapping its arrowhead.
\draw[
    curveorange,
    line width=0.95pt,
    -{Latex[length=2.6mm,width=1.9mm]}
]
({\xwall-0.4},{\ytop-0.6})
.. controls ({\xwall-0.70},{\ytop-0.6}) and ({\xwall-0.75},{\ytop-0.5}) ..
({\xwall-0.8},{\ytop-0.18});

\end{tikzpicture}%
    }
    \caption{Logic of the extended scheme for $\gamma<0$. A single lattice line is shown at two consecutive time levels. Blue arrows show that populations at time $t$ are used to reconstruct $c_{\bar{i}}(t+\delta_t)$, while orange ones indicate that $c_i(t+\delta_t)$ is computed from its opposite and $C_b(t+\delta_t)$.}
    \label{fig:huang_scheme2}
\end{subfigure}
\caption{Schematic representation of the boundary-scheme validity range and its extension. White circles represent solid nodes, black circles represent fluid nodes, while grey markers and lines indicate interface positions. Solid black lines illustrate the lattice structure.}
\label{fig:huang_scheme_validity_extension}
\end{figure}
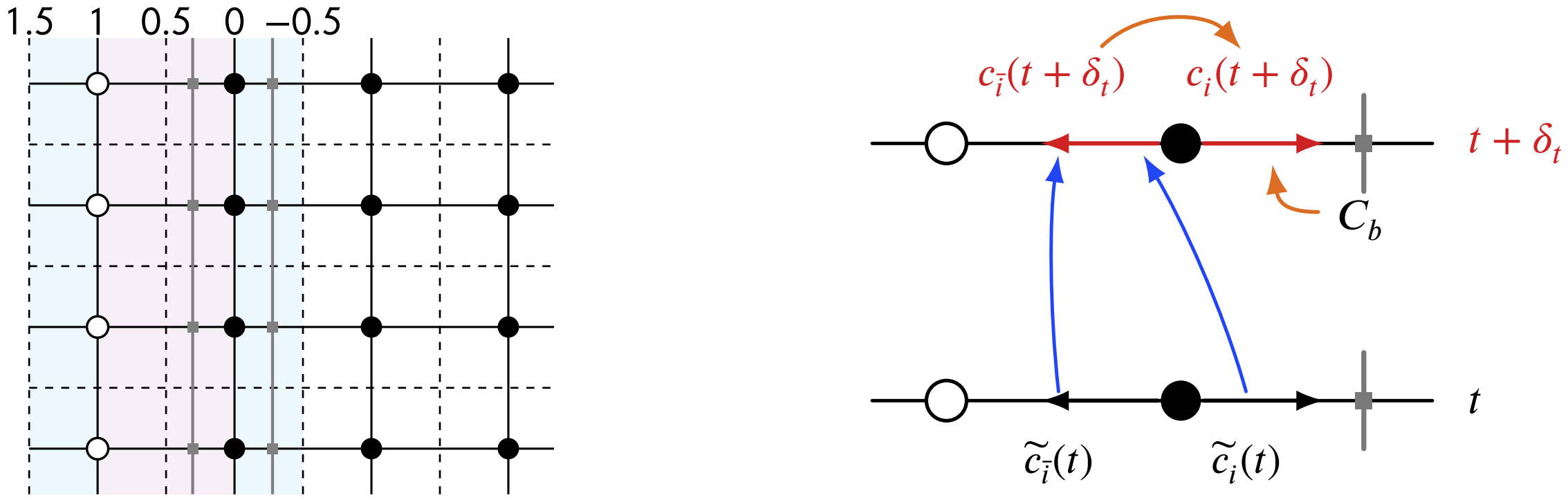
When the boundary is allowed to move as a result of phase change, the $\gamma$ interval of the original Huang scheme, depicted as the light pink region in Fig.~\ref{fig:huang_scheme}, is insufficient to describe the continuous motion of the interface across the entire cell associated with a boundary node. Redefining the interface position as the signed quantity $\gamma=(x-x_b)/\Delta x$ and starting from the initial halfway position $\gamma=0.5$, the interface must be allowed to move up to $\gamma=1.5$ across the solid-node cell during melting (see the left light cyan region in Fig.~\ref{fig:huang_scheme}) and down to $\gamma=-0.5$ across the fluid-node cell during freezing (see the right light cyan region in Fig.~\ref{fig:huang_scheme}), corresponding to a total interface displacement of one lattice spacing $\Delta x$. This extended range is required to allow the interface to traverse an entire lattice cell and therefore to complete the node-conversion process described later in the VOP update. Hence, the Huang boundary scheme~\cite{huang15} is generalized to cover the intervals $-0.5\leq\gamma<0$ and $1<\gamma\leq1.5$. For $\gamma>1$, the existing coefficients can be extended up to $\gamma=1.5$ by directly evaluating them over this additional range. For $\gamma<0$, instead, a new coefficient branch must be derived. Fig.~\ref{fig:huang_scheme2} shows a schematic idea of the extension to negative $\gamma$. The unknown distribution function is the one reflected by the solid node -- i.e. $c_i(\mathbf{x},t+\delta_t)$. Building upon the asymptotic construction in Eq.~\eqref{eq:asymptotic_coefficients}, at leading order $c_i(\mathbf{x},t+\delta_t)=\omega_i C^{(0)}$ and $c_{\bar{i}}(\mathbf{x},t+\delta_t)=\omega_{\bar{i}}C^{(0)}$. Knowing that $\omega_i=\omega_{\bar{i}}$, these two relations are summed up so that $c_i(\mathbf{x},t+\delta_t)$ can be reconstructed as follows
\begin{equation}
c_i(\mathbf{x},t+\delta_t)
+
c_{\bar{i}}(\mathbf{x},t+\delta_t)
=
2\omega_i C^{(0)}(\mathbf{x},t+\delta_t)
+
\mathcal{O}(\Delta x).
\label{eq:huang_leading_reconstruction}
\end{equation}
The two quantities $C^{(0)}(\mathbf{x},t+\delta_t)$ and $c_{\bar{i}}(\mathbf{x},t+\delta_t)$ are evaluated using second-order-accurate reconstructions so as to retain the accuracy of the original scheme as far as possible. Specifically, $c_{\bar{i}}(\mathbf{x},t+\delta_t)$ can be recovered using the classical second-order-accurate scheme in Eq.~\eqref{eq:huang_boundary_scheme}. Exploiting the symmetry of the discrete velocities, $\mathbf{e}_{\bar{i}}=-\mathbf{e}_i$, the population directions in Eq.~\eqref{eq:huang_boundary_scheme} are reversed, giving:
\begin{equation}
c_{\bar{i}}(\mathbf{x},t+\delta_t)
=
p_2^{-}\,\widetilde{c}_{i}(\mathbf{x},t)
+
p_1^{-}\,\widetilde{c}_{\bar{i}}(\mathbf{x},t).
\label{eq:extended_boundary_scheme}
\end{equation}
where
\begin{equation}
p_1^{-}
=
\frac{
\Delta x\,p(2|\gamma|-1)
+
2\omega_i A_1
\left(|\gamma|-\frac{1}{2}\right)
(2\tau_C-1)
}{
2\Delta x\,|\gamma|p
+
4\omega_i A_1|\gamma|(\tau_C-1)
}, \quad
p_2^{-}
=
-
\frac{
\Delta x\,p
+
2\omega_i A_1
\left(|\gamma|-\tau_C+\frac{1}{2}\right)
}{
2\Delta x\,|\gamma|p
+
4\omega_i A_1|\gamma|(\tau_C-1)
}.
\label{eq:extended_p2}
\end{equation}
To account for the negative-$\gamma$ position of the interface, the diffusive flux is mirrored by reversing the sign of $A_2$ in the definition of coefficient
$p^{-}=-A_2\omega_i(\tau_C-1)(2\tau_C-1)$.
Regarding $C^{(0)}(\mathbf{x},t+\delta_t)$, this quantity cannot be obtained directly from the zeroth moment of the distribution functions, since the population $c_i(\mathbf{x},t+\delta_t)$ is the unknown to be reconstructed. It is therefore evaluated independently by expanding the concentration field about the physical interface location $\mathbf{x}_b$.  This expansion is performed along the lattice link $\bar{i}$, corresponding to the direction from the interface at $\gamma<0$ to the fluid node as follows
\begin{equation}
\begin{aligned}
C^{(0)}(\mathbf{x})
&=
C^{(0)}(\mathbf{x}_b)
+
\gamma\Delta x
\left(
\mathbf{e}_{\bar{i}}\cdot\mathbf{\nabla}
\right)
C^{(0)}(\mathbf{x}_b)
+
0.5\,(\gamma\Delta x)^2
\left(
\mathbf{e}_{\bar{i}}\cdot\mathbf{\nabla}
\right)^2
C^{(0)}(\mathbf{x}_b)
+
\mathcal{O}(\Delta x^3).
\end{aligned}
\label{eq:C0_boundary_taylor}
\end{equation}
with the time dependence omitted for simplicity. The first two terms in Eq.~\eqref{eq:C0_boundary_taylor}, namely the boundary concentration and its normal gradient, are obtained by applying the boundary-value reconstruction. Since for $\gamma<0$ the physical interface lies on the opposite side of the fluid node with respect to the original Huang configuration, Eq.~\eqref{eq:huang_cb_gradient} is used with the interchange $i\leftrightarrow\bar{i}$ and $\gamma$ replaced by $|\gamma|$. The last second derivative term in Eq.~\eqref{eq:C0_boundary_taylor} is reconstructed via a linear-combination strategy. Following the derivation presented in the Appendix of~\cite{huang15}, the $\mathcal{O}(\Delta x^1)$ coefficients of two linear combinations of opposite populations are considered:
\begin{align}
(1 / \Delta x)
\left[
c_{\bar{i}}(\mathbf{x},t+\delta_t)
-
\widetilde{c}_{i}(\mathbf{x},t)
\right]
&=
(2\tau_C-1)\omega_{\bar{i}}
\left(
\mathbf{e}_{\bar{i}}\cdot\mathbf{\nabla}
\right)C^{(1)}
-
(\gamma-0.5)(2\tau_C-1)\omega_i
\left(
\mathbf{e}_{\bar{i}}\cdot\mathbf{\nabla}
\right)^2 C^{(0)},
\label{eq:second_derivative_expansion_1}
\\[0.5em]
(1 / \Delta x)
\left[
\widetilde{c}_{\bar{i}}(\mathbf{x},t)
-
\widetilde{c}_{i}(\mathbf{x},t)
\right]
&=
2(\tau_C-1)\omega_{\bar{i}}
\left(
\mathbf{e}_{\bar{i}}\cdot\mathbf{\nabla}
\right)C^{(1)}
-
2\gamma(\tau_C-1)\omega_{\bar{i}}
\left(
\mathbf{e}_{\bar{i}}\cdot\mathbf{\nabla}
\right)^2 C^{(0)}.
\label{eq:second_derivative_expansion_2}
\end{align}
Note that only first-order terms are retained, thus accepting a reduction in the formal order of the reconstruction. At the first order, the two expansions provide the terms required to construct a linear combination that isolates the second directional derivative of $C^{(0)}$. Accordingly, the following linear combination of Eqs.~\eqref{eq:second_derivative_expansion_1} and \eqref{eq:second_derivative_expansion_2} is performed:
\begin{equation}
\begin{aligned}
&2(\tau_C-1)
\frac{1}{\Delta x}
\left[
c_{\bar{i}}(\mathbf{x},t+\delta_t)
-
\widetilde{c}_{i}(\mathbf{x},t)
\right]
-
(2\tau_C-1)
\frac{1}{\Delta x}
\left[
\widetilde{c}_{\bar{i}}(\mathbf{x},t)
-
\widetilde{c}_{i}(\mathbf{x},t)
\right]
\\[0.3em]
&=
2(\tau_C-1)
\left\{
(2\tau_C-1)\omega_{\bar{i}}
\left(
\mathbf{e}_{\bar{i}}\cdot\mathbf{\nabla}
\right)
C^{(1)}
\right.
\left.
-
(\gamma-0.5)
(2\tau_C-1)\omega_{\bar{i}}
\left(
\mathbf{e}_{\bar{i}}\cdot\mathbf{\nabla}
\right)^2
C^{(0)}
\right\}
\\
&\quad
-
(2\tau_C-1)
\left\{
2(\tau_C-1)\omega_{\bar{i}}
\left(
\mathbf{e}_{\bar{i}}\cdot\mathbf{\nabla}
\right)
C^{(1)}
\right.
\left.
-
2\gamma(\tau_C-1)\omega_{\bar{i}}
\left(
\mathbf{e}_{\bar{i}}\cdot\mathbf{\nabla}
\right)^2
C^{(0)}
\right\}.
\end{aligned}
\label{eq:second_derivative_linear_combination}
\end{equation}
After cancellation of the
$\left(\mathbf{e}_{\bar{i}}\cdot\mathbf{\nabla}\right)C^{(1)}$
terms and considering $\omega_{\bar{i}}=\omega_i$, Eq.~\eqref{eq:second_derivative_linear_combination} reduces to
\begin{equation}
\begin{aligned}
\left(
\mathbf{e}_{\bar{i}}\cdot\mathbf{\nabla}
\right)^2
C^{(0)}(\mathbf{x})
&=
\frac{2}
{\omega_i(2\tau_C-1)\Delta x}
\left[
c_{\bar{i}}(t+\delta_t,\mathbf{x})
-
\widetilde{c}_{i}(t,\mathbf{x})
\right]
-
\frac{1}
{\omega_i(\tau_C-1)\Delta x}
\left[
\widetilde{c}_{\bar{i}}(t,\mathbf{x})
-
\widetilde{c}_{i}(t,\mathbf{x})
\right].
\end{aligned}
\label{eq:C_second_derivative}
\end{equation}
Eq.~\eqref{eq:C_second_derivative} gives the required second directional derivative at $\mathbf{x}$, which is used to approximate the corresponding derivative at the interface $\mathbf{x}_b$. Finally, the reconstructed quantities in Eqs.~\eqref{eq:extended_boundary_scheme} and~\eqref{eq:C0_boundary_taylor} are substituted into Eq.~\eqref{eq:huang_leading_reconstruction} to get the unknown as
$c_i(\mathbf{x},t+\delta_t)
=
2\omega_i C^{(0)}(\mathbf{x},t+\delta_t)
-
c_{\bar{i}}(\mathbf{x},t+\delta_t)$.
The final boundary scheme for the negative-$\gamma$ range can also be written in compact form:
\begin{equation}
c_i(\mathbf{x},t+\delta_t)
=
P_2^{-}\,\widetilde{c}_{\bar{i}}(\mathbf{x},t)
+
P_1^{-}\,\widetilde{c}_{i}(\mathbf{x},t).
\label{eq:negative_gamma_boundary_scheme}
\end{equation}
where the coefficients are
\begin{equation}
\begin{aligned}
P_1^{-}
&=
\left(
2|\gamma|
+
2\frac{A_2}{A_1}\gamma|\gamma|
+
\frac{2\gamma^2}{2\tau_C-1}
-
1
\right)
p_2^{-}
+
1
+
\frac{A_2}{A_1}\gamma
-
\frac{2\gamma^2}{2\tau_C-1}
+
\frac{\gamma^2}{\tau_C-1}.
\end{aligned}
\label{eq:negative_gamma_P1}
\end{equation}

\begin{equation}
\begin{aligned}
P_2^{-}
&=
\left(
2|\gamma|
+
2\frac{A_2}{A_1}\gamma|\gamma|
+
\frac{2\gamma^2}{2\tau_C-1}
-
1
\right)
p_1^{-}
-
(2|\gamma|-1)
-
\frac{A_2}{A_1}\gamma(2|\gamma|-1)
-
\frac{\gamma^2}{\tau_C-1}.
\end{aligned}
\label{eq:negative_gamma_P2}
\end{equation}

\begin{figure}[pos=htbp]
\centering

\begin{subfigure}[t]{0.38\textwidth}
    \centering
    \setlength\figurewidth{0.86\linewidth}
    \setlength\figureheight{0.54\linewidth}
    \begin{tikzpicture}[font=\sffamily\sansmath]

\begin{axis}[
width=\figurewidth,
height=\figureheight,
scale only axis,
xmin=-0.50,
xmax=1.50,
ymin=-1.50,
ymax=1.50,
xlabel={$\gamma$},
ylabel={$p$},
axis line style={black},
tick align=inside,
tick pos=both,
grid=none,
tick label style={
    font=\fontsize{11}{13}\selectfont\sffamily\sansmath
},
label style={
    font=\fontsize{13}{15}\selectfont\sffamily\sansmath
},
xlabel style={
    yshift=1pt
},
ylabel style={
    yshift=-2pt
},
legend style={
    draw=none,
    fill=none,
    font=\fontsize{10}{12}\selectfont\sffamily\sansmath,
    at={(0.97,0.97)},
    anchor=north east
},
legend cell align={left},
]

% Extended scheme: P_1^- for negative gamma
\addplot[
black,
dashed,
line width=1.15pt,
unbounded coords=jump
]
table[
x index=0,
y index=1
]{figures/coeff_new_scheme/P1_negative.dat};
\addlegendentry{$P_1^{-}$}

% Original Huang coefficient p_1 for positive gamma
\addplot[
black,
solid,
line width=1.15pt,
unbounded coords=jump
]
table[
x index=0,
y index=1
]{figures/coeff_new_scheme/p1_positive.dat};
\addlegendentry{$p_1$}

\end{axis}

\end{tikzpicture}
    \vspace{-1.05em}
    \caption{First coefficient.}
    \label{fig:p11_new}
\end{subfigure}
\hfill
\begin{subfigure}[t]{0.38\textwidth}
    \centering
    \setlength\figurewidth{0.86\linewidth}
    \setlength\figureheight{0.54\linewidth}
    \begin{tikzpicture}[font=\sffamily\sansmath]

\begin{axis}[
width=\figurewidth,
height=\figureheight,
scale only axis,
xmin=-0.50,
xmax=1.50,
ymin=-1.50,
ymax=1.50,
xlabel={$\gamma$},
axis line style={black},
tick align=inside,
tick pos=both,
grid=none,
tick label style={
    font=\fontsize{11}{13}\selectfont\sffamily\sansmath
},
label style={
    font=\fontsize{13}{15}\selectfont\sffamily\sansmath
},
xlabel style={
    yshift=1pt
},
legend style={
    draw=none,
    fill=none,
    font=\fontsize{10}{12}\selectfont\sffamily\sansmath,
    at={(0.97,0.03)},
    anchor=south east
},
legend cell align={left},
]

% Extended scheme: P_2^- for negative gamma
\addplot[
black,
dashed,
line width=1.15pt,
unbounded coords=jump
]
table[
x index=0,
y index=1
]{figures/coeff_new_scheme/P2_negative.dat};
\addlegendentry{$P_2^{-}$}

% Original Huang coefficient p_2 for positive gamma
\addplot[
black,
solid,
line width=1.15pt,
unbounded coords=jump
]
table[
x index=0,
y index=1
]{figures/coeff_new_scheme/p2_positive.dat};
\addlegendentry{$p_2$}

\end{axis}

\end{tikzpicture}
    \caption{Second coefficient.}
    \label{fig:p22_new}
\end{subfigure}
\caption{Coefficients of the boundary scheme, with $A_2>0$. Dashed lines represent the extension to negative $\gamma$.}
\label{fig:p1_p2_new_melting}
\end{figure}
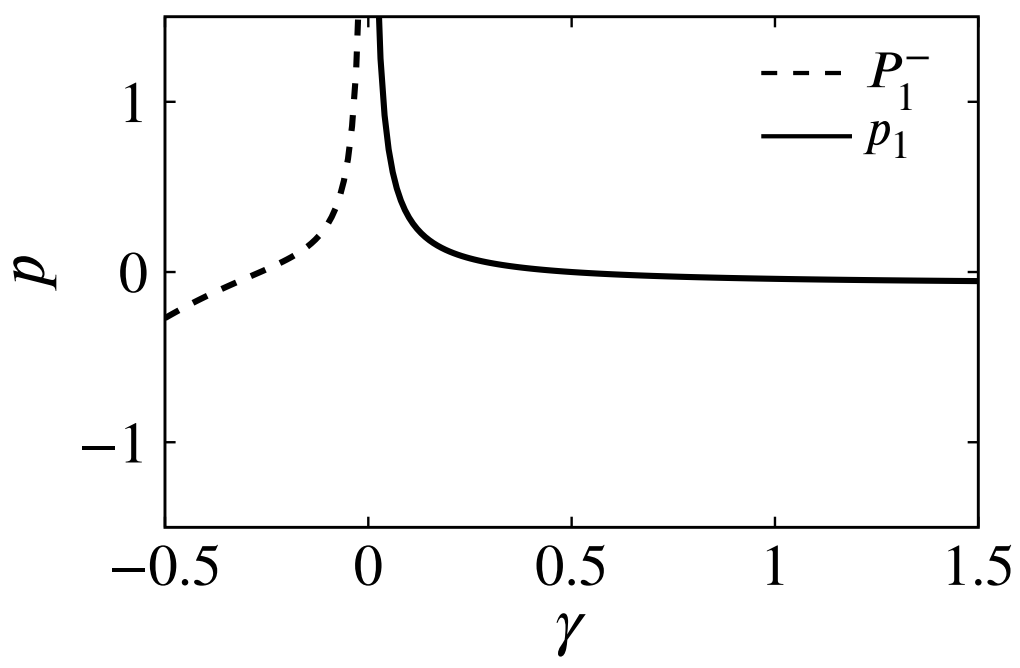
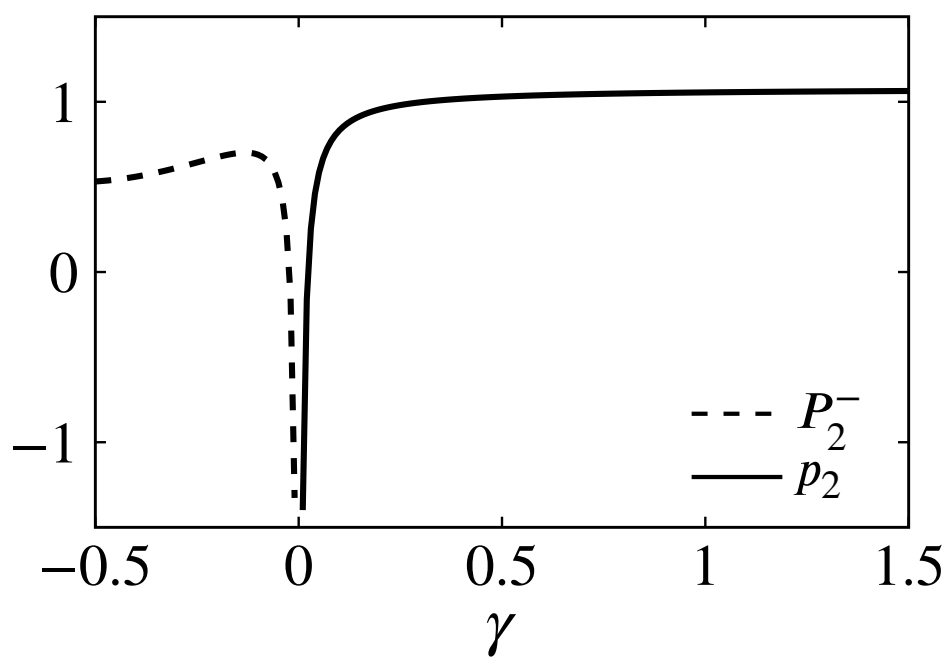

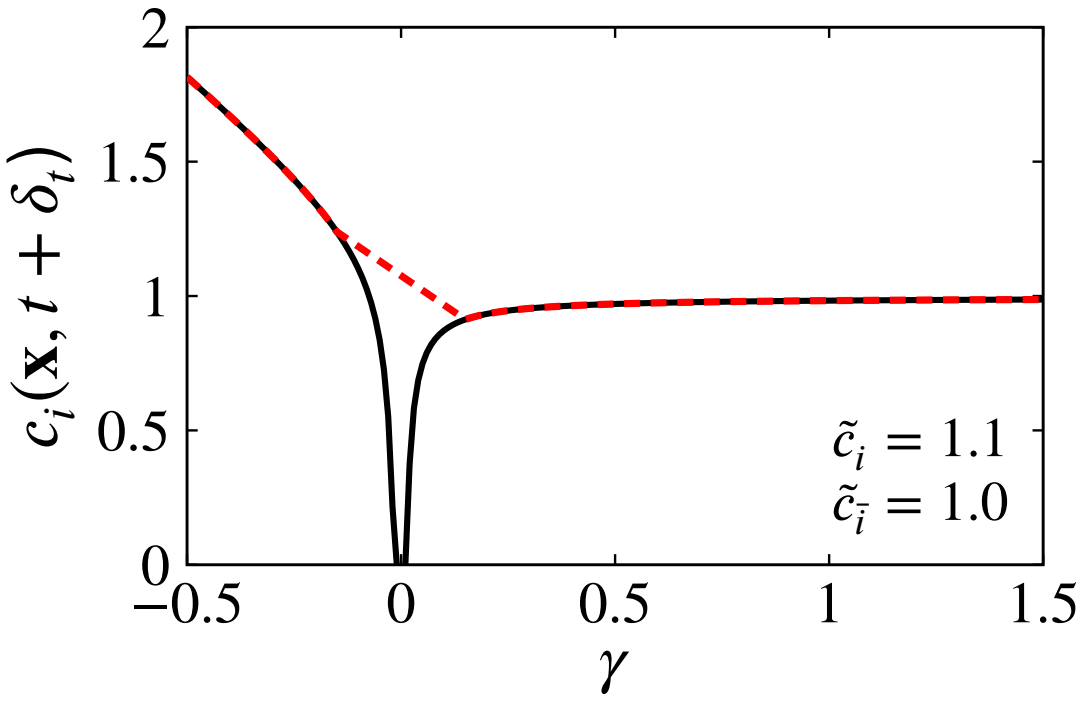
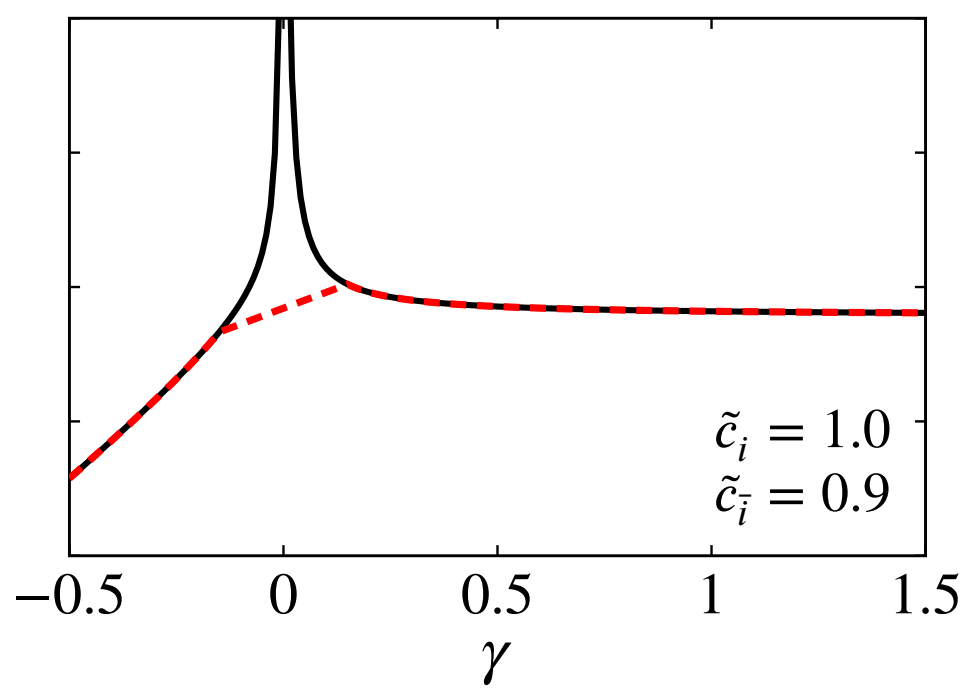
\begin{figure}[pos=htbp]
\centering
\begin{subfigure}[t]{0.40\textwidth}
    \centering
    \setlength\figurewidth{0.86\linewidth}
    \setlength\figureheight{0.54\linewidth}
    \begin{tikzpicture}[font=\sffamily\sansmath]

\begin{axis}[
width=\figurewidth,
height=\figureheight,
scale only axis,
xmin=-0.5,
xmax=1.50,
ymin=0.00,
ymax=2.00,
xlabel={$\gamma$},
ylabel={$c_{i}(\mathbf{x},t+\delta_t$)},
axis line style={black},
tick align=inside,
tick pos=both,
grid=none,
tick label style={
    font=\fontsize{11}{13}\selectfont\sffamily\sansmath
},
label style={
    font=\fontsize{13}{15}\selectfont\sffamily\sansmath
},
xlabel style={
    yshift=1pt
},
ylabel style={
    yshift=-2pt
},
legend style={
    font=\fontsize{9.5}{11}\selectfont\sffamily\sansmath,
    draw=black,
    fill=white,
    inner sep=1.5pt,
    row sep=0pt,
    at={(0.99,0.99)},
    anchor=north east
},
legend cell align=left
]

% Original formulation:
% values below y=-0.5 are discarded and the curve is broken,
% so the singular branch does not reconnect inside the plot.
\addplot[
black,
solid,
line width=1.15pt,
restrict y to domain*=0.00:2.00,
unbounded coords=jump
]
table[
x=gamma,
y=t_original
]{figures/t_regularization/t_regularization_freez/t_regularization_a2_negative.dat};

% Linear regularization
\addplot[
red,
densely dashed,
line width=1.35pt
]
table[
x=gamma,
y=t_regularized
]{figures/t_regularization/t_regularization_freez/t_regularization_a2_negative.dat};

% Parameter annotation
\node[
anchor=south east,
font=\fontsize{10.5}{12}\selectfont\sffamily\sansmath,
align=left,
fill=white,
inner sep=1.5pt
]
at (rel axis cs:0.97,0.04)
{$\tilde{c}_{i}=1.1$\\
 $\tilde{c}_{\bar{i}}=1.0$};

\end{axis}

\end{tikzpicture}
    \vspace{-1.05em}
    \caption{Freezing case, $A_2<0$.}
    \label{fig:t_regularization_freez}
\end{subfigure}
\hfill
\begin{subfigure}[t]{0.4\textwidth}
    \centering
    \setlength\figurewidth{0.86\linewidth}
    \setlength\figureheight{0.54\linewidth}
    \begin{tikzpicture}[font=\sffamily\sansmath]

\begin{axis}[
width=\figurewidth,
height=\figureheight,
scale only axis,
xmin=-0.5,
xmax=1.50,
ymin=0.00,
ymax=2.00,
xlabel={$\gamma$},
axis line style={black},
tick align=inside,
tick pos=both,
grid=none,
yticklabels=\empty,
tick label style={
    font=\fontsize{11}{13}\selectfont\sffamily\sansmath
},
label style={
    font=\fontsize{13}{15}\selectfont\sffamily\sansmath
},
xlabel style={
    yshift=1pt
},
legend style={
    font=\fontsize{9.5}{11}\selectfont\sffamily\sansmath,
    draw=black,
    fill=white,
    inner sep=1.5pt,
    row sep=0pt,
    at={(0.98,0.98)},
    anchor=north east
},
legend cell align=left
]

% Original formulation
\addplot[
black,
solid,
line width=1.15pt,
restrict y to domain*=0.00:2.50,
unbounded coords=jump
]
table[
x=gamma,
y=t_original
]{figures/t_regularization/t_regularization_melt/t_regularization_a2_positive.dat};

% Linear regularization
\addplot[
red,
densely dashed,
line width=1.35pt
]
table[
x=gamma,
y=t_regularized
]{figures/t_regularization/t_regularization_melt/t_regularization_a2_positive.dat};

% Parameter annotation
\node[
anchor=south east,
font=\fontsize{10.5}{12}\selectfont\sffamily\sansmath,
align=left,
fill=white,
inner sep=1.5pt
]
at (rel axis cs:0.97,0.04)
{$\tilde{c}_{i}=1.0$\\
 $\tilde{c}_{\bar{i}}=0.9$};

\end{axis}

\end{tikzpicture}
    \caption{Melting case, $A_2>0$.}
    \label{fig:t_regularization_melt}
\end{subfigure}
\caption{Linear regularization of the boundary scheme adopted over $\gamma\in[-\delta,\delta]=[-0.15,0.15]$.}
\label{fig:t_regularization_freezing_melting}
\end{figure}
Fig.~\ref{fig:p1_p2_new_melting} shows that the combined Huang--extended formulation exhibits a singularity at $\gamma=0$. To address this singularity, several smooth regularization strategies were initially investigated, including the Bercovier--Engelman, Papanastasiou, and Glowinski regularizations~\cite{multiphase_rheology_slides}. Although these techniques removed the formal divergence of the coefficients, the numerical simulations remained sensitive to the regularization parameters and eventually became unstable. A simpler local treatment is therefore adopted. A small interval $[-\delta,\delta]$ is introduced around $\gamma=0$. The reconstructed population at $\gamma = -\delta$ is calculated via the extended boundary scheme, giving $c_i^{-}$, while the one at $\gamma = \delta$ is computed from the original Huang scheme, giving $c_i^{+}$. The populations within the interval are obtained by linear interpolation through $c_i^{\mathrm{reg}}(\gamma)
=
c_i^{-}
+
(\gamma+\delta)
\left(
c_i^{+}-c_i^{-}
\right) / 2\delta,$ as can be seen in Fig.~\ref{fig:t_regularization_freezing_melting}.
\subsection{Link-based VOP model for phase change}
The VOP method is adopted to track the evolution of the solid--fluid interface. The computational domain is discretized into nodes whose phase volumes evolve according to the local interface motion~\cite{chen13}. The present scheme introduces two independent nodal variables, $V_{l,j}$ and $V_{s,j}$, representing the accumulated liquid and solid volumes at the $j$-th lattice node. For an interfacial area $A_{\Gamma,i}=\Delta x^2$ associated with the $i$-th orthogonal solid--fluid link, the phase-volume variation is evaluated from the local normal interface velocity $A_2$. According to the sign convention adopted here, $A_2 > 0$ corresponds to melting, whereas $A_2 < 0$ corresponds to freezing. At boundary nodes, melting contributions increase $V_{l,j}$, whereas freezing contributions increase $V_{s,j}$. Denoting by $\mathcal{I}_j^\Gamma$ the set of orthogonal interfacial links connected to the $j$-th boundary node, the corresponding volume variations are written as
\begin{equation}
\mathrm{d}V_{l,j} \, / \, \mathrm{d}t
=
\sum_{i\in\mathcal{I}_j^\Gamma}
A_{\Gamma,i}
\, \langle
A_2\rangle_{+},
\, \, \, \,
j\in\Omega_l \,;
\qquad
\mathrm{d}V_{s,j} \, / \, \mathrm{d}t
=
\sum_{i\in\mathcal{I}_j^\Gamma}
A_{\Gamma,i}
\, \langle
-A_2
\rangle_{+},
\, \, \, \,
j\in\Omega_s.
\label{eq:vop_liquid_volume}
\end{equation}
where $\langle a\rangle_{+}=\max(a,0)$ denotes the positive part of $a$. When either the accumulated liquid or solid volume exceeds the volume of one lattice cell, a link-wise migration step is performed. During melting, the excess liquid volume accumulated at a boundary liquid node is transferred to the neighbouring solid node along the links associated with a local interface motion, causing the solid node to start melting. Conversely, during freezing, the excess solid volume accumulated at a boundary solid node is transferred to the neighbouring liquid node along the links associated with the interface motion, causing the liquid node to start freezing. The link-wise migration variable $m_{i,j}$ is defined along each orthogonal link of each boundary fluid node as
\begin{equation}
m_{i,j}
=
\begin{cases}
\displaystyle
A_2 \,
\left(V_{l,j}(t)-1\right) \, / \, \left(V_{l,j}(t)-V_{l,j}(t-\delta_t)\right),
&
V_{l,j}(t)>1,
\\[2.0ex]
\displaystyle
A_2 \,
\left(V_{s,j}(t)-1\right) \, / \, \left(V_{s,j}(t)-V_{s,j}(t-\delta_t)\right),
&
V_{s,j}(t)>1,
\\[2.0ex]
0,
&
\text{otherwise},
\end{cases}
\qquad
i\in\mathcal{I}_j^\Gamma,
\quad
j\in\Omega_s\cup\Omega_l.
\label{eq:vop_directional_migration}
\end{equation}
The factor multiplying $A_2$ represents the fraction of the last phase-volume increment that exceeds the unit cell volume. Whenever $m_{i,j}$ is non-zero, the corresponding excess phase content is migrated to the neighbouring node, and the saturated volume at the original node is reset to unity. The updated values of $V_{l,j}$ and $V_{s,j}$ are then used to compute the liquid volume fraction of each node:
\begin{equation}
\alpha_j(t)
=
\begin{cases}
1+V_{l,j}(t)-V_{s,j}(t),
&
\alpha_j(0)=0,
\qquad
j\in\Omega_s,
\\[1.0ex]
V_{l,j}(t)-V_{s,j}(t),
&
\alpha_j(0)=1,
\qquad
j\in\Omega_l.
\end{cases}
\label{eq:liquid_fraction_alpha}
\end{equation}
A freezing liquid node is converted into solid when $\alpha_j(t)\leq0$, whereas a melting solid node is converted into liquid when $\alpha_j(t)\geq1$. During the phase switch, $V_{l,j}$ and $V_{s,j}$ are reassigned according to the state of the newly created node. Particular care is required when a solid node is converted into a liquid node. In this case, the concentration of the newly liquid node is reconstructed by linear interpolation or extrapolation, depending on the interface position.

This link-based approach extends the VOP model by allowing the magnitude and direction of interface propagation to be determined locally and independently along each solid--fluid link, according to the corresponding link-wise interfacial rates.

\subsection{Pore-scale temperature--concentration coupling}
\label{subsec:temperature_concentration_coupling}
The temperature and concentration fields are finally coupled to reproduce
phase-change dynamics in mushy layers through the interface kinetic relation
introduced in Eq.~\eqref{eq:interface_kinetics}. To express the interface kinetics in dimensionless form, temperature and concentration are nondimensionalized as
\begin{equation}
\theta
=
\frac{T-T_0}{T_{\max}-T_0},
\qquad
c
=
\frac{C-C_{\mathrm{eq}}(T_0)}
     {C_{\mathrm{eq}}(T_{\max})-C_{\mathrm{eq}}(T_0)} .
\label{eq:nondim_theta_c}
\end{equation}
Introducing $\Delta T=T_{\max}-T_0$, the linear liquidus relation gives
$T-T_{\mathrm{eq}}(C)=\Delta T(\theta-c)$. The dimensional kinetic relation in Eq.~\eqref{eq:interface_kinetics} therefore becomes
\begin{equation}
\mathrm{d}\widetilde{R} / \mathrm{d}t
=
B\Delta T \,
[ \,
\theta(\widetilde{R})-c(\widetilde{R})
\, ].
\label{eq:kinetic_dim_theta_c}
\end{equation}
The interface position and time are then nondimensionalized using the characteristic length $L$ (the initial geometry diameter) and the diffusive time scale $L^2/D$, 
\begin{equation}
\widetilde{R}'=\widetilde{R} / L ,
\qquad
t'= D \, t /L^2.
\label{eq:nondim_R_t}
\end{equation}
Substitution into Eq.~\eqref{eq:kinetic_dim_theta_c} yields $
\mathrm{d}\widetilde{R}'/ \mathrm{d}t'
=
(B\Delta T L / D)
[ \,
\theta(\widetilde{R}')-c(\widetilde{R}')
\, ]
=
\mathrm{Da}
[ \,
\theta(\widetilde{R}')-c(\widetilde{R}')
\, ],$ where $\mathrm{Da}=B\Delta T L/D$ is the kinetic Damköhler number. The dimensional interface velocity can be written as a function of $\mathrm{Da}$ as
\begin{equation}
\mathrm{d}\widetilde{R} / \mathrm{d}t
=
(D/L) \, \mathrm{Da} \,
[ \,
\theta(\widetilde{R})-c(\widetilde{R})
\, ]
=
k_r
[ \,
\theta(\widetilde{R})-c(\widetilde{R})
\, ],
\label{eq:kinetic_dimensional_Da}
\end{equation}
where the dimensional kinetic coefficient $k_r=(D/L) \, \mathrm{Da}=B\Delta T$ has units $[k_r]=\mathrm{m\,s^{-1}}$. 

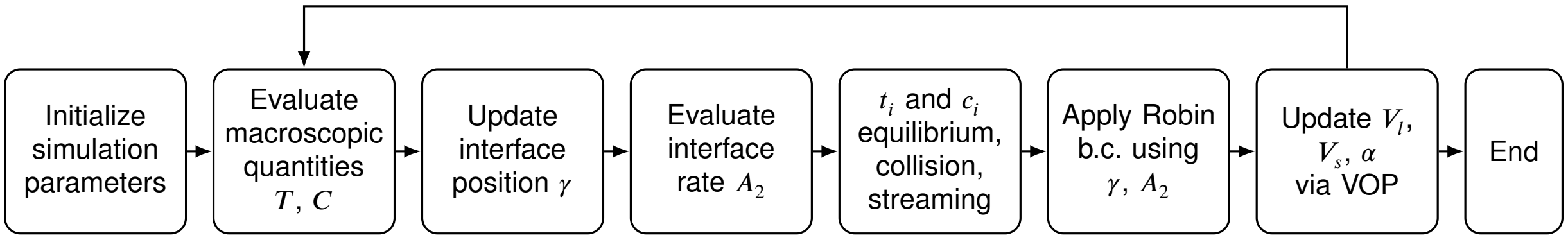
\begin{figure}[pos=htbp]
    \centering
    \resizebox{0.98\textwidth}{!}{%
        \begin{tikzpicture}[
    font=\fontsize{11}{13}\selectfont\sffamily\sansmath,
    >=Latex,
    node distance=0.35cm,
    box/.style={
        draw=black,
        fill=white,
        line width=0.8pt,
        rounded corners=6pt,
        minimum height=2.25cm,
        text width=2.20cm,
        align=center,
        inner xsep=4pt,
        inner ysep=5pt
    },
    smallbox/.style={
        draw=black,
        fill=white,
        line width=0.8pt,
        rounded corners=6pt,
        minimum height=2.25cm,
        text width=1.05cm,
        align=center,
        inner xsep=4pt,
        inner ysep=5pt
    },
    arrow/.style={
        ->,
        black,
        line width=0.8pt
    }
]

% =========================================================
% Nodes: single horizontal row
% =========================================================

\node[box] (init)
{Initialize\\ simulation\\ parameters};

\node[box, right=of init] (macro)
{Evaluate\\ macroscopic\\ quantities $T$, $C$};

\node[box, right=of macro] (gamma)
{Update\\ interface\\ position $\gamma$};

\node[box, right=of gamma] (rate)
{Evaluate\\ interface \\ rate $A_2$};

\node[box, right=of rate] (lbm)
{$t_i$ and $c_i$\\ equilibrium,\\ collision, \\ streaming};

\node[box, right=of lbm] (robin)
{Apply Robin\\ b.c. using\\ $\gamma$, $A_2$};

\node[box, right=of robin] (vop)
{Update $V_{l}$,\\ $V_{s}$, $\alpha$\\ via VOP};

\node[smallbox, right=of vop] (end)
{End};

% =========================================================
% Main arrows
% =========================================================

\draw[arrow] (init) -- (macro);
\draw[arrow] (macro) -- (gamma);
\draw[arrow] (gamma) -- (rate);
\draw[arrow] (rate) -- (lbm);
\draw[arrow] (lbm) -- (robin);
\draw[arrow] (robin) -- (vop);
\draw[arrow] (vop) -- (end);

% =========================================================
% Iteration loop
% =========================================================

\draw[arrow]
(vop.north)
-- ++(0,0.85)
-| (macro.north);

\end{tikzpicture}%
    }
\caption{Algorithmic workflow for temperature, solute concentration, and interface evolution. For clarity, selected subscripts are omitted in the schematic.}
    \label{fig:coupled_algorithm}
\end{figure}

Within the present LBM framework, Eq.~\eqref{eq:kinetic_dimensional_Da} is evaluated for each orthogonal interfacial link $i=1 \dots N_O$ and implemented in two forms. The first is a nodal isotropic form, Eq.~\eqref{eq:rate_nodal}, introduced for simplicity and reduced computational cost. In this case, the dimensionless temperature and concentration are evaluated at the boundary node $\mathbf{x}$, and the same value of $\mathrm{d}\widetilde{R}_i/\mathrm{d}t$ is assigned to all boundary links connected to that node. The second is a link-based form, Eq.~\eqref{eq:rate_link}, used when the directional character of the local phase-change response must be resolved. In this formulation, the kinetic relation is evaluated separately at each interfacial link, using the corresponding interfacial temperature and concentration, $\theta_{b,i}$ and $c_{b,i}$. Different links connected to the same boundary node can therefore undergo different melting or freezing rates. The boundary temperature is obtained by first-order interpolation or extrapolation, whereas the boundary concentration is computed from Eq.~\eqref{eq:huang_cb_gradient}.
\begin{empheq}[left={\displaystyle \frac{\mathrm{d}\widetilde{R}_i}{\mathrm{d}t}= A_2 =
\empheqlbrace}]{align}
&k_r
\left[
\theta(\mathbf{x},t)-c(\mathbf{x},t)
\right],
\quad
i=1,\ldots,N_O,
\label{eq:rate_nodal}
\\[4pt]
&k_r
\left[
\theta_{b,i}(t)-c_{b,i}(t)
\right],
\quad
i=1,\ldots,N_O.
\label{eq:rate_link}
\end{empheq}
The implemented algorithm is schematically illustrated in Fig.~\ref{fig:coupled_algorithm}. After initialization, the macroscopic fields are recomputed at each time step from the distribution functions, with the temperature evaluated according to Eq.~\eqref{eq:eql_t_eval} and the concentration recovered as
$C(\mathbf{x},t)=\sum_{i=1}^{N_Q} c_i(\mathbf{x},t)$. The variable $\gamma_{i,j}$ denotes the normalized interface position associated with boundary fluid node $j$ along the $i$-th orthogonal interfacial link. From the definition of nodes' liquid phase fraction, $\gamma_{i,j}$ is computed relative to the initial halfway position $\gamma_0 = 0.5$ as
\begin{equation}
\gamma_{i,j}(t)
=
\alpha_j(t)
+
\alpha_{j+1}(t)
-
0.5,
\qquad
i=1,\ldots,N_O,
\qquad
j\in\Omega_l,
\label{eq:gamma_update}
\end{equation}
where node $j$ is located at $\mathbf{x}$ and $j+1$ denotes its neighbouring node located at $\mathbf{x}+\mathbf{e}_i$. The interface propagation rate is then evaluated from the local temperature and concentration fields according to Eqs.~\eqref{eq:rate_nodal} or~\eqref{eq:rate_link}. The temperature and concentration distribution functions are subsequently advanced via LBM equilibrium, collision, and streaming steps. At the solid--liquid interface, solute transport is governed by the Robin boundary condition, implemented through either the original or the extended boundary scheme. Finally, the VOP evolution updates the liquid fractions $\alpha_j$ using the local interface propagation rate $A_2$. These fractions are then used to compute the interface position $\gamma_{i,j}$ at the next time step. Overall, the coupling between the solute boundary treatment and the VOP evolution is mediated by $\gamma_{i,j}$ and $A_2$.

\section{Results}
\label{sec:results}
\subsection{Thermal-transport simulations}
\label{subsec:thermal_overview_sim}
In this section, the thermal formulation is assessed independently of solute transport by solving only the temperature field. Two thermal-only cases are considered: the Stefan-problem validation in Sec.~\ref{subsubsec:stefan_problem} and the idealized brine-pore simulation in Sec.~\ref{subsubsec:thermal_pore}.
\subsubsection{Stefan-problem validation}
\label{subsubsec:stefan_problem}
A one-dimensional semi-infinite Stefan diffusion problem is considered. The phase-change temperature is assumed constant and it is denoted by $T_f$. In the original model~\cite{feart}, phase change is described through the evolution of a continuous liquid fraction $\phi(\mathbf{x},t)$, whose rate is approximated as $\phi(\mathbf{x},t)$ as $
\partial \phi / \partial t (\mathbf{x})
\simeq
\left[\phi(\mathbf{x},t+\delta_t)-\phi(\mathbf{x},t)\right] / \delta_t
=
C_p(\mathbf{x})\left[T(\mathbf{x},t)-T_f\right] /(L \,\delta_t).$ In the present work, the same thermodynamic information -- i.e. the liquid-fraction rate $\partial \phi / \partial t$ -- is used instead to compute the normal displacement of the solid--liquid interface, adapting the enthalpy-based estimate to the sharp-interface VOP formulation. Denoting by $\Delta\widetilde{R}$ the signed normal displacement over the finite time interval $\delta_t$, a first-order expansion along the interface normal gives
$\Delta\phi \simeq \nabla\phi\cdot(\Delta\widetilde{R} \, \mathbf{n})
= (\partial\phi/\partial n)\Delta\widetilde{R}$. Therefore, the directional interface velocity
$A_2=\mathrm{d}\widetilde{R}_i/\mathrm{d}t$
is evaluated as
\begin{equation}
A_2
=
f_i \, \partial\phi/\partial t,
\qquad
f_i
=
\left[(\partial\phi/\partial n)_i\right]^{-1},
\qquad
i=1,\ldots,N_O.
\label{eq:thermal_interface_rate}
\end{equation}
In the thermal-only tests Eq.~\eqref{eq:thermal_interface_rate} replaces the thermosolutal kinetic relations in Eq.~\eqref{eq:rate_nodal},~\eqref{eq:rate_link}. The validation is performed in a two-dimensional rectangular geometry discretized on a $200\times10$ lattice. This domain is initially liquid, except for a single solid column at the left boundary, which represents the frozen region from which the solidification front propagates. Physical variables marked with an asterisk are expressed in lattice units. The initial liquid temperature is set to $T_i^*=2$, the phase-change temperature to $T_f^*=1$, and the cold boundary, together with the initial solid column, is kept at $T_w^*=0$. The latent heat is $L^*=0.5$, while the specific heat capacities are $c_{p,s}^*=0.42$ and $c_{p,l}^*=1$. To assess the VOP formulation under different thermal-transport contrasts, the diffusivity ratio $\kappa_s^*/\kappa_l^*$ is varied while keeping $\kappa_s^*$ fixed.
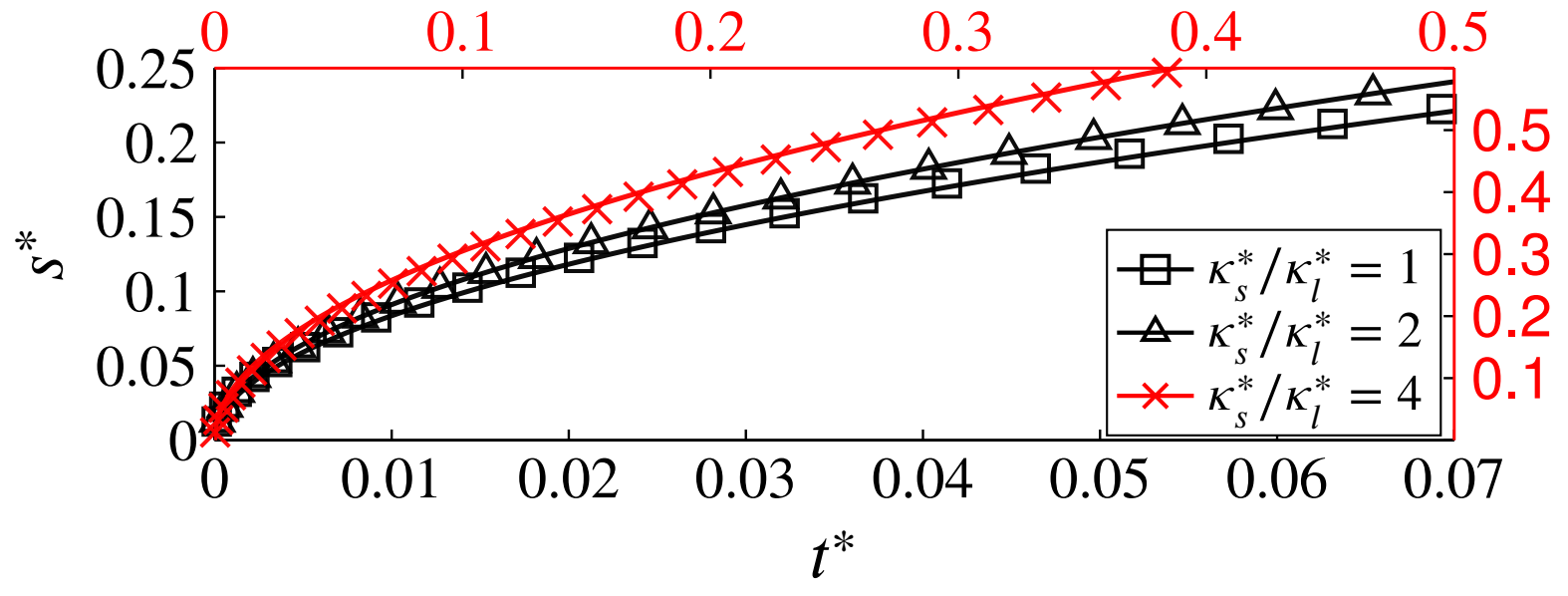
\begin{figure}[pos=htbp]
\centering
\setlength\figurewidth{0.5\textwidth}
\setlength\figureheight{0.15\textwidth}
\begin{tikzpicture}[font=\sffamily\sansmath]

% ============================================================
% Main black axis: alpha_s / alpha_l = 1, 2
% ============================================================

\begin{axis}[
name=mainaxis,
width=\figurewidth,
height=\figureheight,
scale only axis,
xmin=0,
xmax=0.07,
ymin=0,
ymax=0.25,
axis x line*=bottom,
axis y line*=left,
axis line style={black},
xtick pos=bottom,
ytick pos=left,
xticklabel pos=bottom,
yticklabel pos=left,
tick align=inside,
scaled ticks=false,
xtick={0,0.01,0.02,0.03,0.04,0.05,0.06,0.07},
ytick={0,0.05,0.10,0.15,0.20,0.25},
xticklabel style={
    /pgf/number format/fixed,
    /pgf/number format/precision=2,
    font=\fontsize{11}{8}\selectfont\sffamily\sansmath
},
yticklabel style={
    /pgf/number format/fixed,
    /pgf/number format/precision=2,
    font=\fontsize{11}{8}\selectfont\sffamily\sansmath
},
xlabel={$t^*$},
ylabel={$s^*$},
label style={
    font=\fontsize{13}{8}\selectfont\sffamily\sansmath
},
grid=none,
legend style={
    font=\fontsize{9.5}{8}\selectfont\sffamily\sansmath,
    anchor=south east,
    at={(0.99,0.01)},
    draw=black,
    fill=white,
    inner sep=1.5pt,
    row sep=0pt
},
legend image post style={scale=0.85},
legend cell align=left,
clip=true,
]

% alpha_s / alpha_l = 1
\addplot[
black,
solid,
line width=0.95pt,
opacity=0.95,
forget plot
]
table[x index=0,y index=1]
{figures/feart_stefan_test/anal_ratio1.dat};

\addplot[
black,
only marks,
mark=square,
mark size=2.5pt,
mark options={
    solid,
    draw=black,
    fill=white,
    line width=0.75pt
},
opacity=0.95,
forget plot
]
table[x index=0,y index=1]
{figures/feart_stefan_test/vop_ratio1.dat};

\addlegendimage{
black,
solid,
line width=0.95pt,
mark=square,
mark size=3.0pt,
mark options={
    solid,
    draw=black,
    fill=white,
    line width=0.75pt
}
}
\addlegendentry{$\kappa_s^*/\kappa_l^*=1$};

% alpha_s / alpha_l = 2
\addplot[
black,
solid,
line width=0.95pt,
opacity=0.95,
forget plot
]
table[x index=0,y index=1]
{figures/feart_stefan_test/anal_ratio2.dat};

\addplot[
black,
only marks,
mark=triangle,
mark size=3.5pt,
mark options={
    solid,
    draw=black,
    fill=white,
    line width=0.75pt
},
opacity=0.95,
forget plot
]
table[x index=0,y index=1]
{figures/feart_stefan_test/vop_ratio2.dat};

\addlegendimage{
black,
solid,
line width=0.95pt,
mark=triangle,
mark size=3.25pt,
mark options={
    solid,
    draw=black,
    fill=white,
    line width=0.75pt
}
}
\addlegendentry{$\kappa_s^*/\kappa_l^*=2$};

\addlegendimage{
red,
solid,
line width=0.95pt,
mark=x,
mark size=3.8pt,
mark options={
    solid,
    draw=red,
    fill=white,
    line width=0.75pt
}
}
\addlegendentry{$\kappa_s^*/\kappa_l^*=4$};

\end{axis}

% ============================================================
% Secondary red axis: alpha_s / alpha_l = 4
% ============================================================

\begin{axis}[
name=redaxis,
at={(mainaxis.south west)},
anchor=south west,
width=\figurewidth,
height=\figureheight,
scale only axis,
xmin=0,
xmax=0.5,
ymin=0,
ymax=0.6,
axis x line*=top,
axis y line*=right,
axis line style={red},
xtick pos=top,
ytick pos=right,
xticklabel pos=top,
yticklabel pos=right,
tick align=inside,
scaled ticks=false,
xtick={0,0.1,0.2,0.3,0.4,0.5},
ytick={0,0.1,0.2,0.3,0.4,0.5,0.6},
yticklabels={,0.1,0.2,0.3,0.4,0.5,},
xticklabel style={
    /pgf/number format/fixed,
    /pgf/number format/precision=1,
    red,
    font=\fontsize{11}{8}\selectfont\sffamily\sansmath
},
yticklabel style={
    /pgf/number format/fixed,
    /pgf/number format/precision=1,
    red,
    font=\fontsize{11}{8}\selectfont\sffamily\sansmath
},
xlabel style={
    red,
    font=\fontsize{13}{8}\selectfont\sffamily\sansmath
},
ylabel style={
    red,
    font=\fontsize{13}{8}\selectfont\sffamily\sansmath
},
grid=none,
clip=true,
]

\addplot[
red,
solid,
line width=0.95pt,
opacity=0.95,
forget plot
]
table[x index=0,y index=1]
{figures/feart_stefan_test/anal_ratio4.dat};

\addplot[
red,
only marks,
mark=x,
mark size=3.8pt,
mark options={
    solid,
    draw=red,
    fill=white,
    line width=0.75pt
},
opacity=0.95,
forget plot
]
table[
    x index=0,
    y index=1,
    each nth point=2
]
{figures/feart_stefan_test/vop_ratio4.dat};

\end{axis}

\end{tikzpicture}\vspace{-0.25em}
\caption{Evolution of the interface for different thermal-diffusivity ratios. Symbols denote the VOP results and lines the Neumann solutions.}
\label{fig:stef_test_interf}
\end{figure}
Figure~\ref{fig:stef_test_interf} compares the solid--liquid interface evolution predicted by the VOP formulation with the Neumann solution for $\kappa_s/\kappa_l=1$, $2$, and $4$. The analytical interface position is given by $s(t)=2\lambda\sqrt{\kappa_s t}$, where $\lambda$ satisfies the corresponding Stefan transcendental equation~\cite{Ozisik1993}. Results are reported in dimensionless form using $t^*=\kappa_s t/H^2$ and $s^*=s/H$, with $H=200$. The numerical results reproduce the expected $s\propto\sqrt{t}$ behaviour. For $\kappa_s^*/\kappa_l^*=1$, the mean and maximum relative errors are $1.32\%$ and $2.15\%$, respectively. These values decrease to $0.141\%$ and $0.210\%$ for $\kappa_s^*/\kappa_l^*=2$, and become $1.733\%$ and $4.334\%$ for $\kappa_s^*/\kappa_l^*=4$. Overall, the errors remain of the same order as those reported in~\cite{feart,Gao2001}, confirming that the VOP algorithm reproduces the one-dimensional thermal phase-change dynamics within a conjugate heat-transfer framework.
\subsubsection{Thermal simulation of an idealized brine pore}
\label{subsubsec:thermal_pore}
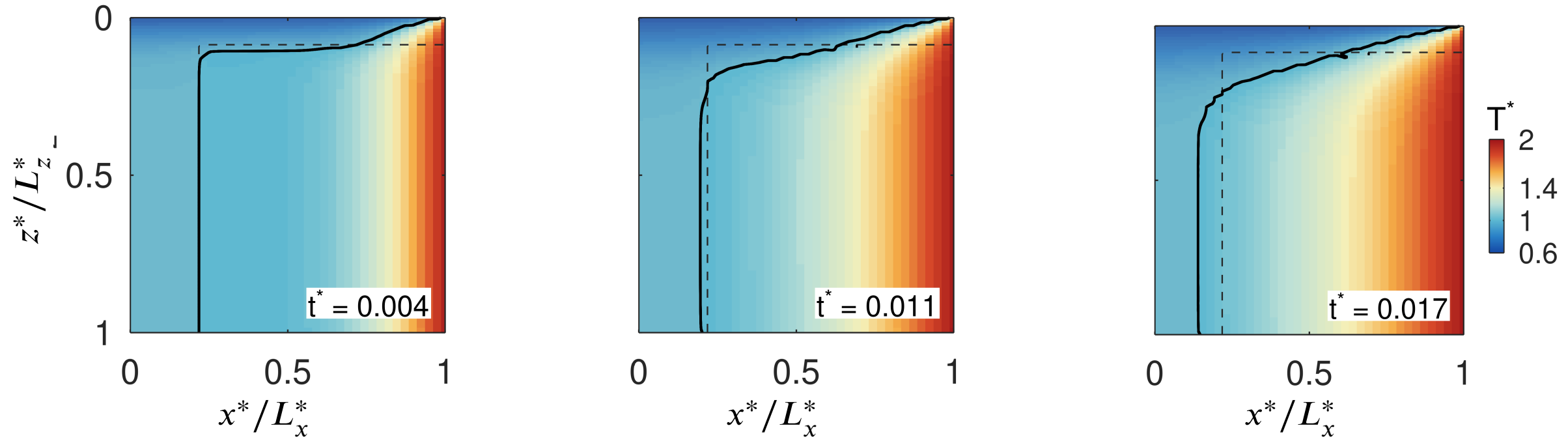
\begin{figure}[pos=htbp]
\centering
\setlength{\tabcolsep}{0pt}
\begin{tabular}{@{}c@{\hspace{-0.35cm}}c@{\hspace{-0.25cm}}c@{}}

% ============================================================
% LEFT PANEL: t* = 0.004
% ============================================================
\begin{overpic}[
    width=0.335\textwidth,
    trim=55 20 85 25,
    clip
]{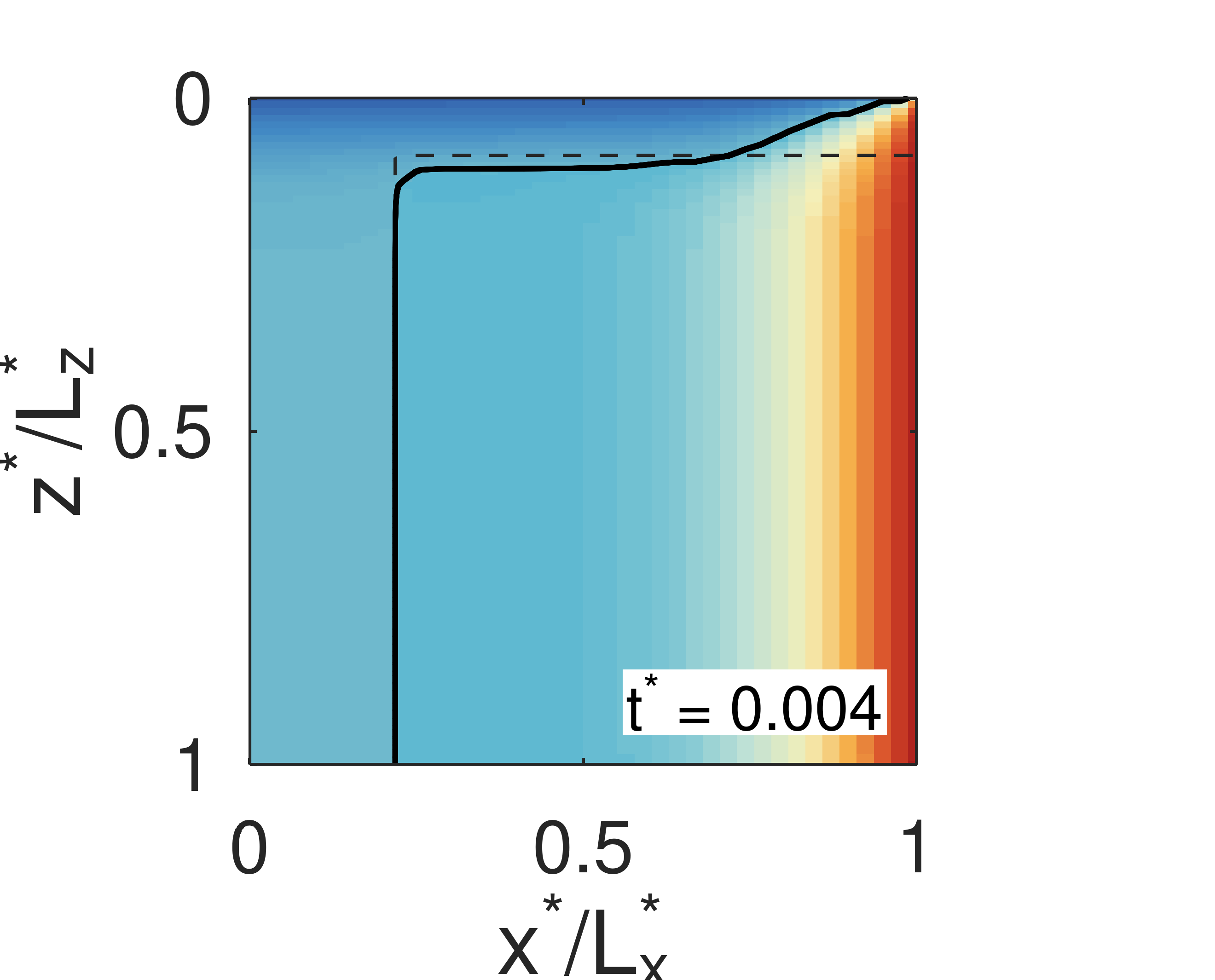}

\put(-5.0,24){%
    \color{white}\rule{1.8em}{7.2em}%
}

\put(-2.0,34){%
    \rotatebox{90}{\figaxislabel{z^*/L_z^*}}%
}

\put(25,1.0){%
    \begingroup
    \setlength{\fboxsep}{3pt}%
    \colorbox{white}{%
        \makebox[6.0em][c]{\figaxislabel{x^*/L_x^*}}%
    }%
    \endgroup
}
\end{overpic}
&
% ============================================================
% CENTER PANEL: t* = 0.011
% ============================================================
\begin{overpic}[
    width=0.335\textwidth,
    trim=55 20 85 25,
    clip
]{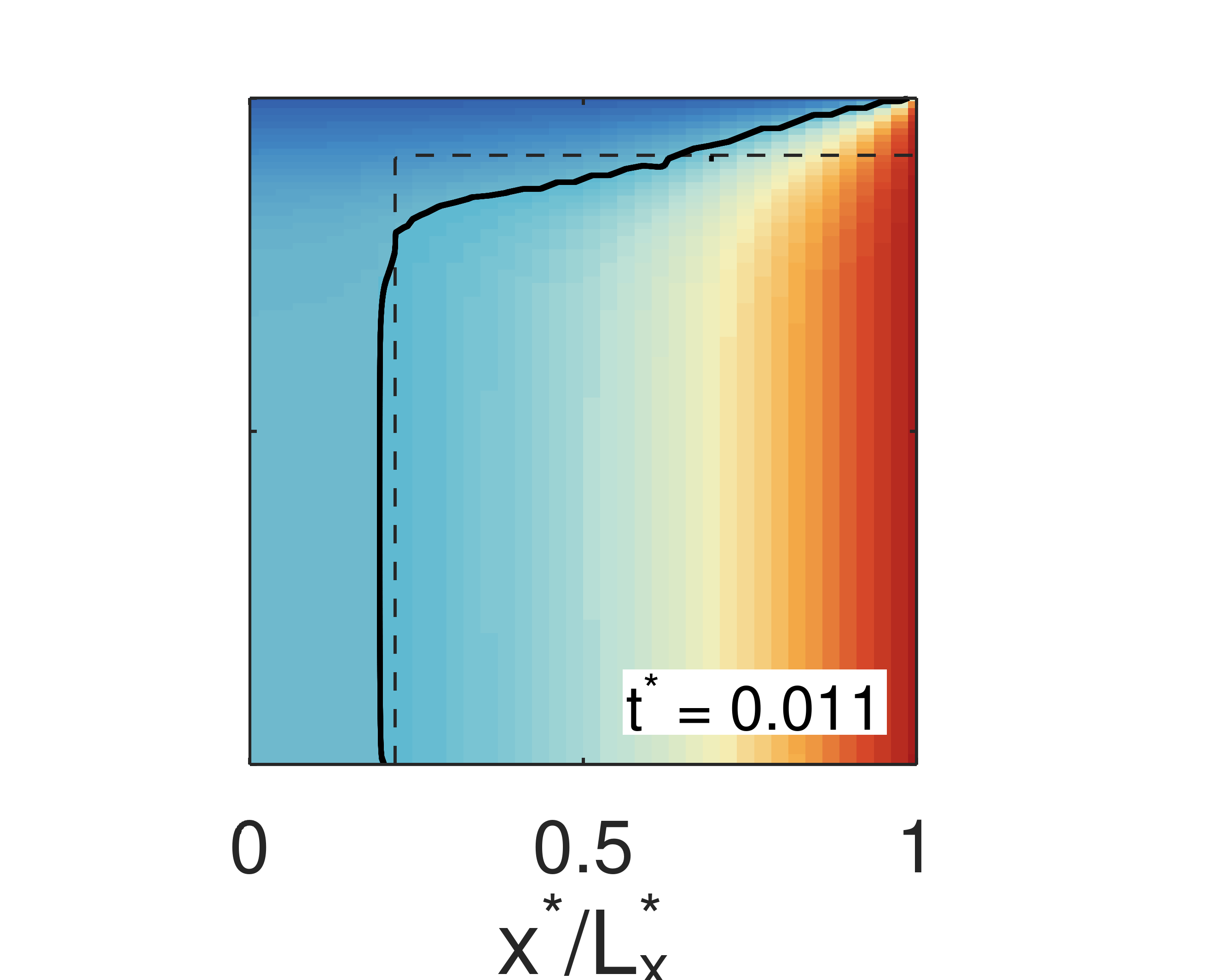}

\put(0.0,27.5){%
    \color{white}\rule{1.0em}{5.6em}%
}

\put(25,1.0){%
    \begingroup
    \setlength{\fboxsep}{3pt}%
    \colorbox{white}{%
        \makebox[6.0em][c]{\figaxislabel{x^*/L_x^*}}%
    }%
    \endgroup
}

\end{overpic}
&
% ============================================================
% RIGHT PANEL: t* = 0.017
% ============================================================
\begin{overpic}[
    width=0.335\textwidth,
    trim=55 20 35 25,
    clip
]{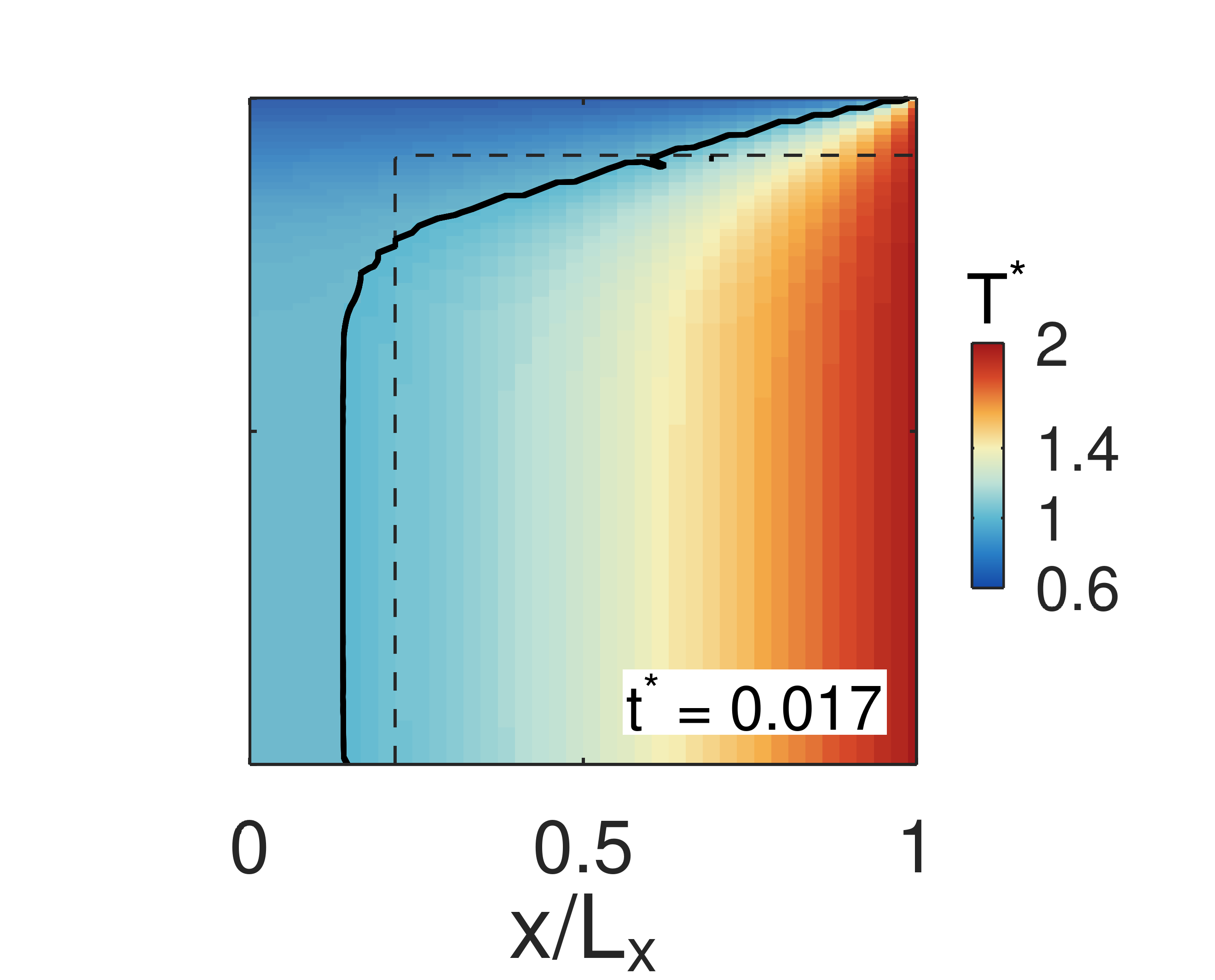}

\put(25,1.0){%
    \begingroup
    \setlength{\fboxsep}{3pt}%
    \colorbox{white}{%
        \makebox[6.0em][c]{\figaxislabel{x^*/L_x^*}}%
    }%
    \endgroup
}

\end{overpic}
\end{tabular}
\vspace{0.2em}
\caption{Evolution of the temperature field and solid--liquid interface in the idealized brine pore. The solid black line denotes the evolving interface, while the dashed line indicates its initial position.}
\label{fig:pore_temperature_evolution}
\end{figure}

Following the one-dimensional Stefan validation, a two-dimensional pore-scale configuration is considered to assess the response of the present VOP algorithm to spatially heterogeneous thermal forcing.
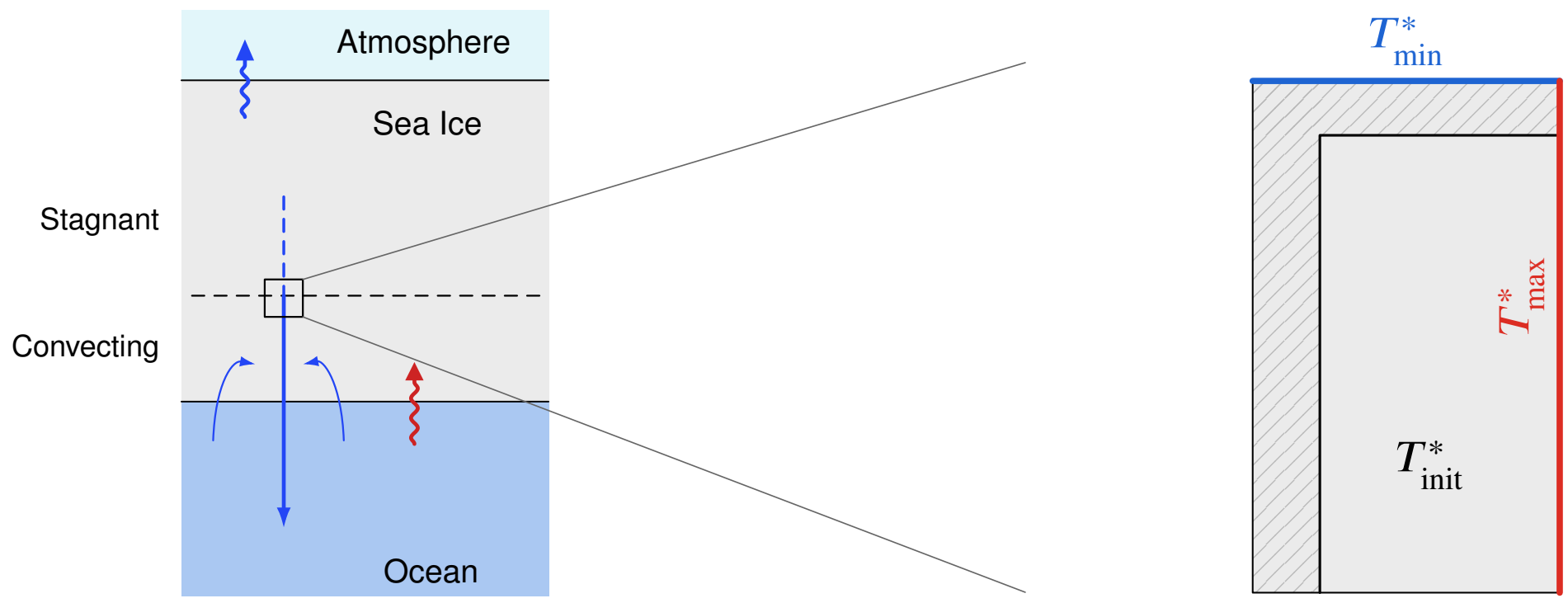
\begin{figure}[pos=htbp]
\centering
\newlength{\porefigheight}
\setlength{\porefigheight}{4.7cm}
\makebox[\textwidth][c]{%
\begin{subfigure}[t]{0.3\textwidth}
\centering
\resizebox{!}{\porefigheight}{%
    \begin{tikzpicture}[
font=\sffamily\sansmath,
>=Latex,
line cap=round,
line join=round
]

% ============================================================
% Dimensions
% ============================================================

\def\W{5.2}

\def\Yzero{0}
\def\Yocean{2.75}
\def\Yinterface{4.25}
\def\YiceTop{7.30}
\def\Ytop{8.30}

% ============================================================
% Colours
% ============================================================

\definecolor{atmospherefill}{RGB}{226,246,250}
\definecolor{seaicefill}{RGB}{235,235,235}
\definecolor{oceanfill}{RGB}{170,200,242}
\definecolor{flowblue}{RGB}{35,70,245}
\definecolor{heatred}{RGB}{205,35,35}

% ============================================================
% Background regions
% ============================================================

\fill[oceanfill]
(0,\Yzero) rectangle (\W,\Yocean);

\fill[seaicefill]
(0,\Yocean) rectangle (\W,\YiceTop);

\fill[atmospherefill]
(0,\YiceTop) rectangle (\W,\Ytop);

% ============================================================
% Horizontal boundaries
% ============================================================

\draw[black, line width=0.65pt]
(0,\YiceTop) -- (\W,\YiceTop);

\draw[black, line width=0.65pt]
(0,\Yocean) -- (\W,\Yocean);

\draw[
black,
dashed,
dash pattern=on 5pt off 4pt,
line width=0.75pt
]
(0.15,\Yinterface) -- (\W,\Yinterface);

% ============================================================
% Main labels
% ============================================================

\node[
font=\fontsize{13}{15}\selectfont\sffamily\sansmath
] at (0.66*\W,7.80)
{Atmosphere};

\node[
font=\fontsize{13}{15}\selectfont\sffamily\sansmath
] at (0.67*\W,6.70)
{Sea Ice};

\node[
font=\fontsize{13}{15}\selectfont\sffamily\sansmath
] at (0.68*\W,0.35)
{Ocean};

% ============================================================
% Brine channels
% ============================================================

% Left channel: dashed part in stagnant layer
\draw[
flowblue,
dashed,
dash pattern=on 5pt off 4pt,
line width=1.15pt
]
(1.45,5.65) -- (1.45,\Yinterface);

% Left channel: downward brine flow
\draw[
flowblue,
line width=1.55pt,
-{Latex[length=3.0mm,width=2.0mm]}
]
(1.45,\Yinterface) -- (1.45,0.95);

% ============================================================
% Brine-channel annotation
% ============================================================

%\node[
%anchor=west,
%align=left,
%font=\fontsize{12}{14}\selectfont\sffamily\sansmath
%] at (2.05,5.35)
%{Brine\\channel};

%\draw[
%black,
%dashed,
%line width=0.65pt,
%-{Latex[length=2.2mm,width=1.5mm]}
%]
%(2.02,4.98) -- (1.53,4.50);

% ============================================================
% Convective circulation
% Only two arrows around the left brine channel
% ============================================================

% Left curved arrow, moved closer to the brine channel
\draw[
flowblue,
line width=0.75pt,
-{Latex[length=2.2mm,width=1.5mm]}
]
(0.45,2.20)
.. controls (0.48,3.00) and (0.72,3.34) ..
(1.05,3.28);

% Right curved arrow, moved closer to the brine channel
\draw[
flowblue,
line width=0.75pt,
-{Latex[length=2.2mm,width=1.5mm]}
]
(2.30,2.20)
.. controls (2.27,3.00) and (2.02,3.34) ..
(1.72,3.28);

% ============================================================
% Incoming heat flux
% Crosses the ocean–ice interface
% ============================================================

% Wavy heat-flux line
\draw[
heatred,
line width=1.55pt,
decorate,
decoration={
snake,
amplitude=0.48mm,
segment length=2.8mm
}
]
(3.30,2.15) -- (3.30,3.09);

% Independent triangular arrowhead
\fill[heatred]
(3.30,3.31)
-- (3.17,3.05)
-- (3.43,3.05)
-- cycle;

% ============================================================
% Outgoing heat flux
% Crosses the ice–atmosphere interface
% ============================================================

% Wavy heat-flux line
\draw[
flowblue,
line width=1.55pt,
decorate,
decoration={
snake,
amplitude=0.48mm,
segment length=2.8mm
}
]
(0.90,6.78) -- (0.90,7.60);

% Independent triangular arrowhead
\fill[flowblue]
(0.90,7.88)
-- (0.77,7.58)
-- (1.03,7.58)
-- cycle;

% ============================================================
% Layer labels
% ============================================================

\node[
anchor=east,
font=\fontsize{12}{14}\selectfont\sffamily\sansmath
] at (-0.20,5.30)
{Stagnant};

\node[
anchor=east,
font=\fontsize{12}{14}\selectfont\sffamily\sansmath
] at (-0.20,3.50)
{Convecting};

% ============================================================
% Zoom box around the brine channel
% ============================================================

\draw[
    black,
    line width=0.75pt
]
    (1.18,3.95) rectangle (1.72,4.48);

% ============================================================
% Zoom connector lines
% ============================================================

% Upper connector:
% zoom-box upper-right corner -> numerical-scheme upper-left corner
\draw[
    overlay,
    black!60,
    line width=0.55pt
]
    (1.72,4.48) -- (11.95,7.55);

% Lower connector:
% zoom-box lower-right corner -> numerical-scheme lower-left corner
\draw[
    overlay,
    black!60,
    line width=0.55pt
]
    (1.72,3.95) -- (11.95,0.05);

\end{tikzpicture}%
}
\vspace{0.5em}
\caption{Sea ice schematic.}
\label{fig:sea_ice_convection}
\end{subfigure}%
\hspace{0.25\textwidth}%
\begin{subfigure}[t]{0.3\textwidth}
\centering
\resizebox{!}{\porefigheight}{%
    \begin{tikzpicture}[
font=\sffamily\sansmath,
line cap=round,
line join=round
]

% ============================================================
% Geometry
% ============================================================

\def\W{4.8}
\def\H{8.0}

% Solid--liquid interface
\def\Xinterface{1.05}
\def\Zinterface{0.85}

% Horizontal interface measured from the bottom in TikZ coordinates
\pgfmathsetmacro{\Yinterface}{\H-\Zinterface}

% ============================================================
% Colours
% ============================================================

% Same light grey used for sea ice in the left-hand schematic
\definecolor{Tinitcolor}{RGB}{235,235,235}

% Imposed thermal boundaries
\definecolor{Tcoldcolor}{RGB}{30,100,210}
\definecolor{Thotcolor}{RGB}{220,45,35}

% ============================================================
% Uniform initial temperature field
% ============================================================

\fill[Tinitcolor]
    (0,0) rectangle (\W,\H);

% ============================================================
% Solid-phase hatching
% Same temperature colour, but different material
% ============================================================

% Left solid wall
\fill[
    pattern={
        Lines[
            angle=45,
            distance=3.0pt,
            line width=0.25pt
        ]
    },
    pattern color=black!28
]
    (0,0) rectangle (\Xinterface,\H);

% Upper solid layer
\fill[
    pattern={
        Lines[
            angle=45,
            distance=3.0pt,
            line width=0.25pt
        ]
    },
    pattern color=black!28
]
    (0,\Yinterface) rectangle (\W,\H);

% ============================================================
% Outer boundary
% ============================================================

\draw[
    black,
    line width=0.75pt
]
    (0,0) rectangle (\W,\H);

% ============================================================
% Solid--liquid interface
% Vertical segment extends downward from the upper solid layer
% ============================================================

\draw[
    black,
    line width=1.15pt
]
    (\Xinterface,0)
    -- (\Xinterface,\Yinterface)
    -- (\W,\Yinterface);

% ============================================================
% Thermal boundary conditions
% ============================================================

% Cold upper boundary
\draw[
    Tcoldcolor,
    line width=2.8pt
]
    (0,\H) -- (\W,\H);

% Hot right boundary
\draw[
    Thotcolor,
    line width=2.8pt
]
    (\W,0) -- (\W,\H);

% ============================================================
% Boundary-condition labels
% ============================================================

% Cold upper boundary label
\node[
    anchor=south,
    text=Tcoldcolor,
    font=\fontsize{20}{22}\selectfont\sffamily\sansmath
] at (0.50*\W,\H+0.14)
{$T_{\min}^{*}$};

% Hot right boundary label
\node[
    anchor=center,
    text=Thotcolor,
    rotate=90,
    font=\fontsize{20}{22}\selectfont\sffamily\sansmath
] at (\W-0.6,0.58*\H)
{$T_{\max}^{*}$};

% ============================================================
% Uniform initial-temperature label
% ============================================================

\node[
    font=\fontsize{20}{22}\selectfont\sffamily\sansmath
] at (0.58*\W,0.25*\H)
{$T_{\mathrm{init}}^{*}$};

\end{tikzpicture}%
}
\vspace{0.2em}
\caption{Initial and boundary conditions.}
\label{fig:pore_het_initial_configuration}
\end{subfigure}%
}
\vspace{-0.4em}
\caption{Conceptual sea ice configuration and corresponding pore-scale setup under heterogeneous thermal forcing.}
\label{fig:sea_ice_and_pore_configuration}
\end{figure}
The upper portion of a brine channel is simulated, with the solid--liquid interface subjected to thermal gradients induced by thermal boundary forcing along the vertical and horizontal directions. The setup could be motivated by the environment of sea ice during winter, where the ice cover is cooled by the atmosphere while receiving heat from the underlying ocean, as depicted in Fig.~\ref{fig:sea_ice_convection}. As shown in Fig.~\ref{fig:pore_het_initial_configuration}, only one half of a symmetric two-dimensional brine pore is simulated to reduce the computational cost. The domain is initially at the equilibrium temperature $T^*_{\mathrm{init}}=T^*_{\mathrm{f}}=1.0$. At $t^*=0$, a cold Dirichlet condition $T^*_{\min}=0.6$ is imposed at the upper boundary, whereas a hot Dirichlet condition $T^*_{\max}=2.0$ is imposed at the right boundary. The remaining external boundaries are treated as thermally insulating. To isolate the effect of heterogeneous thermal forcing on the VOP response, identical thermal properties are assigned to the two phases, with $c^*_{p,l}=c^*_{p,s}=1$ and $\kappa_l^*=\kappa_s^*=0.25$, while the latent heat is set to $L^*=0.5$. The local interface velocity is computed using the same enthalpy-based relation adopted in the Stefan validation, Eq.~\eqref{eq:thermal_interface_rate}. Figure~\ref{fig:pore_temperature_evolution} shows the resulting evolution of the temperature field and solid--liquid interface. Starting from the initial equilibrium state, the imposed boundary temperatures generate two competing thermal perturbations. Cooling from the upper boundary produces $T^*<T^*_{\mathrm{f}}$ and promotes local solidification, whereas heating from the right boundary produces $T^*>T^*_{\mathrm{f}}$ and drives melting. The interface therefore evolves in response to the spatially heterogeneous temperature field, illustrating how the VOP formulation translates directional thermal forcing into a locally non-uniform interface displacement.
\subsection{Solute-transport simulations}
\label{subsec:solut_sim}
In this section, the solute-transport formulation is assessed independently of the thermal field by solving only the concentration equation. The analysis includes a two-dimensional Robin-boundary concentration test, Sec.~\ref{subsec:solut_sim_2d}, diffusion in a ring geometry, Sec.~\ref{subsec:ring_diff}, and an idealized brine-channel simulation, Sec.~\ref{subsec:pore_solutal_diff}.
\subsubsection{Two-dimensional concentration test}
\label{subsec:solut_sim_2d}
\begin{figure}[pos=htbp]
\centering
\begin{subfigure}[t]{0.40\textwidth}
\centering
\setlength\figurewidth{0.84\linewidth}
\setlength\figureheight{0.46\linewidth}
\begin{tikzpicture}[font=\sffamily\sansmath]

\begin{axis}[
width=\figurewidth,
height=\figureheight,
scale only axis,
xmin=0,
xmax=1,
ymin=0,
ymax=1,
xlabel={$x^*/L_x^*$},
ylabel={$y^*/L_y^*$},
grid=none,
axis line style={black},
tick align=inside,
tick pos=both,
xtick={0,0.5,1},
ytick={0,0.5,1},
xticklabels={$0$,$0.5$,$1$},
yticklabels={$0$,$0.5$,$1$},
tick label style={
    font=\fontsize{11}{8}\selectfont\sffamily\sansmath
},
label style={
    font=\fontsize{13}{8}\selectfont\sffamily\sansmath
},
xlabel style={
    font=\fontsize{13}{8}\selectfont\sffamily\sansmath
},
ylabel style={
    font=\fontsize{13}{8}\selectfont\sffamily\sansmath
},
colormap={dario}{
    rgb255=(0,0,255)
    rgb255=(255,0,255)
    rgb255=(255,0,0)
    rgb255=(255,220,0)
},
point meta min=0,
point meta max=10,
]

% Analytical contours: coloured solid lines
\addplot[
mesh,
solid,
line width=1.10pt,
point meta=\thisrowno{2},
unbounded coords=jump,
forget plot
]
table[
x expr=\thisrowno{0}/124,
y expr=\thisrowno{1}/100
] {figures/da48/contour_data_M04/contour_analytical_gamma_-0.40_NX125_NZ102.dat};

% Numerical contours: coloured dashed lines
\addplot[
mesh,
dash pattern=on 5pt off 3pt,
line width=0.95pt,
point meta=\thisrowno{2},
unbounded coords=jump,
forget plot
]
table[
x expr=\thisrowno{0}/124,
y expr=\thisrowno{1}/100
] {figures/da48/contour_data_M04/contour_numerical_gamma_-0.40_NX125_NZ102.dat};

% Numerical markers: coloured hollow triangles
\addplot[
scatter,
only marks,
scatter src=explicit,
scatter/use mapped color={draw=mapped color, fill=white},
point meta=\thisrowno{2},
mark=triangle,
mark repeat=16,
mark size=4pt,
mark options={solid, line width=0.65pt},
unbounded coords=jump,
forget plot
]
table[
x expr=\thisrowno{0}/124,
y expr=\thisrowno{1}/100
] {figures/da48/contour_data_M04/contour_numerical_gamma_-0.40_NX125_NZ102.dat};

\end{axis}

\end{tikzpicture}
\vspace{-1.05em}
\caption{Extended scheme, $\gamma = -0.4.$}
\label{fig:contour_gamma_m04}
\end{subfigure}
\hfill
\begin{subfigure}[t]{0.40\textwidth}
\centering
\setlength\figurewidth{0.84\linewidth}
\setlength\figureheight{0.46\linewidth}
\begin{tikzpicture}[font=\sffamily\sansmath]

\begin{axis}[
width=\figurewidth,
height=\figureheight,
scale only axis,
xmin=0,
xmax=1,
ymin=0,
ymax=1,
xlabel={$x^*/L_x^*$},
ylabel={},
grid=none,
axis line style={black},
tick align=inside,
tick pos=both,
xtick={0,0.5,1},
ytick={0,0.5,1},
xticklabels={$0$,$0.5$,$1$},
yticklabels={,,},
tick label style={
    font=\fontsize{11}{8}\selectfont\sffamily\sansmath
},
label style={
    font=\fontsize{13}{8}\selectfont\sffamily\sansmath
},
xlabel style={
    font=\fontsize{13}{8}\selectfont\sffamily\sansmath
},
ylabel style={
    font=\fontsize{13}{8}\selectfont\sffamily\sansmath
},
colormap={dario}{
    rgb255=(0,0,255)
    rgb255=(255,0,255)
    rgb255=(255,0,0)
    rgb255=(255,220,0)
},
point meta min=0,
point meta max=10,
colorbar,
colorbar style={
    width=2.0mm,
    height=0.36\figureheight,
    at={(1.08,0.00)},
    anchor=south west,
    ytick={0,10},
    yticklabels={$0$,$10$},
    ylabel={$C^*$},
    ylabel style={
        rotate=0,
        at={(2.35,0.50)},
        anchor=center,
        font=\fontsize{13}{8}\selectfont\sffamily\sansmath
    },
    tick label style={
        font=\fontsize{11}{8}\selectfont\sffamily\sansmath
    },
},
]

% Analytical contours: coloured solid lines
\addplot[
mesh,
solid,
line width=1.10pt,
point meta=\thisrowno{2},
unbounded coords=jump,
forget plot
]
table[
x expr=\thisrowno{0}/124,
y expr=\thisrowno{1}/100
] {figures/da48/contour_data_P06/contour_analytical_gamma_0.60_NX125_NZ101.dat};

% Numerical contours: coloured dashed lines
\addplot[
mesh,
dash pattern=on 5pt off 3pt,
line width=0.95pt,
point meta=\thisrowno{2},
unbounded coords=jump,
forget plot
]
table[
x expr=\thisrowno{0}/124,
y expr=\thisrowno{1}/100
] {figures/da48/contour_data_P06/contour_numerical_gamma_0.60_NX125_NZ101.dat};

% Numerical markers: coloured hollow triangles
\addplot[
scatter,
only marks,
scatter src=explicit,
scatter/use mapped color={draw=mapped color, fill=white},
point meta=\thisrowno{2},
mark=triangle,
mark repeat=16,
mark size=4pt,
mark options={solid, line width=0.65pt},
unbounded coords=jump,
forget plot
]
table[
x expr=\thisrowno{0}/124,
y expr=\thisrowno{1}/100
] {figures/da48/contour_data_P06/contour_numerical_gamma_0.60_NX125_NZ101.dat};

\end{axis}

\end{tikzpicture}
\vspace{-1.05em}
\caption{Huang scheme, $\gamma = +0.6.$}
\label{fig:contour_gamma_p06}
\end{subfigure}
\vspace{-0.25em}
\caption{Analytical (line) and numerical (triangles) concentration profiles, $\mathrm{Da} = 48$.}
\label{fig:contour_gamma_comparison_da48}
\end{figure}
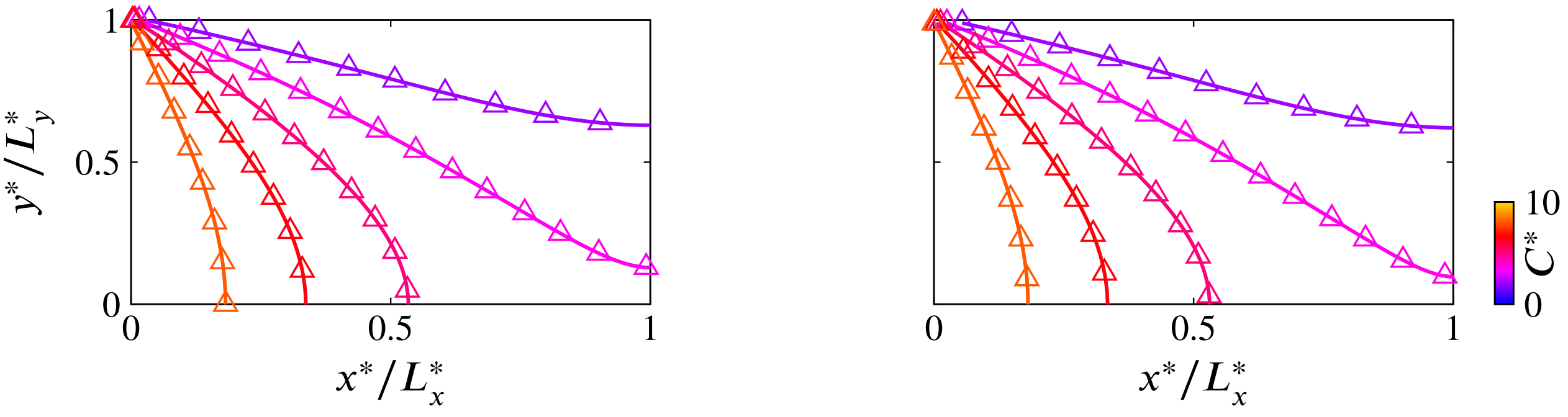

The first concentration benchmark considers the two-dimensional steady diffusion--reaction problem proposed by Zhang \emph{et al.}~\cite{zhang12}. The rectangular domain, defined by $0\leq x^*\leq L_x^*$ and $0\leq y^*\leq L_y^*$, is discretized using a $100\times80$ D2Q5 lattice. A prescribed concentration $C_0^*=10$ is imposed at the left boundary $x^*=0$, while homogeneous Neumann conditions are applied at $y^*=0$ and $x^*=L_x^*$. At the upper boundary, $y^*=L_y^*$, a first-order Robin condition is imposed to model a surface reaction at a stationary interface, with $A_1=0.1$ and $A_3= C^*_s=0$. The reaction rate $A_2$ is chosen so as to set a Damk\"ohler number $
\mathrm{Da}
=
A_2 L^* / \, A_1 = 48$. The resulting steady-state concentration field satisfies the two-dimensional Laplace equation,
$\nabla^2 C^* = 0$, with boundary conditions:
\begin{equation}
\begin{aligned}
-A_1 (\partial C^* / \, \partial y^*)
&=
A_2 C^*
\quad &&\text{at } y^*=L_y^*,
&& \qquad
\partial C^* / \, \partial y^*
&=0
\quad &&\text{at } y^*=0,
\\[1mm]
C^*
&=C_0^*
\quad &&\text{at } x^*=0,
&&\qquad
\partial C^* / \, \partial x^*
&=0
\quad &&\text{at } x^*=L_x^* .
\end{aligned}
\label{eq:zhang_bc}
\end{equation}
The analytical solution is obtained by separation of variables, following the classical treatment of two-dimensional diffusion problems by Carslaw and Jaeger~\cite{carslaw1986}, as
\begin{equation}
C^*(x^*,y^*)
=
C_s^*
+
\left(C^*_0-C^*_s\right)
\sum_{n=1}^{\infty}
\frac{\sin(\beta_n L^*_y)}
     {N_n^2\beta_n}
\,
\frac{\cosh[\beta_n(x^*-L^*_x)]}
     {\cosh(\beta_nL^*_x)}
\cos(\beta_n y^*),
\label{eq:zhang_analytical}
\end{equation}
where $ N_n^2 = (L^*_y \, / \, 2) 
\left[
1+
(\, \sin(2\beta_nL^*_y) \, / \,
     2\beta_nL^*_y \, )
\right],
%\label{eq:zhang_Nn}
%\end{equation}
$ and $(\beta_nL^*_y)\tan(\beta_nL^*_y)
=
A_2 L^*_y \, / \, D
=
\mathrm{Da}.$ 
To compare the original and extended branches of the boundary formulation, pairs of simulations are constructed so as to represent the same physical domain while placing the Robin boundary at different positions relative to the boundary node. Each negative-$\gamma$ configuration therefore contains one additional lattice row; the contribution of this extra row is exactly compensated by the negative sub-grid boundary offset. As a result, the effective domain length $L_y^*$ is identical within each positive-/negative-$\gamma$ pair. Preliminary D3Q19 calculations showed a localized discrepancy at the corner where the Dirichlet and Robin boundaries meet. Since this point represents a boundary singularity, the diagonal populations allow the associated error to propagate to neighbouring nodes. The following analysis is therefore performed using the D2Q5 lattice, which provides a cleaner assessment of the grid-convergence behaviour. Figure~\ref{fig:contour_gamma_comparison_da48} shows the results for the investigated Damk\"ohler number, with $\gamma = -0.4$ and $\gamma = +0.6$. These results are in close agreement with the analytical solution, consistently with previous validations~\cite{Maggiolo2023}. In particular, the negative-$\gamma$ configuration reproduces the same analytical solution as their geometrically equivalent positive-$\gamma$ counterparts, confirming that the extended boundary treatment preserves the prescribed macroscopic Robin condition beyond the original Huang range.

\begin{figure}[pos=htbp]
\begin{subfigure}[t]{0.44\textwidth}
\centering
\setlength\figurewidth{0.76\textwidth}
\setlength\figureheight{0.54\textwidth}
\begin{tikzpicture}[font=\sffamily\sansmath]

\begin{axis}[
width=\figurewidth,
height=\figureheight,
scale only axis,
xmode=log,
ymode=log,
xmin=5e-3,
xmax=7e-2,
ymin=1.8e-5,
ymax=1.5e-3,
xlabel={$h$},
ylabel={$E_{\mathrm{rel},L_2}$},
grid=none,
axis line style={black},
tick align=inside,
tick pos=both,
xtick={5e-3,1e-2,2e-2,5e-2},
xticklabels={
{$5\,{\cdot}\,10^{-3}$},
{$10^{-2}$},
{$2\,{\cdot}\,10^{-2}$},
{$5\,{\cdot}\,10^{-2}$}
},
ytick={5e-5,1e-4,2e-4,5e-4,1e-3},
yticklabels={
{$5\,{\cdot}\,10^{-5}$},
{$10^{-4}$},
{$2\,{\cdot}\,10^{-4}$},
{$5\,{\cdot}\,10^{-4}$},
{$10^{-3}$}
},
legend style={
    font=\fontsize{9.5}{8}\selectfont\sffamily\sansmath,
    anchor=north east,
    at={(0.99,0.99)},
    draw=black,
    fill=white,
    inner sep=1.5pt,
    row sep=0pt
},
legend cell align=left,
tick label style={
    font=\fontsize{11}{8}\selectfont\sffamily\sansmath
},
label style={
    font=\fontsize{13}{8}\selectfont\sffamily\sansmath
},
xlabel style={
    font=\fontsize{13}{8}\selectfont\sffamily\sansmath,
    at={(0.5,-0.12)}
},
clip=true,
]

% ============================================================
% gamma = 0.6
% ============================================================
\addplot[
orange,
solid,
mark=o,
mark size=2.7pt,
line width=1.1pt,
mark options={
    solid,
    draw=orange,
    fill=white,
    line width=0.65pt
}
]
table[x index=0,y index=1]
{figures/grid_conv_zhang/grid_convergence_full2D_p06.dat};
\addlegendentry{$\gamma=0.6$};

% ============================================================
% gamma = 0.7
% ============================================================
\addplot[
cyan!70!blue,
solid,
mark=square,
mark size=2.6pt,
line width=1.1pt,
mark options={
    solid,
    draw=cyan!70!blue,
    fill=white,
    line width=0.65pt
}
]
table[x index=0,y index=1]
{figures/grid_conv_zhang/grid_convergence_full2D_p07.dat};
\addlegendentry{$\gamma=0.7$};

% ============================================================
% gamma = 0.8
% ============================================================
\addplot[
green!60!black,
solid,
mark=triangle,
mark size=3.4pt,
line width=1.1pt,
mark options={
    solid,
    draw=green!60!black,
    fill=white,
    line width=0.55pt
}
]
table[x index=0,y index=1]
{figures/grid_conv_zhang/grid_convergence_full2D_p08.dat};
\addlegendentry{$\gamma=0.8$};

% ============================================================
% Upper second-order guide: 1:2
% ============================================================
\addplot[
black,
dash pattern=on 4pt off 2pt,
line width=1.1pt,
forget plot
]
coordinates {
(5.0e-3,9.0e-5)
(7.0e-2,1.764e-2)
}
node[
pos=0.12,
sloped,
above=2pt,
fill=white,
inner sep=1pt,
font=\fontsize{11}{8}\selectfont\sffamily\sansmath
] {$1:2$};

% ============================================================
% Lower first-order guide: 1:1
% ============================================================
\addplot[
black,
dash pattern=on 4pt off 2pt,
line width=1.1pt,
forget plot
]
coordinates {
(5.0e-3,1.8e-5)
(7.0e-2,2.52e-4)
}
node[
pos=0.82,
sloped,
below=2pt,
fill=white,
inner sep=1pt,
font=\fontsize{11}{8}\selectfont\sffamily\sansmath
] {$1:1$};

\end{axis}

\end{tikzpicture}
\caption{Original Huang branch, positive $\gamma$.}
\label{fig:grid_conv_zhang_positive}
\end{subfigure}
\hfill
\begin{subfigure}[t]{0.44\textwidth}
\centering
\setlength\figurewidth{0.76\textwidth}
\setlength\figureheight{0.54\textwidth}
\begin{tikzpicture}[font=\sffamily\sansmath]

\begin{axis}[
width=\figurewidth,
height=\figureheight,
scale only axis,
xmode=log,
ymode=log,
xmin=5e-3,
xmax=7e-2,
ymin=5e-4,
ymax=3e-2,
xlabel={$h$},
ylabel={},
grid=none,
axis line style={black},
tick align=inside,
tick pos=both,
xtick={5e-3,1e-2,2e-2,5e-2},
xticklabels={
{$5\,{\cdot}\,10^{-3}$},
{$10^{-2}$},
{$2\,{\cdot}\,10^{-2}$},
{$5\,{\cdot}\,10^{-2}$}
},
ytick={5e-4,1e-3,2e-3,5e-3,1e-2,2e-2},
yticklabels={
{$5\,{\cdot}\,10^{-4}$},
{$10^{-3}$},
{$2\,{\cdot}\,10^{-3}$},
{$5\,{\cdot}\,10^{-3}$},
{$10^{-2}$},
{$2\,{\cdot}\,10^{-2}$}
},
legend style={
    font=\fontsize{9.5}{8}\selectfont\sffamily\sansmath,
    anchor=north east,
    at={(0.99,0.99)},
    draw=black,
    fill=white,
    inner sep=1.5pt,
    row sep=0pt
},
legend cell align=left,
tick label style={
    font=\fontsize{11}{8}\selectfont\sffamily\sansmath
},
label style={
    font=\fontsize{13}{8}\selectfont\sffamily\sansmath
},
xlabel style={
    font=\fontsize{13}{8}\selectfont\sffamily\sansmath,
    at={(0.5,-0.12)}
},
ylabel style={
    font=\fontsize{13}{8}\selectfont\sffamily\sansmath,
    at={(-0.25,0.5)}
},
clip=true,
]

% ============================================================
% gamma = -0.2
% ============================================================
\addplot[
orange,
solid,
mark=o,
mark size=2.7pt,
line width=1.1pt,
mark options={
    solid,
    draw=orange,
    fill=white,
    line width=0.65pt
}
]
table[x index=0,y index=1]
{figures/grid_conv_zhang/grid_convergence_full2D_m02.dat};
\addlegendentry{$\gamma=-0.2$};

% ============================================================
% gamma = -0.3
% ============================================================
\addplot[
cyan!70!blue,
solid,
mark=square,
mark size=2.6pt,
line width=1.1pt,
mark options={
    solid,
    draw=cyan!70!blue,
    fill=white,
    line width=0.65pt
}
]
table[x index=0,y index=1]
{figures/grid_conv_zhang/grid_convergence_full2D_m03.dat};
\addlegendentry{$\gamma=-0.3$};

% ============================================================
% gamma = -0.4
% ============================================================
\addplot[
green!60!black,
solid,
mark=triangle,
mark size=3.4pt,
line width=1.1pt,
mark options={
    solid,
    draw=green!60!black,
    fill=white,
    line width=0.55pt
}
]
table[x index=0,y index=1]
{figures/grid_conv_zhang/grid_convergence_full2D_m04.dat};
\addlegendentry{$\gamma=-0.4$};

% ============================================================
% Upper second-order guide: 1:2
% Starts exactly from the left boundary and is shifted downward.
% The label is attached to the plotted line, so it is exactly parallel.
% ============================================================
\addplot[
black,
dash pattern=on 4pt off 2pt,
line width=1.1pt,
forget plot
]
coordinates {
(5.0e-3,1.60e-3)
(7.0e-2,3.136e-1)
}
node[
pos=0.12,
sloped,
above=2pt,
fill=white,
inner sep=1pt,
font=\fontsize{11}{8}\selectfont\sffamily\sansmath
] {$1:2$};

% ============================================================
% Lower first-order guide: 1:1
% Crosses the complete x-domain and is shifted downward.
% The label is attached to the plotted line, so it is exactly parallel.
% ============================================================
\addplot[
black,
dash pattern=on 4pt off 2pt,
line width=1.1pt,
forget plot
]
coordinates {
(5.0e-3,3.8e-4)
(7.0e-2,5.32e-3)
}
node[
pos=0.82,
sloped,
above=2pt,
fill=white,
inner sep=1pt,
font=\fontsize{11}{8}\selectfont\sffamily\sansmath
] {$1:1$};

\end{axis}

\end{tikzpicture}
\caption{Extended branch, negative $\gamma$.}
\label{fig:grid_conv_zhang_negative}
\end{subfigure}
\caption{Grid convergence of the full-domain relative $L_2$ error, D2Q5 model, $\tau_c = 0.8$.}
\label{fig:grid_conv_zhang_full2d_split}
\end{figure}
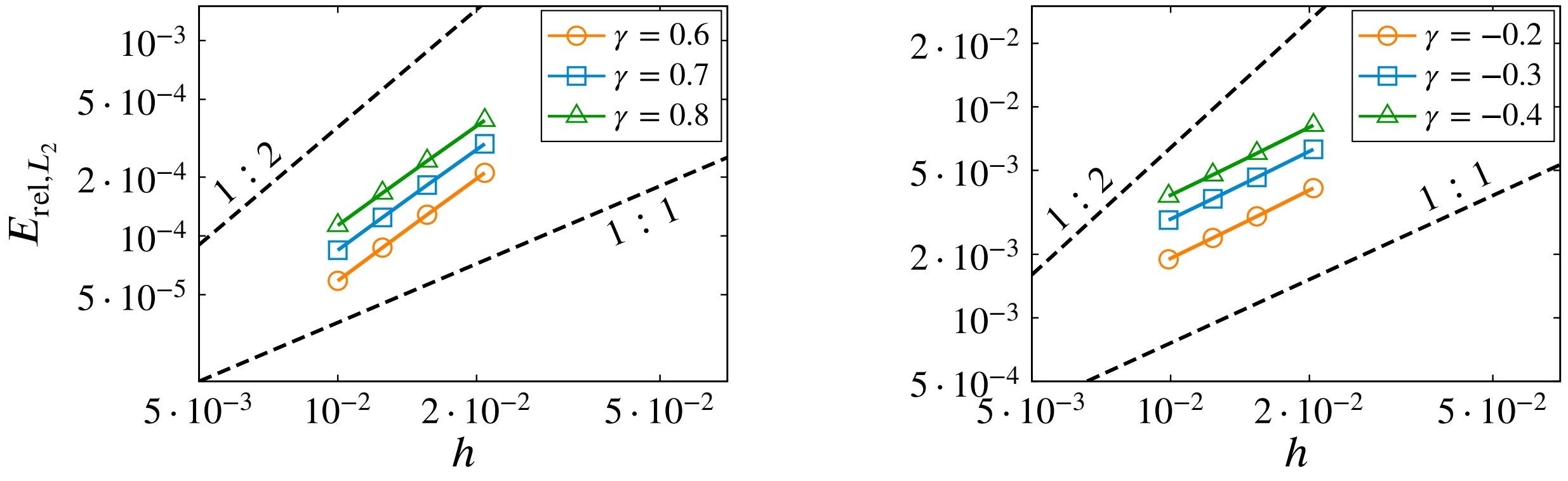

Subsequently,, a grid-convergence analysis is performed while preserving the same physical problem at each lattice resolution. In the present formulation, the lattice spacing is fixed to $\Delta x=1$ and grid refinement is achieved by representing the same physical domain with an increasing number of lattice nodes. Hence, the domain size in lattice units, $L^*$, increases proportionally with the resolution, while the geometrical aspect ratio is kept unchanged as $L^*_x / L^*_y = 1.25$. In addition to geometric similarity, the relative strength of interfacial reaction and solute diffusion must also be conserved. To this end, the Robin boundary condition is nondimensionalized following the procedure adopted by Pettersson \emph{et al.}~\cite{pettersson2026}, yielding
$-\partial C^*/\partial n^*=\mathrm{Da}\,C^*$, where $n^*=n/L^*$ and $\mathrm{Da} = A_2 L^* / A_1$. Since $L^*$ increases under grid refinement and $A_1 = \mathrm{const}= 0.1$, $A_2$ is decreased accordingly in order to keep $\mathrm{Da} \simeq 4.8$.  Note that the effective domain length also depends on the fractional boundary position $\gamma$, thus $L^*$ and the corresponding $A_2$ values are evaluated separately for each investigated $\gamma$. The convergence analysis is based on the relative $L_2$ error evaluated over the entire two-dimensional fluid domain,
\begin{equation}
E_{\mathrm{rel},L_2}
=
\sqrt{
\sum_{\mathbf{x}\in\Omega_l}
\left(
C_{\mathrm{num}}(\mathbf{x})
-
C_{\mathrm{an}}(\mathbf{x})
\right)^2
}
\bigg/
\sqrt{
\sum_{\mathbf{x}\in\Omega_l}
C_{\mathrm{an}}^2(\mathbf{x})
},
\qquad
\mathbf{x}\in\Omega_l,
\label{eq:zhang_L2_error}
\end{equation}
where $C_{\mathrm{num}}(\mathbf{x})$ and $C_{\mathrm{an}}(\mathbf{x})$ denote the numerical and analytical concentration fields at the lattice nodes' positions. The characteristic mesh size is defined as $h=1/N_y$, where $N_y$ is the number of nodes along the $y$ direction. As shown in Fig.~\ref{fig:grid_conv_zhang_positive}, the relative errors obtained with the original Huang scheme follow a slope close to the second-order reference. For the extended scheme at negative $\gamma$, Fig.~\ref{fig:grid_conv_zhang_negative} shows a systematic error reduction with convergence rates between first and second order. This reduced accuracy is consistent with the local approximation introduced in Sec.~\ref{subsec:extended_robin}. Nevertheless, convergence is retained throughout the investigated negative-$\gamma$ range, with nearly constant slopes over the considered resolutions. Moreover, the errors for negative $\gamma$ are approximately one order of magnitude larger than those of positive-$\gamma$ formulation, but remain small, ranging from $10^{-3}$ to $10^{-2}$. The extended formulation therefore retains satisfactory quantitative accuracy despite the reduction in convergence order.

\subsubsection{Concentration diffusion in an annular geometry}
\label{subsec:ring_diff}
\begin{figure}[pos=htbp]
\centering
\captionsetup[subfigure]{skip=2pt}
\begin{subfigure}[t]{0.36\textwidth}
\centering
\setlength\figurewidth{0.86\linewidth}
\setlength\figureheight{0.86\linewidth}
\parbox[t][0.1\textwidth][t]{\linewidth}{%
\centering
\raisebox{0.175\figureheight}{%
\begin{tikzpicture}[font=\sffamily\sansmath]

% Main geometry: no axes, same clean typography as the other figures.
\begin{axis}[
name=ringmain,
width=\figurewidth,
height=\figureheight,
scale only axis,
xmin=95,
xmax=185,
ymin=95,
ymax=185,
axis equal image,
hide axis,
clip=true,
]

% Ideal circular boundaries, retained only as a light reference.
\addplot[
black,
densely dashed,
line width=0.7pt,
domain=0:360,
samples=720,
forget plot
]
({100.5 + 20*cos(x)}, {100.5 + 20*sin(x)});

\addplot[
black,
densely dashed,
line width=0.7pt,
domain=0:360,
samples=720,
forget plot
]
({100.5 + 80*cos(x)}, {100.5 + 80*sin(x)});

% Numerical stair-step boundaries: both black.
\addplot[
black,
solid,
line width=1.15pt,
line join=round,
line cap=round,
no markers,
forget plot
]
table[x=x,y=z]
{figures/ring/ring_geom/ring_inner_stair_edges.dat};

\addplot[
black,
solid,
line width=1.15pt,
line join=round,
line cap=round,
no markers,
forget plot
]
table[x=x,y=z]
{figures/ring/ring_geom/ring_outer_stair_edges.dat};

% External square domain boundary.
\draw[black, line width=0.65pt]
(axis cs:96.0,96.0) rectangle (axis cs:180.5,180.5);

% Small source windows corresponding to the two magnifications.
\draw[black, line width=0.65pt]
(axis cs:113.4,113.4) rectangle (axis cs:116.8,116.8);

\draw[black, line width=0.65pt]
(axis cs:107.5,178.6) rectangle (axis cs:113.5,181.0);

% Save the corners of the selected regions and the inset positions.
% Two corners per source window are used to create the two-line zoom callouts.
\coordinate (innerSourceNE) at (axis cs:116.8,116.8);
\coordinate (innerSourceSE) at (axis cs:116.8,113.4);
\coordinate (outerSourceSW) at (axis cs:107.5,178.6);
\coordinate (outerSourceSE) at (axis cs:113.5,178.6);

% Insets positioned approximately as indicated in the annotated sketch.
% Outer zoom: below the selected stair at the top of the outer boundary.
% Inner zoom: lower-right, as in the previous version.
\coordinate (innerInsetPos) at (rel axis cs:0.58,0.19);
\coordinate (outerInsetPos) at (rel axis cs:0.22,0.64);

% Material-region labels.
\node[
    font=\fontsize{11}{13}\selectfont\sffamily\sansmath,
    anchor=center,
    inner sep=1.2pt
] at (axis cs:165.0,170.0) {solid};

\node[
    font=\fontsize{11}{13}\selectfont\sffamily\sansmath,
    anchor=center,
    inner sep=1.2pt
] at (axis cs:141.5,137.5) {fluid};

\node[
    font=\fontsize{11}{13}\selectfont\sffamily\sansmath,
    anchor=center,
    inner sep=1.2pt
] at (axis cs:104.0,105.0) {solid};

\end{axis}

% --------------------------------------------------------------------------
% Inner-boundary magnification, placed in the empty annular region.
% --------------------------------------------------------------------------
\begin{axis}[
name=innerzoom,
at={(innerInsetPos)},
anchor=center,
width=0.34\figurewidth,
height=0.22\figureheight,
scale only axis,
xmin=113.4,
xmax=116.8,
ymin=113.4,
ymax=116.8,
axis equal image,
axis line style={black, line width=0.65pt},
tick align=inside,
tick pos=both,
xtick=\empty,
ytick=\empty,
clip=true,
]

\addplot[
black,
densely dashed,
line width=0.7pt,
domain=0:360,
samples=720,
forget plot
]
({100.5 + 20*cos(x)}, {100.5 + 20*sin(x)});

\addplot[
black,
solid,
line width=1.4pt,
line join=miter,
line cap=butt,
no markers,
forget plot
]
table[x=x,y=z]
{figures/ring/ring_geom/ring_inner_stair_edges.dat};

\end{axis}

% --------------------------------------------------------------------------
% Outer-boundary magnification.
% --------------------------------------------------------------------------
\begin{axis}[
name=outerzoom,
at={(outerInsetPos)},
anchor=center,
width=0.30\figurewidth,
height=0.18\figureheight,
scale only axis,
xmin=107.5,
xmax=113.5,
ymin=178.6,
ymax=181.0,
axis equal image,
axis line style={black, line width=0.65pt},
tick align=inside,
tick pos=both,
xtick=\empty,
ytick=\empty,
clip=true,
]

\addplot[
black,
densely dashed,
line width=0.7pt,
domain=0:360,
samples=720,
forget plot
]
({100.5 + 80*cos(x)}, {100.5 + 80*sin(x)});

\addplot[
black,
solid,
line width=1.4pt,
line join=miter,
line cap=butt,
no markers,
forget plot
]
table[x=x,y=z]
{figures/ring/ring_geom/ring_outer_stair_edges.dat};

\end{axis}

% Two thinner oblique connectors for each zoom.
% Inner source window -> upper-left side of the lower-right inset.
\draw[black, line width=0.40pt]
(innerSourceNE) -- (innerzoom.north west);
\draw[black, line width=0.40pt]
(innerSourceSE) -- (innerzoom.south west);

% Outer source window at the top -> upper corners of the zoom below.
\draw[black, line width=0.40pt]
(outerSourceSW) -- (outerzoom.north west);
\draw[black, line width=0.40pt]
(outerSourceSE) -- (outerzoom.north east);

\end{tikzpicture}%
}
}
\vspace{-1.8em}
\caption{Stair-step ring geometry.}
\label{fig:ring_stair_step_geometry}
\end{subfigure}
\hspace{0.075\textwidth}
\begin{subfigure}[t]{0.40\textwidth}
\centering
\setlength\figurewidth{0.82\linewidth}
\setlength\figureheight{0.62\linewidth}
\parbox[t][0.1\textwidth][t]{\linewidth}{%
\centering
\raisebox{0.9em}{%
\begin{tikzpicture}[font=\sffamily\sansmath]

\begin{axis}[
width=\figurewidth,
height=\figureheight,
scale only axis,
xmode=log,
ymode=log,
xmin=1.2e-2,
xmax=7.0e-2,
ymin=7.0e-4,
ymax=8.0e-2,
xlabel={h},
ylabel={$E_{\mathrm{rel},L_2}$},
grid=none,
axis line style={black},
tick align=inside,
tick pos=both,
xtick={2e-2,5e-2},
xticklabels={
{$2\,{\cdot}\,10^{-2}$},
{$5\,{\cdot}\,10^{-2}$}
},
minor x tick num=8,
ytick={1e-3,1e-2,1e-1},
yticklabels={
{$10^{-3}$},
{$10^{-2}$},
{$10^{-1}$}
},
legend style={
    font=\fontsize{7}{6}\selectfont\sffamily\sansmath,
    anchor=south east,
    at={(0.99,0.01)},
    draw=black,
    fill=white,
    inner sep=1.3pt,
    row sep=0pt
},
legend cell align=left,
tick label style={
    font=\fontsize{11}{8}\selectfont\sffamily\sansmath
},
label style={
    font=\fontsize{13}{8}\selectfont\sffamily\sansmath
},
xlabel style={
    font=\fontsize{13}{8}\selectfont\sffamily\sansmath,
    at={(0.5,-0.10)}
},
clip=true,
]

% ============================================================
% D3Q19, gamma = -0.4
% ============================================================
\addplot[
orange,
solid,
mark=o,
mark size=2.7pt,
line width=1.1pt,
mark options={
    solid,
    draw=orange,
    fill=white,
    line width=0.65pt
}
]
table[x=h,y=E_L2_rel]
{figures/ring/ring_gridconv/plot_grid_convergence_D3Q19_gammaM04.dat};
\addlegendentry{D3Q19, $\gamma=-0.4$};

% ============================================================
% D3Q19, gamma = +0.5
% ============================================================
\addplot[
cyan!70!blue,
solid,
mark=square,
mark size=2.6pt,
line width=1.1pt,
mark options={
    solid,
    draw=cyan!70!blue,
    fill=white,
    line width=0.65pt
}
]
table[x=h,y=E_L2_rel]
{figures/ring/ring_gridconv/plot_grid_convergence_D3Q19_gammaP05.dat};
\addlegendentry{D3Q19, $\gamma=0.5$};

% ============================================================
% D2Q5, gamma = +0.5
% ============================================================
\addplot[
green!60!black,
solid,
mark=triangle,
mark size=3.4pt,
line width=1.1pt,
mark options={
    solid,
    draw=green!60!black,
    fill=white,
    line width=0.55pt
}
]
table[x=h,y=E_L2_rel]
{figures/ring/ring_gridconv/plot_grid_convergence_D2Q5_gammaP05.dat};
\addlegendentry{D2Q5, $\gamma=0.5$};

% ============================================================
% Upper first-order guide: shifted upward and spanning full x-domain
% ============================================================
\addplot[
black,
dash pattern=on 4pt off 2pt,
line width=1.1pt,
forget plot
]
coordinates {
(1.2e-2,1.90e-2)
(7.0e-2,1.108e-1)
}
node[
pos=0.20,
sloped,
above=2pt,
fill=white,
inner sep=1pt,
font=\fontsize{11}{8}\selectfont\sffamily\sansmath
] {$1:1$};

% ============================================================
% Lower second-order guide: shifted downward and spanning full x-domain
% ============================================================
\addplot[
black,
dash pattern=on 4pt off 2pt,
line width=1.1pt,
forget plot
]
coordinates {
(1.2e-2,4.50e-4)
(7.0e-2,1.531e-2)
}
node[
pos=0.20,
sloped,
above=2pt,
fill=white,
inner sep=1pt,
font=\fontsize{11}{8}\selectfont\sffamily\sansmath
] {$1:2$};

\end{axis}

\end{tikzpicture}
}%
}
\vspace{-1.8em}
\caption{Grid convergence analysis.}
\label{fig:ring_grid_convergence}
\end{subfigure}
\vspace{-0.1em}
\caption{Ring grid convergence study, $\tau_c = 2.0$.}
\label{fig:ring_geometry_grid_convergence}
\end{figure}
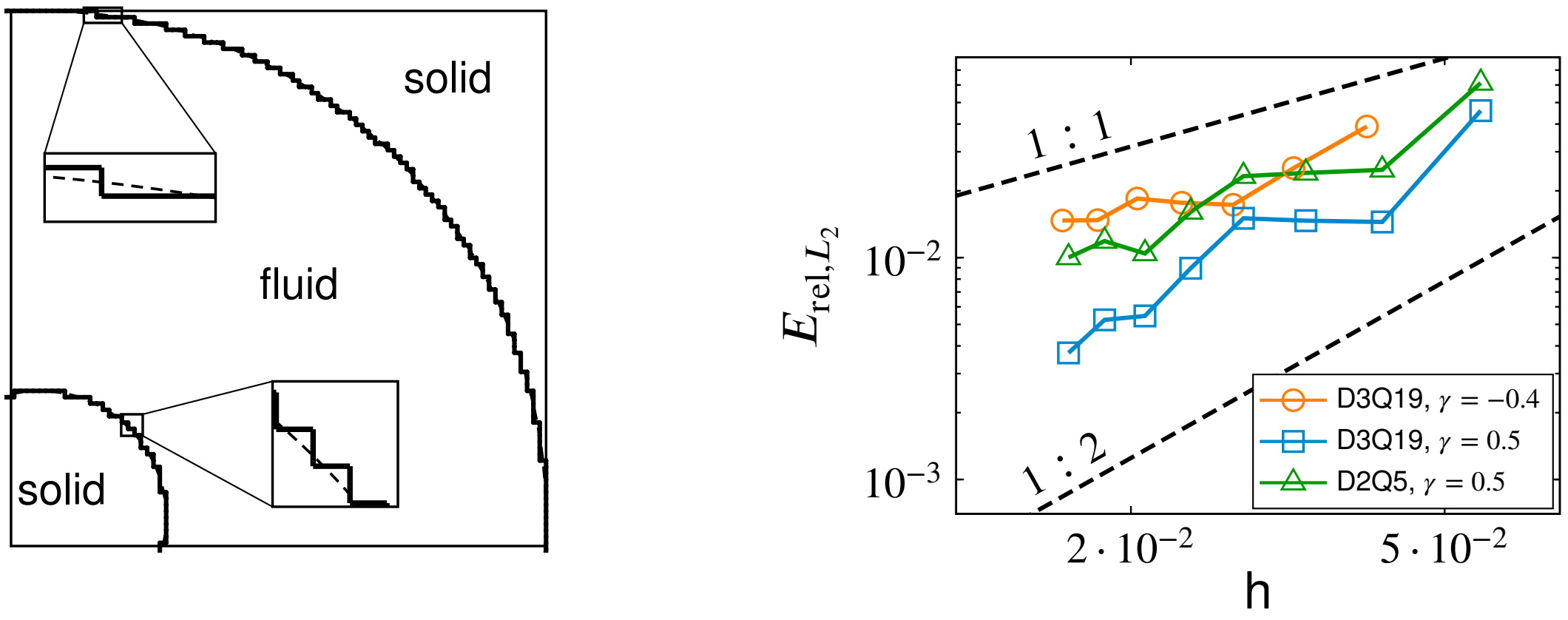

A further validation is performed on a curved geometry, following the annular benchmark originally considered by Huang \emph{et al.}~\cite{huang15}. The circular boundaries are represented on the Cartesian lattice through a stair-step approximation and enclosed within a square domain, as illustrated in Fig.~\ref{fig:ring_stair_step_geometry}. The local boundary orientation varies along the circular interfaces, so that the Robin condition is enforced through different lattice-link directions. This test is performed using both D2Q5 and D3Q19 lattices in order to assess the generalization of the boundary formulation to a three-dimensional discrete-velocity stencil.
At steady state, the concentration field satisfies Laplace's equation in $R_{\mathrm{in}} < r^* < R_{\mathrm{out}}$. A prescribed concentration is imposed at the inner boundary $r^*=R_{\mathrm{in}}$, namely $C^*=C^*_{\mathrm{in}}=4$, while the outer boundary $r^*=R_{\mathrm{out}}$ is governed by the Robin condition $A_2 C^* + A_1 \, \partial C^* \, / \, \partial n^* = A_3$, where $A_1=0.5$ and $A_3=0$. The corresponding radial analytical solution can be written as $
C^*(r^*)
=
C^*_{\mathrm{in}}
+
K_1
\ln\!\left(r^* \, / \, R_{\mathrm{in}}\right),$ where
$K_1 =
\left(A_3-A_2C^*_{\mathrm{in}}\right)
\, / \,
\left[
A_2\ln\!\left(R_{\mathrm{out}} \, / \, R_{\mathrm{in}}\right)
+
A_1 \, / \, R_{\mathrm{out}}
\right]$.
Denoting by $N$ the number of lattice nodes along each Cartesian direction of the square external domain, the geometrical ratios are fixed to $R_{\mathrm{in}}/N=0.1$ and $R_{\mathrm{out}}/N=0.4$, while the lattice resolution is progressively increased from $N=80$ to $N=200$. The effective radial width is denoted by $H$ and accounts for the halfway position of the inner Dirichlet boundary and the $\gamma$-dependent position of the outer Robin boundary. The characteristic mesh size is then defined as $h=1/H$. The coefficient $A_2$ is rescaled at each resolution to preserve $\mathrm{Da}=A_2H/A_1\simeq10$. The numerical solution is compared pointwise with the analytical one using the radial distance $r^*$ of each fluid lattice node from the centre of the ring. The relative $L_2$ error is then evaluated over the annular fluid domain according to Eq.~\eqref{eq:zhang_L2_error}, where $\mathbf{x}$ is substituted by the coordinate $r^*$. 

As shown in Fig.~\ref{fig:ring_grid_convergence}, the error decreases with mesh refinement for all the investigated configurations. The convergence order generally lies between first and second order, consistently with the behaviour reported by Huang~\cite{huang15}. In contrast to the straight-boundary case, the convergence curves exhibit some oscillations, which have also been observed in previous lattice-Boltzmann studies of curved boundaries~\cite{LiMeiKlausner2013}. The results further show that the negative-$\gamma$ formulation retains convergence when applied to curved geometries. Moreover, the D3Q19 implementation generally yields lower relative errors than the D2Q5 lattice over the investigated resolutions. The additional diagonal lattice links provide a richer directional representation of the local boundary orientation and can therefore improve the approximation of curved interfaces. These results support the use of both the original Huang boundary treatment and the negative-$\gamma$ formulation on the D3Q19 lattice.

\subsubsection{Solutal simulation of an idealized brine channel}
\label{subsec:pore_solutal_diff}
\begin{figure}[pos=htbp]
\centering
\resizebox{0.30\textwidth}{!}{%
\begin{tikzpicture}[font=\sffamily\sansmath]

% ============================================================
% Symmetric pore section
% Solid: gray!28
% Liquid: cyan!22
% ============================================================

% Geometry
\def\W{8.0}
\def\H{3.2}

% Central liquid-channel half width
\def\xL{3.25}
\def\xR{4.75}

% Highlighted simulated region
\definecolor{simregionred}{RGB}{130,0,0}
\def\xBoxL{2.25}
\def\xBoxR{4.05}

% ============================================================
% SOLID REGIONS
% ============================================================

% Left solid
\fill[gray!28]
(0,0) rectangle (\xL,\H);

% Right solid
\fill[gray!28]
(\xR,0) rectangle (\W,\H);

% ============================================================
% LIQUID PORE / BRINE CHANNEL
% ============================================================

\fill[cyan!22]
(\xL,0) rectangle (\xR,\H);

% ============================================================
% SOLID--LIQUID INTERFACES
% ============================================================

\draw[
black,
line width=1.4pt
]
(\xL,0) -- (\xL,\H);

\draw[
black,
line width=1.4pt
]
(\xR,0) -- (\xR,\H);

% ============================================================
% OUTER DOMAIN BOUNDARY
% ============================================================

\draw[
black,
line width=0.8pt
]
(0,0) rectangle (\W,\H);

% ============================================================
% HIGHLIGHTED SIMULATED REGION
% ============================================================

\draw[
    simregionred,
    line width=2.2pt
]
(\xBoxL,-0.05) rectangle (\xBoxR,{\H+0.05});

\end{tikzpicture}%
}
\vspace{-0.35em}
\caption{Idealized brine-channel geometry. The red box marks the simulated section.}
\label{fig:idealized_pore_section}
\end{figure}
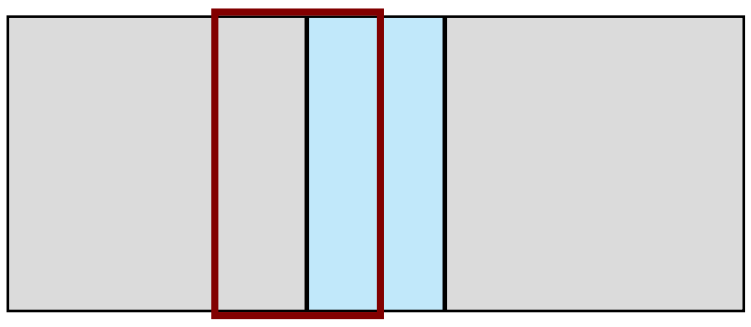

To assess the behaviour of the moving Robin boundary treatment during interface motion, a two-dimensional solutal test is considered in an idealized planar brine channel bounded by solid walls, as illustrated in Fig.~\ref{fig:idealized_pore_section}. 
In this configuration, the evolved concentration field represents the local salinity of the brine and is defined only within the liquid phase. The domain is discretized on a $125\times125$ lattice. Phase change is imposed by prescribing a constant interface migration rate normal to the walls, with $A_2=+0.00025$ for melting and $A_2=-0.00025$ for freezing. The remaining Robin coefficients are $A_1=0.003333$ and $A_3=0$, corresponding to $|\mathrm{Da}|\simeq4.65$. Since no solute enters or leaves the system through the external boundaries, any variation of the concentration field is caused solely by the redistribution of salt induced by the moving solid--liquid interface, Eq.~\eqref{eq:robin_general_A}. 

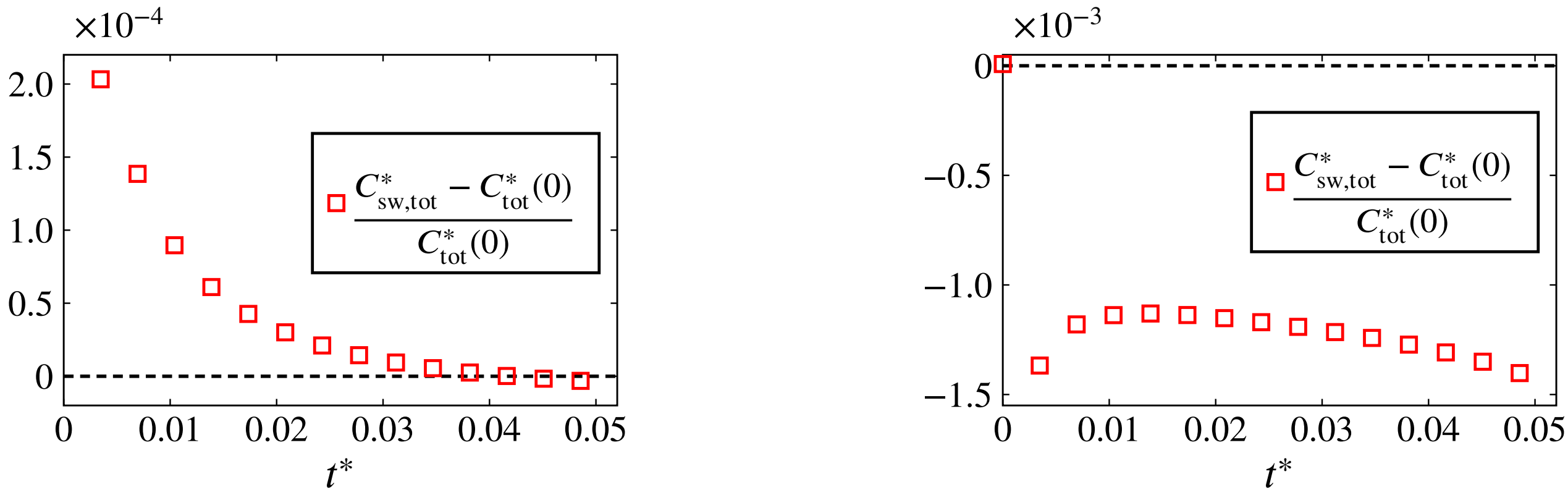
\begin{figure}[pos=htbp]
\centering

\begin{subfigure}[t]{0.40\textwidth}
\centering
\setlength\figurewidth{0.90\linewidth}
\setlength\figureheight{0.57\linewidth}
\begin{tikzpicture}[font=\sffamily\sansmath]

% Diffusive dimensionless time:
% t^* = alpha_L t_L / L_L^2
\def\alphaL{0.003333}
\def\LL{62}

\begin{axis}[
width=\figurewidth,
height=\figureheight,
scale only axis,
xmin=0,
xmax=0.052,
ymin=-0.00002,
ymax=0.00022,
xlabel={$t^*$},
clip=false,
axis line style={black},
tick align=inside,
tick pos=both,
grid=none,
tick label style={
    font=\fontsize{11}{13}\selectfont\sffamily\sansmath
},
label style={
    font=\fontsize{13}{15}\selectfont\sffamily\sansmath
},
xlabel style={yshift=1pt},
scaled x ticks=false,
scaled y ticks=false,
xtick={0,0.01,0.02,0.03,0.04,0.05},
xticklabels={$0$,$0.01$,$0.02$,$0.03$,$0.04$,$0.05$},
ytick={0,0.00005,0.00010,0.00015,0.00020},
yticklabels={$0$,$0.5$,$1.0$,$1.5$,$2.0$},
legend style={
    font=\fontsize{10.5}{12.5}\selectfont\sffamily\sansmath,
    draw=black,
    line width=0.9pt,
    fill=white,
    inner xsep=5pt,
    inner ysep=10pt,
    at={(0.97,0.78)},
    anchor=north east
},
legend cell align=left
]

% ============================================================
% y-axis multiplier
% ============================================================

\node[
anchor=south west,
font=\fontsize{11}{13}\selectfont\sffamily\sansmath,
yshift=2pt
] at (rel axis cs:0,1)
{$\times 10^{-4}$};

% ============================================================
% LOCAL SOLUTE-CONTENT ERROR PER SWITCH
% ============================================================

\addplot[
red,
only marks,
mark=square,
mark size=2.3pt,
mark options={
    draw=red,
    fill=none,
    line width=0.9pt
}
]
table[
x expr=\thisrow{t}*\alphaL/(\LL*\LL),
y=local_switch_error
]
{figures/1d_mass_check_melting/melting_local_switch_error.dat};

\addlegendentry{%
$\displaystyle
\frac{
C_{\mathrm{sw,tot}}^*-C_{\mathrm{tot}}^*(0)
}{
C_{\mathrm{tot}}^*(0)
}$%
}

% ============================================================
% ZERO-ERROR REFERENCE
% ============================================================

\addplot[
black,
densely dashed,
line width=1.05pt,
domain=0:0.052,
samples=2,
forget plot
]
{0};

\end{axis}

\end{tikzpicture}
\vspace{-1.4em}
\caption{Melting, $A_2>0$.}
\label{fig:melting_switch_error}
\end{subfigure}
\hfill
\begin{subfigure}[t]{0.40\textwidth}
\centering
\setlength\figurewidth{0.90\linewidth}
\setlength\figureheight{0.57\linewidth}
\begin{tikzpicture}[font=\sffamily\sansmath]

% Diffusive dimensionless time:
% t^* = alpha_L t_L / L_L^2
\def\alphaL{0.003333}
\def\LL{62}

\begin{axis}[
width=\figurewidth,
height=\figureheight,
scale only axis,
xmin=0,
xmax=0.052,
ymin=-0.00155,
ymax=0.00005,
xlabel={$t^*$},
clip=false,
axis line style={black},
tick align=inside,
tick pos=both,
grid=none,
tick label style={
    font=\fontsize{11}{13}\selectfont\sffamily\sansmath
},
label style={
    font=\fontsize{13}{15}\selectfont\sffamily\sansmath
},
xlabel style={yshift=1pt},
scaled x ticks=false,
scaled y ticks=false,
xtick={0,0.01,0.02,0.03,0.04,0.05},
xticklabels={$0$,$0.01$,$0.02$,$0.03$,$0.04$,$0.05$},
ytick={-0.0015,-0.0010,-0.0005,0},
yticklabels={$-1.5$,$-1.0$,$-0.5$,$0$},
legend style={
    font=\fontsize{10.5}{12.5}\selectfont\sffamily\sansmath,
    draw=black,
    line width=0.9pt,
    fill=white,
    inner xsep=5pt,
    inner ysep=10pt,
    at={(0.97,0.84)},
    anchor=north east
},
legend cell align=left
]

% ============================================================
% y-axis multiplier
% ============================================================

\node[
anchor=south west,
font=\fontsize{11}{13}\selectfont\sffamily\sansmath,
yshift=2pt
] at (rel axis cs:0,1)
{$\times 10^{-3}$};

% ============================================================
% RELATIVE SOLUTE-CONTENT ERROR AFTER EACH SWITCH
% ============================================================

\addplot[
red,
only marks,
mark=square,
mark size=2.3pt,
mark options={
    draw=red,
    fill=none,
    line width=0.9pt
}
]
table[
x expr=\thisrow{t}*\alphaL/(\LL*\LL),
y=local_switch_error
]
{figures/1d_mass_check_freezing/freezing_local_switch_error.dat};

\addlegendentry{%
$\displaystyle
\frac{
C_{\mathrm{sw,tot}}^*-C_{\mathrm{tot}}^*(0)
}{
C_{\mathrm{tot}}^*(0)
}$%
}

% ============================================================
% ZERO-ERROR REFERENCE
% ============================================================

\addplot[
black,
densely dashed,
line width=1.05pt,
domain=0:0.052,
samples=2,
forget plot
]
{0};

\end{axis}

\end{tikzpicture}
\vspace{-1.4em}
\caption{Freezing, $A_2<0$.}
\label{fig:freezing_switch_error}
\end{subfigure}
\vspace{-0.2em}
\caption{Post-switch solute discrepancies during planar-interface phase change.}
\label{fig:switch_error_melting_freezing}
\end{figure}

During melting of salt-free ice, fresh liquid is introduced into the brine channel, generating a low-salinity layer next to the moving interface, as shown in Fig.~\ref{fig:scalar_contour_melting}. This behaviour is consistent with the solutally controlled regime described by Wells and Worster~\cite{WellsWorster2011}. From a numerical perspective, each descending segment in Fig.~\ref{fig:melting_solute_sawtooth} shows that the total discrete solute content carried by the active liquid lattice nodes,
$C^*_{\mathrm{tot}}(t^*)=\sum_{\mathbf{x}\in\Omega_l} C^*(\mathbf{x},t^*)$,
decreases between successive node-conversion events, where $t^*=Dt/L^2$ is the dimensionless diffusive time. Once the interface has traversed the entire boundary cell, the adjacent solid node is converted into a liquid node. Each jump in Fig.~\ref{fig:melting_solute_sawtooth} therefore corresponds to a node-conversion event and to the associated update of the liquid domain. During freezing, instead, the salt-free solid grows into the brine channel and the available liquid volume decreases. Hence, the brine becomes locally enriched, forming the high-salinity layer next to the moving interface shown in Fig.~\ref{fig:scalar_contour_freezing}. Consistently, Fig.~\ref{fig:freezing_solute_sawtooth} illustrates that the total discrete solute content increases between successive node-conversion events. Once the interfacial liquid node has completed its transformation, it is removed from the liquid domain and converted into a solid node, corresponding to a downward jump in Fig.~\ref{fig:freezing_solute_sawtooth}.
\begin{figure}[pos=htbp]
\centering
\captionsetup[subfigure]{skip=1pt}

\begin{subfigure}[t]{0.35\textwidth}
\centering
\setlength\figurewidth{0.75\linewidth}
\setlength\figureheight{0.65\linewidth}
\parbox[t][0.215\textwidth][t]{\linewidth}{%
\centering
\begin{tikzpicture}[font=\sffamily\sansmath]

% ============================================================
% Custom concentration colormap
% ============================================================

\pgfplotsset{
    colormap={dario}{
        rgb255=(0,0,255)
        rgb255=(255,0,255)
        rgb255=(255,0,0)
        rgb255=(255,220,0)
    }
}

% ============================================================
% MAIN AXIS:
% concentration field + midline profile
% ============================================================

\begin{axis}[
name=fieldaxis,
width=\figurewidth,
height=\figureheight,
scale only axis,
xmin=0.2,
xmax=0.8,
ymin=0,
ymax=1,
xlabel={$x^*/L_x^*$},
ylabel={$y^*/L_y^*$},
axis line style={black},
tick align=inside,
tick pos=both,
grid=none,
clip=true,
scaled x ticks=false,
scaled y ticks=false,
xtick={0.3,0.4,0.5,0.6,0.7},
ytick={0,0.2,0.4,0.6,0.8,1},
tick label style={
    font=\fontsize{11}{13}\selectfont\sffamily\sansmath
},
label style={
    font=\fontsize{13}{15}\selectfont\sffamily\sansmath
},
xlabel style={yshift=1pt},
ylabel style={yshift=-2pt},
]

% ============================================================
% RASTER CONCENTRATION FIELD
% ============================================================

\addplot graphics[
xmin=0.2,
xmax=0.8,
ymin=0,
ymax=1
]
{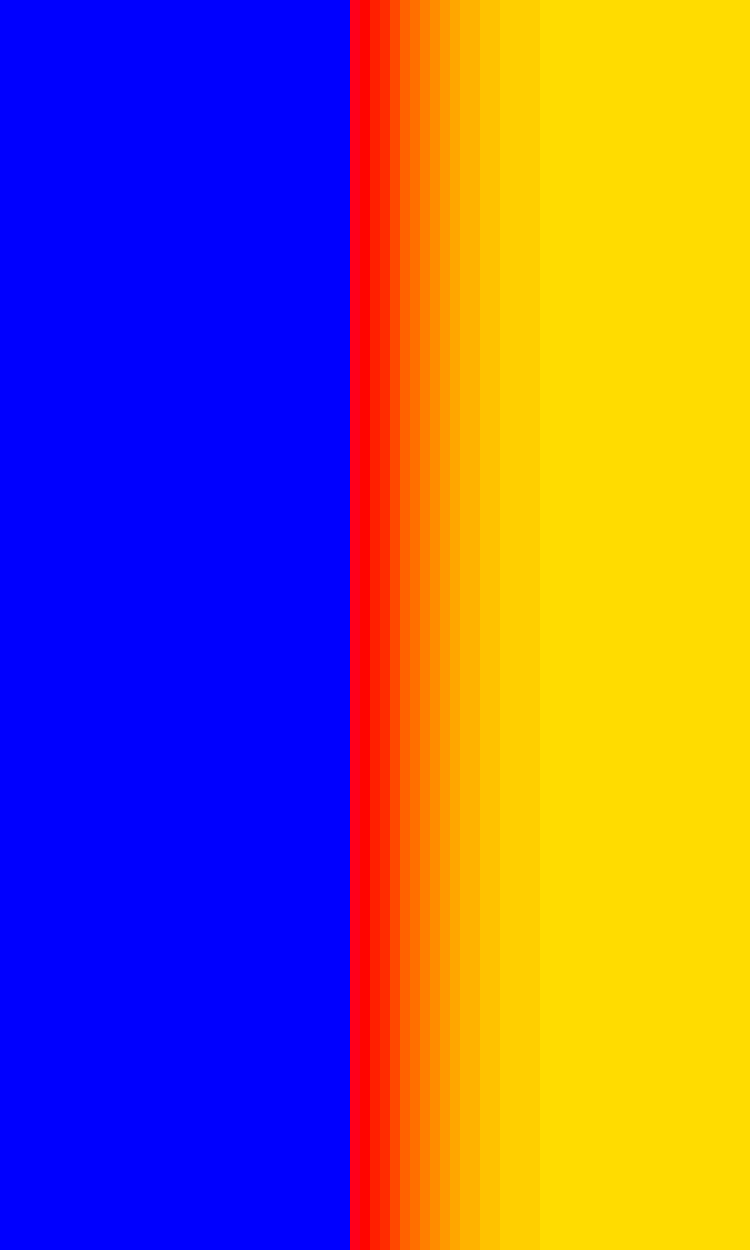};

\end{axis}

% ============================================================
% COMPACT COLORBAR - BOTTOM LEFT
% ============================================================

% ============================================================
% COMPACT COLORBAR - BOTTOM LEFT
% ============================================================

\begin{axis}[
at={(fieldaxis.south west)},
anchor=south east,
xshift=-38pt,
yshift=5pt,
width=4pt,
height=0.20\figureheight,
scale only axis,
hide axis,
colormap name=dario,
point meta min=0,
point meta max=1,
colorbar,
colorbar style={
    width=5pt,
    height=0.20\figureheight,
    at={(0,0)},
    anchor=south west,
    ytick={0,1},
    yticklabels={$0$,$1$},
    tick align=outside,
    tick label style={
        font=\fontsize{9}{10}\selectfont\sffamily\sansmath
    },
    ylabel={$C^*$},
    ylabel style={
        font=\fontsize{11}{13}\selectfont\sffamily\sansmath,
        at={(axis description cs:-1.3,0.5)},
        anchor=south
    }
}
]

% Dummy plot only to generate the color scale
\addplot[
scatter,
only marks,
mark size=0pt,
scatter src=explicit
]
coordinates {
    (0,0) [0]
    (0,1) [1]
};

\end{axis}

\end{tikzpicture}
}
\vspace{-1.05em}
\caption{Melting concentration field.}
\label{fig:scalar_contour_melting}
\end{subfigure}
\hfill
\begin{subfigure}[t]{0.35\textwidth}
\centering
\setlength\figurewidth{0.75\linewidth}
\setlength\figureheight{0.65\linewidth}
\parbox[t][0.215\textwidth][t]{\linewidth}{%
\centering
\begin{tikzpicture}[font=\sffamily\sansmath]

% ============================================================
% Custom concentration colormap
% ============================================================

\pgfplotsset{
    colormap={dario}{
        rgb255=(0,0,255)
        rgb255=(255,0,255)
        rgb255=(255,0,0)
        rgb255=(255,220,0)
    }
}

% ============================================================
% MAIN AXIS:
% freezing concentration field
% ============================================================

\begin{axis}[
name=fieldaxis,
width=\figurewidth,
height=\figureheight,
scale only axis,
xmin=0.2,
xmax=0.8,
ymin=0,
ymax=1,
xlabel={$x^*/L_x^*$},
ylabel={},
axis line style={black},
tick align=inside,
tick pos=both,
grid=none,
clip=true,
scaled x ticks=false,
scaled y ticks=false,
xtick={0.3,0.4,0.5,0.6,0.7},
ytick={0,0.2,0.4,0.6,0.8,1},
tick label style={
    font=\fontsize{11}{13}\selectfont\sffamily\sansmath
},
label style={
    font=\fontsize{13}{15}\selectfont\sffamily\sansmath
},
xlabel style={yshift=1pt},
ylabel style={yshift=-2pt}
]

% ============================================================
% RASTER CONCENTRATION FIELD
% ============================================================

\addplot graphics[
xmin=0.2,
xmax=0.8,
ymin=0,
ymax=1
]
{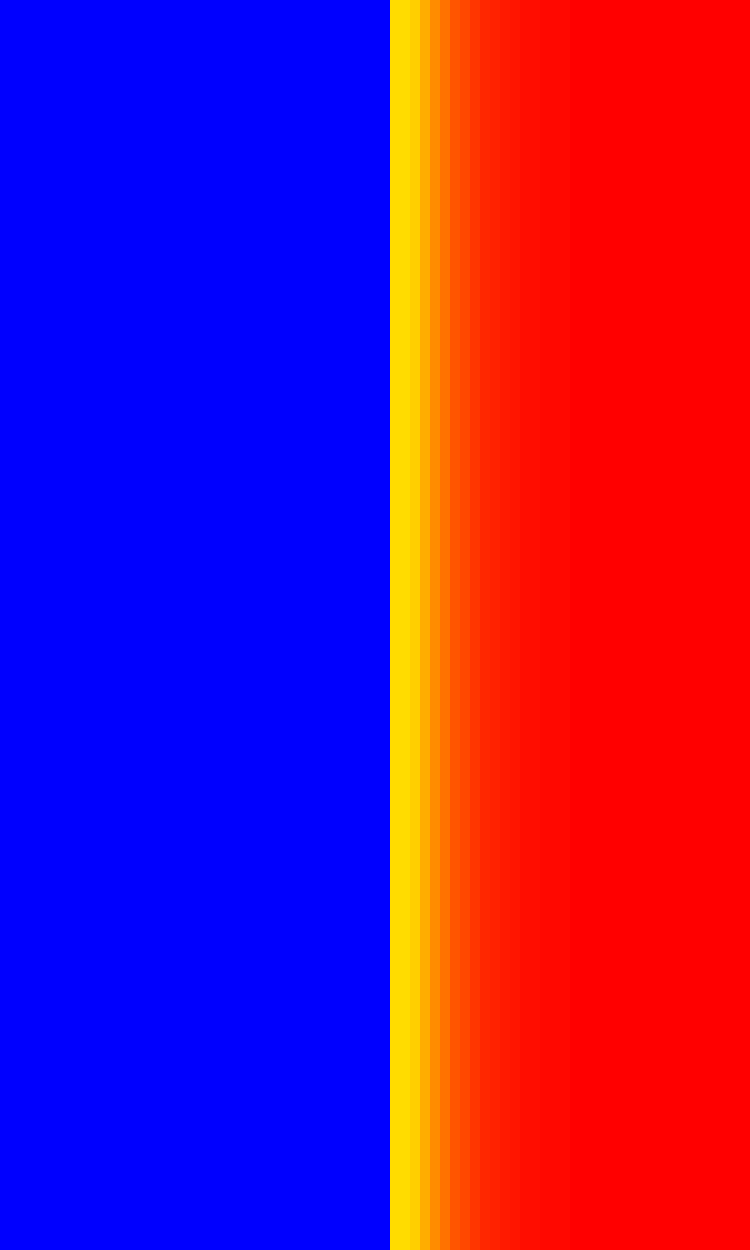};

\end{axis}

\end{tikzpicture}
}
\vspace{-1.05em}
\caption{Freezing concentration field.}
\label{fig:scalar_contour_freezing}
\end{subfigure}

\vspace{-0.45em}
\caption{Solutal response of the idealized brine channel during planar-interface motion at
$t^* = Dt/L^2 \simeq 0.0104$.}
\label{fig:scalar_contours_melting_freezing}
\end{figure}
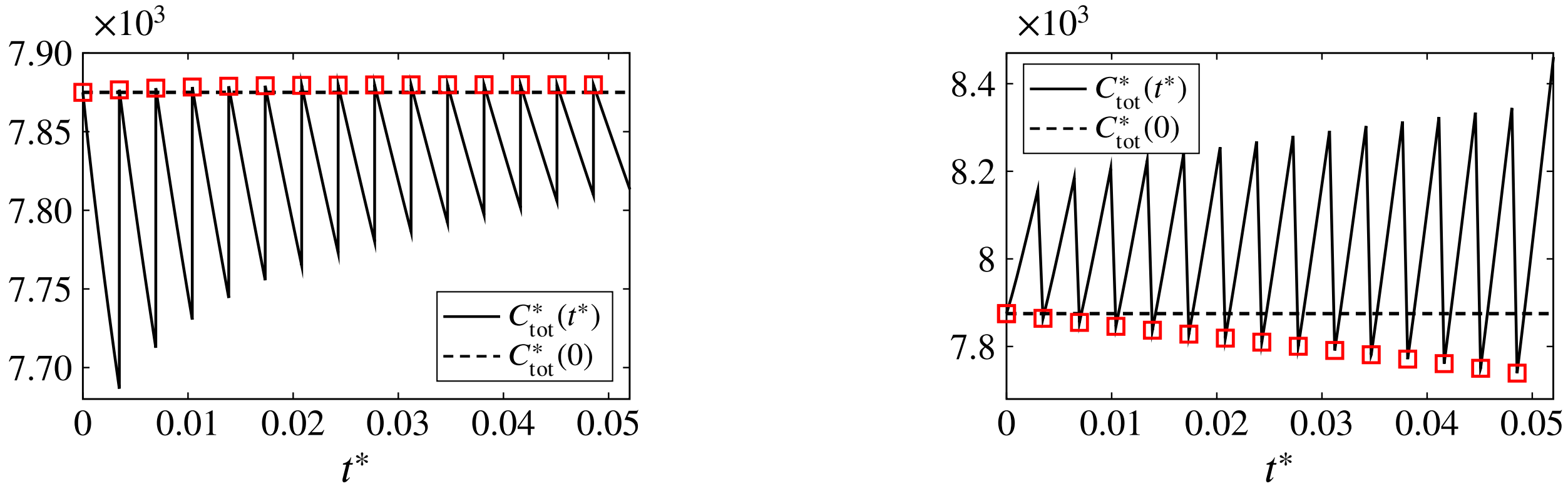
\begin{figure}[pos=htbp]
\centering
\begin{subfigure}[t]{0.39\textwidth}
\centering
\setlength\figurewidth{0.90\linewidth}
\setlength\figureheight{0.57\linewidth}
\begin{tikzpicture}[font=\sffamily\sansmath]

% Diffusive dimensionless time:
% t^* = alpha_L t_L / L_L^2
\def\alphaL{0.003333}
\def\LL{62}

\begin{axis}[
width=\figurewidth,
height=\figureheight,
scale only axis,
xmin=0,
xmax=0.052,
ymin=7.68,
ymax=7.90,
xlabel={$t^*$},
clip=false,
axis line style={black},
tick align=inside,
tick pos=both,
grid=none,
tick label style={
    font=\fontsize{11}{13}\selectfont\sffamily\sansmath
},
label style={
    font=\fontsize{13}{15}\selectfont\sffamily\sansmath
},
xlabel style={yshift=1pt},
scaled x ticks=false,
scaled y ticks=false,
xtick={0,0.01,0.02,0.03,0.04,0.05},
xticklabels={$0$,$0.01$,$0.02$,$0.03$,$0.04$,$0.05$},
ytick={7.70,7.75,7.80,7.85,7.90},
yticklabels={$7.70$,$7.75$,$7.80$,$7.85$,$7.90$},
legend style={
    font=\fontsize{9.5}{11}\selectfont\sffamily\sansmath,
    draw=black,
    fill=white,
    inner sep=1.5pt,
    row sep=0pt,
    at={(0.97,0.05)},
    anchor=south east
},
legend cell align=left
]

% ============================================================
% y-axis multiplier
% ============================================================

\node[
anchor=south west,
font=\fontsize{11}{13}\selectfont\sffamily\sansmath,
yshift=2pt
] at (rel axis cs:0,1)
{$\times 10^{3}$};

% ============================================================
% TOTAL DISCRETE SOLUTE CONTENT
% ============================================================

\addplot[
black,
solid,
line width=0.85pt,
mark=none
]
table[
x expr=\thisrow{t}*\alphaL/(\LL*\LL),
y expr=\thisrow{C}/1000
]
{figures/1d_mass_check_melting/melting_mass_check_wall_motion_reduced.dat};

\addlegendentry{$C^*_{\mathrm{tot}}(t^*)$}

% ============================================================
% INITIAL SOLUTE CONTENT
% ============================================================

\addplot[
black,
densely dashed,
line width=1.05pt,
mark=none
]
table[
x expr=\thisrow{t}*\alphaL/(\LL*\LL),
y expr=\thisrow{C_init}/1000
]
{figures/1d_mass_check_melting/melting_mass_check_wall_motion_reduced.dat};

\addlegendentry{$C^*_{\mathrm{tot}}(0)$}

% ============================================================
% POST-SWITCH PEAKS
% ============================================================

\addplot[
red,
only marks,
mark=square,
mark size=2.4pt,
mark options={
    draw=red,
    fill=none,
    line width=0.9pt
},
forget plot
]
table[
x expr=\thisrow{t}*\alphaL/(\LL*\LL),
y expr={
    \thisrow{switch} > 0.5
    ? \thisrow{C}/1000
    : nan
}
]
{figures/1d_mass_check_melting/melting_mass_check_wall_motion_reduced.dat};

\end{axis}

\end{tikzpicture}
\vspace{-1.4em}
\caption{Melting, $A_2>0$.}
\label{fig:melting_solute_sawtooth}
\end{subfigure}
\hfill
\begin{subfigure}[t]{0.39\textwidth}
\centering
\setlength\figurewidth{0.90\linewidth}
\setlength\figureheight{0.57\linewidth}
\begin{tikzpicture}[font=\sffamily\sansmath]

% Diffusive dimensionless time:
% t^* = alpha_L t_L / L_L^2
\def\alphaL{0.003333}
\def\LL{62}

\begin{axis}[
width=\figurewidth,
height=\figureheight,
scale only axis,
xmin=0,
xmax=0.052,
ymin=7.68,
ymax=8.47,
xlabel={$t^*$},
clip=false,
axis line style={black},
tick align=inside,
tick pos=both,
grid=none,
tick label style={
    font=\fontsize{11}{13}\selectfont\sffamily\sansmath
},
label style={
    font=\fontsize{13}{15}\selectfont\sffamily\sansmath
},
xlabel style={yshift=1pt},
scaled x ticks=false,
scaled y ticks=false,
xtick={0,0.01,0.02,0.03,0.04,0.05},
xticklabels={$0$,$0.01$,$0.02$,$0.03$,$0.04$,$0.05$},
yticklabel style={
    /pgf/number format/fixed,
    /pgf/number format/precision=1
},
legend style={
    font=\fontsize{9.5}{11}\selectfont\sffamily\sansmath,
    draw=black,
    fill=white,
    inner sep=1.5pt,
    row sep=0pt,
    at={(0.03,0.97)},
    anchor=north west
},
legend cell align=left
]

% ============================================================
% y-axis scale
% ============================================================

\node[
anchor=south west,
font=\fontsize{11}{13}\selectfont\sffamily\sansmath,
yshift=2pt
] at (rel axis cs:0,1)
{$\times 10^{3}$};

% ============================================================
% TOTAL DISCRETE SOLUTE CONTENT
% ============================================================

\addplot[
black,
solid,
line width=0.85pt,
mark=none
]
table[
x expr=\thisrow{t}*\alphaL/(\LL*\LL),
y expr=\thisrow{C}/1000
]
{figures/1d_mass_check_freezing/mass_check_wall_motion_reduced.dat};

\addlegendentry{$C^*_{\mathrm{tot}}(t^*)$}

% ============================================================
% INITIAL SOLUTE CONTENT
% ============================================================

\addplot[
black,
densely dashed,
line width=1.05pt,
mark=none
]
table[
x expr=\thisrow{t}*\alphaL/(\LL*\LL),
y expr=\thisrow{C_init}/1000
]
{figures/1d_mass_check_freezing/mass_check_wall_motion_reduced.dat};

\addlegendentry{$C^*_{\mathrm{tot}}(0)$}

% ============================================================
% VOP SWITCHING POINTS
% ============================================================

\addplot[
red,
only marks,
mark=square,
mark size=2.4pt,
mark options={
    draw=red,
    fill=none,
    line width=0.9pt
},
forget plot
]
table[
x expr=\thisrow{t}*\alphaL/(\LL*\LL),
y expr={
    \thisrow{base} > 0.5
    ? \thisrow{C}/1000
    : nan
}
]
{figures/1d_mass_check_freezing/mass_check_wall_motion_reduced.dat};

\end{axis}

\end{tikzpicture}
\vspace{-1.4em}
\caption{Freezing, $A_2<0$.}
\label{fig:freezing_solute_sawtooth}
\end{subfigure}
\vspace{-0.2em}
\caption{Evolution of the total discrete solute content during planar-interface phase change.}
\label{fig:total_solute_content_melting_freezing}
\end{figure}
Figures~\ref{fig:melting_switch_error} and~\ref{fig:freezing_switch_error} show the normalised discrepancy between the total discrete solute content after each node-conversion event, $C^*_{\mathrm{sw,tot}}$, and its initial value, $C^*_{\mathrm{tot}}(0)$. Since the system is closed to external solute fluxes and the solid phase is salt-free, $C^*_{\mathrm{sw,tot}}$ should return to a value close to $C^*_{\mathrm{tot}}(0)$ after each switch if the moving Robin condition and the VOP migration procedure are coupled consistently. For melting, the discrepancy remains very small and progressively approaches zero. During freezing, a small systematic drift persists, but the error remains limited over repeated node-conversion events, showing that the global solute balance is retained to a good approximation. This larger sensitivity is consistent with the fact that freezing progressively reduces the active liquid domain and repeatedly involves the negative-$\gamma$ branch of the boundary scheme, including crossings of the regularized region around $\gamma=0$. Overall, the present coupling with the VOP migration procedure provides a near-conservative solute-redistribution treatment, also in the more sensitive freezing case.

\subsection{Coupled temperature--concentration simulations}
\label{subsec:coupled_sim}
Having independently assessed the thermal and solutal components of the model in Secs.~\ref{subsec:thermal_overview_sim} and~\ref{subsec:solut_sim}, the two transport processes are now coupled to study the thermodynamic response of a mushy layer. In Sec.~\ref{subsec:coupled_sim_anal}, approximate analytical solutions are derived for sinusoidal temperature and concentration variations in an ideal diffusive mushy layer, providing reference trends for the subsequent numerical comparison. The coupled LBM formulation is then applied in Sec.~\ref{subsec:coupled_sim_num} to an idealized pore-scale segment of a sea ice brine channel with a planar solid--liquid interface.

\subsubsection{Approximate analytical solutions for sinusoidal variations in a diffusive ideal mushy layer}
\label{subsec:coupled_sim_anal}
Based on the derivations of~\cite{worster92} and~\cite{butler11}, an analytical description of mushy layers in a diffusive regime can be obtained under the ideal continuum mushy-layer assumption. In this limit, the one-dimensional energy equation for an ideal mushy layer reads $
\overline{\rho c_p} \, \partial T / \partial t
=
\partial /\partial z
\left(
\overline{k} \, \partial T / \partial z
\right)
-
\rho_s L \, \partial\phi / \partial t$, where $\phi$ denotes the local liquid volume fraction (or porosity),
$\overline{\rho c_p}
=
\phi\,\rho_l c_{p,l}
+
(1-\phi)\,\rho_s c_{p,s}$
is the phase-averaged volumetric heat capacity, and
$\overline{k}
=
\phi\,k_l
+
(1-\phi)\,k_s$ is the phase-averaged thermal conductivity. Dimensionless variables are then introduced as in Eqs.~\eqref{eq:nondim_theta_c} and~\eqref{eq:nondim_R_t}, together with $z^*=z / H$, where $H$ is the characteristic axial length. Substituting these variables into the one-dimensional energy equation and dividing by $\rho_l c_{p,l}\Delta T D/H^2$, with $\Delta T=T_{\max}-T_0$, one obtains
\begin{equation}
\overline{\rho c_p}^{\,\prime}
\frac{\partial\theta}{\partial t^*}
=
\frac{\overline{k}'}{\epsilon}
\frac{\partial^2  \theta}{\partial z^{*2}}
-
S\frac{\partial\phi}{\partial t^*},
\label{eq:temperature_dimensionless}
\end{equation}
where $
\overline{\rho c_p}^{\,\prime}
= \overline{\rho c_p} / \rho_l c_{p,l},$ $\overline{k}'=\overline{k} / k_l,$ $
\kappa_l=k_l / \rho_l c_{p,l},$ $
\epsilon= D / \kappa_l,$ $S= \rho_s L / \rho_l c_{p,l}\Delta T.$ Here, $\epsilon$ is the inverse Lewis number and $S$ is the Stefan number. For the solute field, the one-dimensional ideal mushy-layer conservation equation in the diffusive limit is $
\phi \, \partial C / \partial t
=
\partial / \partial z
\left(
\overline{D} \, \partial C / \partial z
\right)
+
( \rho_s / \rho_l )
\left(C_s-C\right)
\partial\phi / \partial t,$
with $\overline{D} = \phi D$. Substituting the dimensionless variables and dividing by $\Delta C D/H^2$, with $\Delta C=C_{\max}-C_0$, one obtains
\begin{equation}
\phi
\frac{\partial c}{\partial t^*}
=
\phi
\frac{\partial^2 c}{\partial z^{*2}}
+
\rho_s'
\left(
\mathcal{C}-c
\right)
\frac{\partial\phi}{\partial t^*},
\label{eq:concentration_dimensionless}
\end{equation}
where $D'=\overline{D}/D=\phi$, $\rho_s'=\rho_s/\rho_l$, and
$\mathcal{C}=(C_s-C_0)/\Delta C$. The variable $\mathcal{C}$ is referred to as the concentration ratio. In this analytical formulation, axial gradients of the porosity are neglected, so that $\phi$ enters the transport coefficients locally but does not contribute through additional spatial-derivative terms. The ideal mushy-layer model further assumes local thermodynamic equilibrium, $\theta=c$, throughout the liquid phase. This relation can be used to rearrange the governing equations into two coupled evolution equations for $\theta=c$ and $\phi$. Defining
$X=\partial\theta/\partial t^*$ and
$Y=\partial\phi/\partial t^*$, the conservation equations become
Eq.~\eqref{eq:theta_c_system}. Solving this system for $X$ gives Eq.~\eqref{eq:theta_analytical_diff}.
\begin{gather}
\begin{cases}
\overline{\rho c_p}'X + SY
=
\overline{k}'
(\partial^2 \theta / \partial z^{*2}) \, / \, \epsilon,
\\[6pt]
\phi X
-
\rho_s'(\mathcal{C}-\theta)Y
=
\phi\,
(\partial^2 \theta / \partial z^{*2}) .
\end{cases}
\label{eq:theta_c_system}
\\[0.2em]
\frac{\partial \theta}{\partial t^*}
=
A_{1,2}^{(\mathrm{diff})}
\frac{\partial^2 \theta}{\partial z^{*2}},
\qquad
\text{with}
\qquad
A_{1,2}^{(\mathrm{diff})}
=
\frac{
(\overline{k}'/\epsilon)\rho_s'(\mathcal{C}-\theta)
+
S\phi
}{
\overline{\rho c_p}'\rho_s'(\mathcal{C}-\theta)
+
S\phi
}.
\label{eq:theta_analytical_diff}
\end{gather}
Since temperature and concentration evolve together under the local-equilibrium assumption, the system is governed by a single effective diffusivity, denoted by $A_{1,2}^{(\mathrm{diff})}$. If $\theta \ll \mathcal{C}$ and the variations in $\phi$ remain small, this coefficient can be approximated as constant. In this limit, $\mathcal{C}-\theta \simeq \mathcal{C}$ and $\phi\simeq\phi_0$, so that
\begin{equation}
A_{1,2}^{(\mathrm{diff})}
\simeq
[(\overline{k}' / \epsilon)\rho'_s\mathcal{C}
+
S\phi_0
] \, / \, [
\overline{\rho c_p}'\rho'_s\mathcal{C}
+
S\phi_0].
\label{eq:A12_diff_constant}
\end{equation}
Following the approach in~\cite{amiri24}, the resulting diffusion equation admits a decaying periodic solution. For a sinusoidal initial condition and defining $\eta=4\pi^2A_{1,2}^{(\mathrm{diff})}$, one obtains
\begin{equation}
\theta(z^*,t^*)
=
\sin\left(2\pi z^*\right)
\exp\left(
-\eta t^*
\right)
\quad \Longrightarrow \quad
\theta(z,t)
=
\sin\left(2\pi z / H\right)
\exp\left(
-\eta Dt/H^2
\right).
\label{eq:analytical_theta}
\end{equation}
The porosity evolution can be derived from the second equation in~\eqref{eq:theta_c_system} by substituting the temperature equation~\eqref{eq:theta_analytical_diff}:
\begin{equation}
\frac{\partial \phi}{\partial t^*}
=
\left[(A_{1,2}^{(\mathrm{diff})}-1)\phi \, / 
\, ( \, \rho_s'(\mathcal{C}-\theta) \,)\right]
\frac{\partial^2\theta}{\partial z^{*2}}.
\label{eq:phi_eq2}
\end{equation}
Under the assumptions $\mathcal{C}\gg\theta$ and
\(\phi\simeq\phi_0\), the coefficient multiplying the temperature curvature can be approximated as a constant $K =
(A_{1,2}^{(\mathrm{diff})}-1)\phi_0/
\rho_s' \mathcal{C}$. Taking the second derivative of the temperature solution~\eqref{eq:analytical_theta} and integrating in time with \(\phi(z^*,0)=\phi_0\), the porosity field becomes:
\begin{equation}
\phi(z^*,t^*)
=
\phi_0
-
\left( K \, / \, A_{1,2}^{(\mathrm{diff})} \right)
\sin\left(2\pi z^*\right)
\left[
\, 1-
\exp\left(
-\eta t^*
\right)
\, \right].
\label{eq:phi_sol_anal}
\end{equation}
The expression~\eqref{eq:phi_sol_anal} can also be written in terms of the dimensional variables $z$ and $t$, following the same transformation used above for $\theta$.

\subsubsection{Coupled LBM simulations for sea ice mushy layers}
\label{subsec:coupled_sim_num}
\begin{figure}[pos=htbp]
\centering
\scalebox{0.7}[0.7]{%
    \begin{tikzpicture}[
    font=\sffamily\large,
    line join=round,
    line cap=round,
    scale=0.95
]

% ================================================================
% Geometry
% ================================================================

% Front face
\coordinate (A) at (0,0);
\coordinate (B) at (6,0);
\coordinate (C) at (6,3);
\coordinate (D) at (0,3);

% Back face
\coordinate (Ab) at (1.6,1.1);
\coordinate (Bb) at (7.6,1.1);
\coordinate (Cb) at (7.6,4.1);
\coordinate (Db) at (1.6,4.1);

% Solid-liquid interface: x = 3
\coordinate (M0)  at (3,0);
\coordinate (M1)  at (3,3);
\coordinate (M0b) at (4.6,1.1);
\coordinate (M1b) at (4.6,4.1);

% ================================================================
% Filled regions
% ================================================================

\fill[gray!25] (A) -- (M0) -- (M1) -- (D) -- cycle;
\fill[cyan!20] (M0) -- (B) -- (C) -- (M1) -- cycle;

\fill[gray!25] (D) -- (M1) -- (M1b) -- (Db) -- cycle;
\fill[cyan!20] (M1) -- (C) -- (Cb) -- (M1b) -- cycle;

\fill[cyan!20] (B) -- (Bb) -- (Cb) -- (C) -- cycle;

% ================================================================
% Homogeneous cubic lattice
%
% Logical lattice spacing = 1 in x, y, z.
%
% x-nodes:
%   solid = 0.5, 1.5, 2.5   --> 3 columns
%   fluid = 3.5, 4.5, 5.5   --> 3 columns
%
% The interface x=3 is exactly halfway between
% the last solid column (2.5) and first fluid column (3.5).
%
% z-nodes:
%   0.5, 1.5, 2.5
%
% y/depth planes:
%   w = 0,1,2
%
% Since the total projected depth is (1.6,1.1),
% one depth lattice step is (0.8,0.55).
% ================================================================

% Lattice style: deliberately light to avoid visual overcrowding
\tikzset{
    latticeedge/.style={draw=black!55, line width=0.45pt},
    latticenode/.style={fill=black}
}

% --- Edges parallel to x -----------------------------------------
\foreach \zz in {0.5,1.5,2.5}{
    \foreach \ww in {0,1,2}{
        \draw[latticeedge]
        ({0.5+0.8*\ww},{\zz+0.55*\ww}) --
        ({5.5+0.8*\ww},{\zz+0.55*\ww});
    }
}

% --- Edges parallel to z -----------------------------------------
\foreach \xx in {0.5,1.5,2.5,3.5,4.5,5.5}{
    \foreach \ww in {0,1,2}{
        \draw[latticeedge]
        ({\xx+0.8*\ww},{0.5+0.55*\ww}) --
        ({\xx+0.8*\ww},{2.5+0.55*\ww});
    }
}

% --- Edges parallel to y/depth -----------------------------------
\foreach \xx in {0.5,1.5,2.5,3.5,4.5,5.5}{
    \foreach \zz in {0.5,1.5,2.5}{
        \draw[latticeedge]
        (\xx,\zz) --
        ({\xx+0.8},{\zz+0.55}) --
        ({\xx+1.6},{\zz+1.10});
    }
}

% --- Nodes --------------------------------------------------------
\foreach \xx in {0.5,1.5,2.5,3.5,4.5,5.5}{
    \foreach \zz in {0.5,1.5,2.5}{
        \foreach \ww in {0,1,2}{
            \fill[latticenode]
            ({\xx+0.8*\ww},{\zz+0.55*\ww})
            circle[radius=1.8pt];
        }
    }
}

% ================================================================
% Outer geometry edges
% ================================================================

\draw[black, line width=1.15pt] (A) -- (B);
\draw[black, line width=1.15pt] (A) -- (D);
\draw[black, line width=1.15pt] (D) -- (Db);
\draw[black, line width=1.15pt] (Db) -- (Cb);
\draw[black, line width=1.15pt] (Cb) -- (Bb);
\draw[black, line width=1.15pt] (Bb) -- (B);
\draw[black, line width=1.15pt] (B) -- (C);
\draw[black, line width=1.15pt] (C) -- (Cb);
\draw[black, line width=1.15pt] (D) -- (C);

% ================================================================
% Solid-liquid interface
% Drawn last, so it is clearly visible above the lattice.
% No lattice node lies on the interface.
% ================================================================

\draw[black, line width=1.45pt]
(M0) -- (M1) -- (M1b);

% ================================================================
% Coordinate axes -- moved farther to the left
% ================================================================

\coordinate (Oaxis) at (-3.15,-0.45);

\draw[->, black, line width=0.95pt]
(Oaxis) -- ++(1.45,0)
node[below right=-1pt] {$x^{*}$};

\draw[->, black, line width=0.95pt]
(Oaxis) -- ++(0,1.35)
node[above left=-1pt] {$z^{*}$};

\draw[->, black, line width=0.95pt]
(Oaxis) -- ++(0.95,0.95)
node[above right=-1pt] {$y^{*}$};

% ================================================================
% Dimension lines
% ================================================================

\tikzset{
    dimline/.style={<->, >=latex, black, line width=0.85pt},
    dimext/.style={black, line width=0.55pt}
}

% ------------------------------------------------
% Length L*: front horizontal dimension
% ------------------------------------------------
\coordinate (Lleft)  at (0,-0.58);
\coordinate (Lright) at (6,-0.58);

% Extension lines
\draw[dimext] (A) -- (0,-0.72);
\draw[dimext] (B) -- (6,-0.72);

% Dimension arrow
\draw[dimline] (Lleft) -- (Lright)
node[midway, below=3pt] {$L^{*}$};

% ------------------------------------------------
% Depth W*: parallel to the y/depth direction
% Shifted outside the lower-right corner
% ------------------------------------------------
\coordinate (Wfront) at (6.28,-0.30);
\coordinate (Wback)  at (7.88,0.80);

% Extension lines from the corresponding front/back corners
\draw[dimext] (B)  -- (Wfront);
\draw[dimext] (Bb) -- (Wback);

% Dimension arrow
\draw[dimline] (Wfront) -- (Wback)
node[midway, below right=1pt] {$W^{*}$};

% ------------------------------------------------
% Height H*: vertical dimension on the far right
% ------------------------------------------------
\coordinate (Hbottom) at (8.12,1.10);
\coordinate (Htop)    at (8.12,4.10);

% Extension lines
\draw[dimext] (Bb) -- (Hbottom);
\draw[dimext] (Cb) -- (Htop);

% Dimension arrow
\draw[dimline] (Hbottom) -- (Htop)
node[midway, right=4pt] {$H^{*}$};

\end{tikzpicture}%
}
\caption{Schematic representation of the three-dimensional sea ice domain and its lattice discretization, with ghost nodes omitted for clarity and solid and liquid regions separated by a planar interface. The dimensions shown along $x^*$ and $z^*$ are schematic: the actual values of $L^*$ and $H^*$ vary among different test cases.}
\label{fig:amiri_geom_3d}
\end{figure}
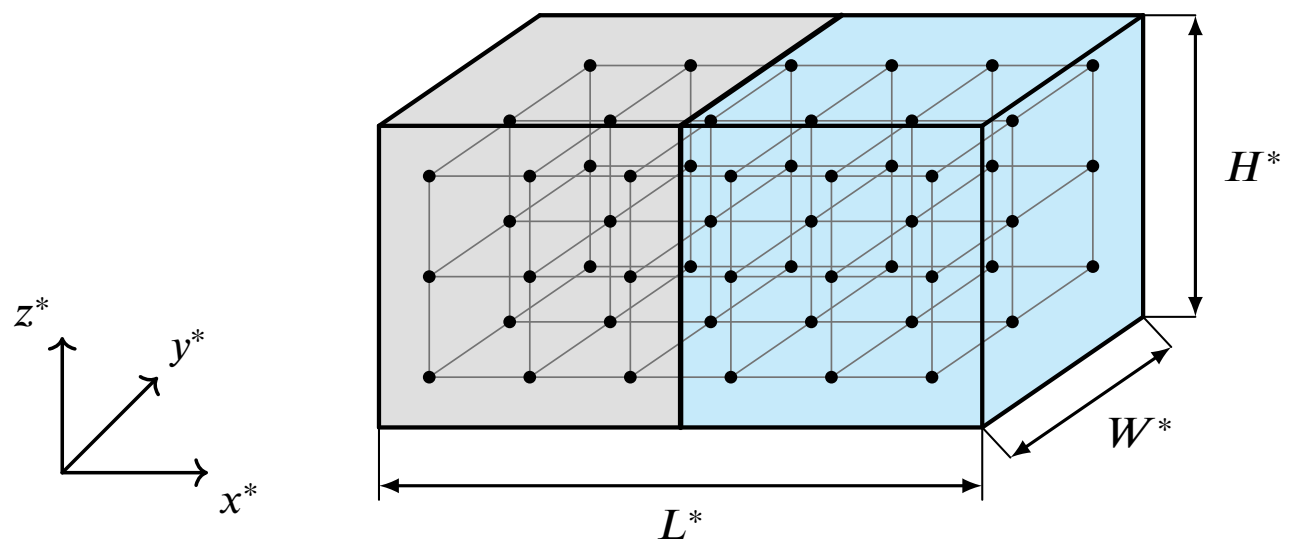

LBM simulations are performed to compare the coupled numerical results with the approximate analytical solution derived in Sec.~\ref{subsec:coupled_sim_anal}. The computational domain represents an idealized pore-scale segment of a sea ice brine channel, as illustrated in Fig.~\ref{fig:amiri_geom_3d}. It is initially half solid and half liquid, corresponding to $\phi_0=0.5$. Periodic boundary conditions are imposed along the $z$- and $y$-directions, while symmetry conditions, i.e. zero normal gradients of $T^*$ and $C^*$, are applied at the external boundaries along the $x$-direction.
The dimensionless parameters adopted in the simulations are summarized in Table~\ref{tab:high_diff_parameters}. These values are derived from the data reported by Butler~\cite{butler11}, Peppin \emph{et al.}~\cite{peppin08}, and Hallworth \emph{et al.}~\cite{hallworth05}. Each column corresponds to a different simulated configuration, selected to investigate distinct regimes and test cases.
\begin{table}[htbp]
\centering
\caption{Dimensionless parameters adopted for the aqueous NH$_4$Cl configurations.}
\label{tab:high_diff_parameters}
\begin{tabular}{@{}lccc@{}}
\toprule
Parameter
& NH$_4$Cl 
& Modified NH$_4$Cl 
& Modified NH$_4$Cl (2)\\
\midrule
$\rho_s'$        & 1.0      & 1.0      & 1.0 \\
$c_{p,s}'$       & 0.42     & 0.42     & 0.42 \\
$S$              & 4.01     & 4.01     & 4.01 \\
$\mathcal{C}$    & 11.42    & 11.42    & 1.00 \\
$\phi_0$         & 0.50     & 0.50     & 0.50 \\
$k_s'$           & 4.07     & 1.05     & 1.05 \\
$\epsilon$       & 0.01     & 0.10     & 0.10 \\
\bottomrule
\end{tabular}
\end{table}
The reference lattice temperature is set to $T_0^*=1$, while the maximum value of the imposed sinusoidal perturbation is $T_{\max}^*=1.01$, giving $\Delta T^*=T_{\max}^*-T_0^*=0.01$. The numerical liquidus slope is prescribed as $m^*=-514.286$, with $m^*=-\Delta T_{\mathrm{phys}}/\Delta C_{\mathrm{phys}}$, $\Delta T_{\mathrm{phys}}=36\,^{\circ}\mathrm{C}$, and $\Delta C_{\mathrm{phys}}=0.07$~\cite{amiri24}. This slope maps the imposed temperature variation into the corresponding concentration variation through $\Delta C^*=\Delta T^*/m^*$. Because $m^*<0$, the concentration perturbation is anticorrelated with the imposed temperature perturbation. Since the solid phase is assumed to be salt-free, $C_s^*=0$, the reference liquid concentration is determined from the prescribed concentration ratio $\mathcal{C}=(C_s^*-C_0^*)/\Delta C^*$, yielding $C_0^*=-\mathcal{C}\Delta C^*$. Two values of $\mathcal{C}$ are considered: $\mathcal{C}=11.42$, i.e. the large-concentration-ratio regime associated with a NH$_4$Cl reference configuration~\cite{amiri24}, and $\mathcal{C}=1$, representative of the lower-concentration-ratio behaviour expected in sea-ice-like systems~\cite{butler11}. In all configurations, the fields are initialized on the linear liquidus by prescribing $T^*(z,0)=T_0^*+\Delta T^*\sin(2\pi z^*)$ and computing the liquid concentration as $C^*(z,0)=(T^*(z,0)-T_m^*)/m^*=C_0^*+\Delta C^*\sin(2\pi z^*)$, with $T_m^*=T_0^*-m^*C_0^*$.

\medskip
\noindent\textit{Scale-separation analysis}\par
\smallskip
\noindent
The ideal mushy-layer formulation assumes local thermodynamic equilibrium, so that the dimensionless temperature and concentration satisfy $\theta=c$ throughout the mushy region. In the present pore-scale configuration, this continuum assumption can be recovered only if transverse diffusion is sufficiently fast compared with axial diffusion, so that temperature and concentration remain nearly uniform across the pore and evolve predominantly along the axial direction. Since diffusive time scales scale with the square of the characteristic length, this separation is favoured by a high-aspect-ratio geometry $H^* \gg L^*$, where $L^*$ is the characteristic transverse pore scale and $H^*$ is the axial length scale in lattice units. A preliminary aspect-ratio study is therefore performed to assess the sensitivity of the numerical solution to the domain dimensions and to identify a configuration consistent with the one-dimensional analytical assumption. Two domains are considered: a $N_X \times N_Y \times N_Z = 20 \times 200 \times 20$ lattice and a longer $20 \times 800 \times 20$ lattice. For this scale-separation analysis, the Modified NH$_4$Cl configuration reported in Table~\ref{tab:high_diff_parameters} is adopted. In particular, the transport coefficients are changed with respect to the reference NH$_4$Cl configuration by setting $k_s'=1.05$ and $\epsilon=0.10$, while preserving faster thermal than solutal diffusion. These values accelerate the thermal and solutal transients, reducing the computational time required to reach the regime of interest. The interface evolution is computed using the nodal form of the kinetic relation, Eq.~\eqref{eq:rate_nodal}, in order to reduce the computational cost compared with the link-based implementation. The kinetic coefficient $k_r$ is assigned a sufficiently large value to accelerate interface motion while retaining numerical stability; further details on this parameter are provided in the following paragraph.
\begin{figure}[pos=htbp]
\centering
\begin{subfigure}[t]{0.47\textwidth}
\centering
\setlength\figurewidth{0.78\linewidth}
\setlength\figureheight{0.56\linewidth}
\begin{tikzpicture}[font=\sffamily\sansmath]

\begin{axis}[
width=\figurewidth,
height=\figureheight,
scale only axis,
xmin=0,
xmax=1,
ymin=-1.05,
ymax=1.05,
xlabel={$z^*/H^*$},
ylabel={$c$ and $\theta$},
axis line style={black},
tick align=inside,
tick pos=both,
xtick={0,0.2,0.4,0.6,0.8,1.0},
ytick={-1.0,-0.5,0,0.5,1.0},
xticklabels={$0$,$0.2$,$0.4$,$0.6$,$0.8$,$1.0$},
yticklabels={$-1.0$,$-0.5$,$0$,$0.5$,$1.0$},
tick label style={
    font=\fontsize{11}{8}\selectfont\sffamily\sansmath
},
label style={
    font=\fontsize{13}{8}\selectfont\sffamily\sansmath
},
legend style={
    font=\fontsize{8.7}{8}\selectfont\sffamily\sansmath,
    anchor=south west,
    at={(0.03,0.03)},
    draw=black,
    fill=white,
    inner sep=1.5pt,
    row sep=0pt
},
legend cell align=left,
clip=true,
]

% theta all dump 00000
%\addplot[
%black,
%dotted,
%line width=0.95pt,
%mark=none
%]
%table[x index=0,y index=1] {figures/amiri_scales/narrow/interface_equilibrium_narrow.dat};
%\addlegendentry{$\overline{\theta}(t^*=0$};

% theta all dump 18000
\addplot[
orange,
solid,
line width=1.05pt,
mark=o,
mark repeat=120,
mark size=2.4pt,
mark options={solid, draw=orange, fill=white, line width=0.65pt}
]
table[x index=0,y index=2] {figures/amiri_scales/narrow/interface_equilibrium_narrow.dat};
\addlegendentry{$\overline{\theta}$};

% c liquid mean dump 18000
\addplot[
blue,
dotted,
line width=1.00pt,
mark=diamond,
mark repeat=120,
mark size=2.4pt,
mark options={solid, draw=blue, fill=white, line width=0.65pt}
]
table[x index=0,y index=3] {figures/amiri_scales/narrow/interface_equilibrium_narrow.dat};
\addlegendentry{$\overline{c}$};

% theta interface dump 18000
\addplot[
green!60!black,
solid,
line width=1.05pt,
mark=triangle,
mark repeat=120,
mark size=2.6pt,
mark options={solid, draw=green!60!black, fill=green!60!black, line width=0.55pt}
]
table[x index=0,y index=4] {figures/amiri_scales/narrow/interface_equilibrium_narrow.dat};
\addlegendentry{$\overline{\theta}_{\mathrm{interf}}$};

% c interface dump 18000
\addplot[
red,
dash pattern=on 4pt off 2pt,
line width=1.05pt,
mark=x,
mark repeat=120,
mark size=2.8pt,
mark options={solid, draw=red, line width=0.65pt}
]
table[x index=0,y index=5] {figures/amiri_scales/narrow/interface_equilibrium_narrow.dat};
\addlegendentry{$\overline{c}_{\mathrm{interf}}$};

% theta analytic modo 2
\addplot[
cyan!70!blue,
solid,
line width=1.10pt,
mark=square,
mark repeat=120,
mark size=2.2pt,
mark options={solid, draw=cyan!70!blue, fill=white, line width=0.65pt}
]
table[x index=0,y index=6] {figures/amiri_scales/narrow/interface_equilibrium_narrow.dat};
\addlegendentry{$\theta_{\mathrm{analytic}}$};

% quadratino sul picco
\coordinate (zoomA) at (axis cs:0.220,0.880);
\coordinate (zoomB) at (axis cs:0.255,1.030);
\coordinate (zoomC) at (axis cs:0.220,1.030);
\coordinate (zoomD) at (axis cs:0.255,0.880);

\draw[black, line width=0.4pt]
(axis cs:0.220,0.880) rectangle (axis cs:0.255,1.030);

% scritte dPhi/dt
\node[font=\fontsize{9}{6}\selectfont\sffamily\sansmath, anchor=center]
at (axis cs:0.245,0.42) {$d\phi/dt < 0$};

\node[font=\fontsize{9}{6}\selectfont\sffamily\sansmath, anchor=center]
at (axis cs:0.755,-0.28) {$d\phi/dt > 0$};

\end{axis}

% -------------------------------------------------------------------------
% Zoom inset sul picco, senza labels né numeri sugli assi
% -------------------------------------------------------------------------
\begin{axis}[
    at={(rel axis cs:0.95,0.92)},
    anchor=north east,
    width=0.38\figurewidth,
    height=0.34\figureheight,
    scale only axis,
    xmin=0.225,
    xmax=0.275,
    ymin=0.920,
    ymax=0.995,
    grid=major,
    axis background/.style={fill=white},
    axis line style={black},
    tick align=inside,
    tick pos=both,
    xtick=\empty,
    ytick=\empty,
    xlabel={},
    ylabel={},
    clip=true,
]

\addplot[
orange,
solid,
line width=1.05pt,
mark=o,
mark repeat=50,
mark size=2.4pt,
mark options={solid, draw=orange, fill=white, line width=0.65pt}
]
table[
x index=0,
y index=2,
restrict x to domain=0.215:0.275,
each nth point=1
] {figures/amiri_scales/narrow/interface_equilibrium_narrow.dat};

\addplot[
blue,
dotted,
line width=1.00pt,
mark=diamond,
mark repeat=50,
mark size=2.4pt,
mark options={solid, draw=blue, fill=white, line width=0.65pt}
]
table[
x index=0,
y index=3,
restrict x to domain=0.215:0.275,
each nth point=1
] {figures/amiri_scales/narrow/interface_equilibrium_narrow.dat};

\addplot[
green!60!black,
solid,
line width=1.05pt,
mark=triangle,
mark repeat=50,
mark size=2.6pt,
mark options={solid, draw=green!60!black, fill=green!60!black, line width=0.55pt}
]
table[
x index=0,
y index=4,
restrict x to domain=0.215:0.275,
each nth point=1
] {figures/amiri_scales/narrow/interface_equilibrium_narrow.dat};

\addplot[
red,
dash pattern=on 4pt off 2pt,
line width=1.05pt,
mark=x,
mark repeat=50,
mark size=2.8pt,
mark options={solid, draw=red, line width=0.65pt}
]
table[
x index=0,
y index=5,
restrict x to domain=0.215:0.275,
each nth point=1
] {figures/amiri_scales/narrow/interface_equilibrium_narrow.dat};

\addplot[
cyan!70!blue,
solid,
line width=1.10pt,
mark=square,
mark repeat=50,
mark size=2.2pt,
mark options={solid, draw=cyan!70!blue, fill=white, line width=0.65pt}
]
table[
x index=0,
y index=6,
restrict x to domain=0.215:0.275,
each nth point=1
] {figures/amiri_scales/narrow/interface_equilibrium_narrow.dat};

% angoli del riquadro zoom
\coordinate (insetSW) at (rel axis cs:0,0);
\coordinate (insetNW) at (rel axis cs:0,1);

\end{axis}

% linee oblique di connessione
\draw[black, line width=0.4pt] (zoomB) -- (insetNW);
\draw[black, line width=0.4pt] (zoomD) -- (insetSW);

\end{tikzpicture}
\vspace{-1.25em}
\caption{$H^*/L^*=10$.}
\label{fig:narrow_amiri_scales}
\end{subfigure}
\hfill
\begin{subfigure}[t]{0.47\textwidth}
\centering
\setlength\figurewidth{0.78\linewidth}
\setlength\figureheight{0.56\linewidth}
\begin{tikzpicture}[font=\sffamily\sansmath]

\begin{axis}[
width=\figurewidth,
height=\figureheight,
scale only axis,
xmin=0,
xmax=1,
ymin=-1.05,
ymax=1.05,
xlabel={$z^*/H^*$},
ylabel={},
axis line style={black},
tick align=inside,
tick pos=both,
xtick={0,0.2,0.4,0.6,0.8,1.0},
ytick={-1.0,-0.5,0,0.5,1.0},
xticklabels={$0$,$0.2$,$0.4$,$0.6$,$0.8$,$1.0$},
yticklabels={$-1.0$,$-0.5$,$0$,$0.5$,$1.0$},
tick label style={
    font=\fontsize{11}{8}\selectfont\sffamily\sansmath
},
label style={
    font=\fontsize{13}{8}\selectfont\sffamily\sansmath
},
]

% theta all dump 00000
%\addplot[
%black,
%dotted,
%line width=0.95pt,
%mark=none
%]
%table[x index=0,y index=1] {figures/amiri_scales/large/interface_equilibrium_large.dat};
%\addlegendentry{$\overline{\theta}(t^*=0$};

% theta all dump 18000
\addplot[
orange,
solid,
line width=1.05pt,
mark=o,
mark repeat=120,
mark size=2.4pt,
mark options={solid, draw=orange, fill=white, line width=0.65pt}
]
table[
x index=0,
y index=2,
each nth point=4
] {figures/amiri_scales/large/interface_equilibrium_large.dat};

% c liquid mean dump 18000
\addplot[
blue,
dotted,
line width=1.00pt,
mark=diamond,
mark repeat=120,
mark size=2.4pt,
mark options={solid, draw=blue, fill=white, line width=0.65pt}
]
table[
x index=0,
y index=3,
each nth point=4
] {figures/amiri_scales/large/interface_equilibrium_large.dat};

% theta interface dump 18000
\addplot[
green!60!black,
solid,
line width=1.05pt,
mark=triangle,
mark repeat=120,
mark size=2.6pt,
mark options={solid, draw=green!60!black, fill=green!60!black, line width=0.55pt}
]
table[
x index=0,
y index=4,
each nth point=4
] {figures/amiri_scales/large/interface_equilibrium_large.dat};

% c interface dump 18000
\addplot[
red,
dash pattern=on 4pt off 2pt,
line width=1.05pt,
mark=x,
mark repeat=120,
mark size=2.8pt,
mark options={solid, draw=red, line width=0.65pt}
]
table[
x index=0,
y index=5,
each nth point=4
] {figures/amiri_scales/large/interface_equilibrium_large.dat};

% theta analytic modo 2
\addplot[
cyan!70!blue,
solid,
line width=1.10pt,
mark=square,
mark repeat=120,
mark size=2.2pt,
mark options={solid, draw=cyan!70!blue, fill=white, line width=0.65pt}
]
table[
x index=0,
y index=6,
each nth point=4
] {figures/amiri_scales/large/interface_equilibrium_large.dat};

% quadratino sul picco
\coordinate (zoomA) at (axis cs:0.220,0.880);
\coordinate (zoomB) at (axis cs:0.255,1.030);
\coordinate (zoomC) at (axis cs:0.220,1.030);
\coordinate (zoomD) at (axis cs:0.255,0.880);

\draw[black, line width=0.4pt]
(axis cs:0.220,0.880) rectangle (axis cs:0.255,1.030);

% scritte dPhi/dt
\node[font=\fontsize{9}{6}\selectfont\sffamily\sansmath, anchor=center]
at (axis cs:0.245,0.42) {$d\phi/dt < 0$};

\node[font=\fontsize{9}{6}\selectfont\sffamily\sansmath, anchor=center]
at (axis cs:0.755,-0.28) {$d\phi/dt > 0$};

\end{axis}

% -------------------------------------------------------------------------
% Zoom inset sul picco, senza labels né numeri sugli assi
% -------------------------------------------------------------------------
\begin{axis}[
    at={(rel axis cs:0.95,0.92)},
    anchor=north east,
    width=0.38\figurewidth,
    height=0.34\figureheight,
    scale only axis,
    xmin=0.225,
    xmax=0.275,
    ymin=0.920,
    ymax=0.995,
    grid=major,
    axis background/.style={fill=white},
    axis line style={black},
    tick align=inside,
    tick pos=both,
    xtick=\empty,
    ytick=\empty,
    xlabel={},
    ylabel={},
    clip=true,
]

\addplot[
orange,
solid,
line width=1.05pt,
mark=o,
mark repeat=50,
mark size=2.4pt,
mark options={solid, draw=orange, fill=white, line width=0.65pt}
]
table[
x index=0,
y index=2,
restrict x to domain=0.215:0.275,
each nth point=1
] {figures/amiri_scales/large/interface_equilibrium_large.dat};

\addplot[
blue,
dotted,
line width=1.00pt,
mark=diamond,
mark repeat=50,
mark size=2.4pt,
mark options={solid, draw=blue, fill=white, line width=0.65pt}
]
table[
x index=0,
y index=3,
restrict x to domain=0.215:0.275,
each nth point=1
] {figures/amiri_scales/large/interface_equilibrium_large.dat};

\addplot[
green!60!black,
solid,
line width=1.05pt,
mark=triangle,
mark repeat=50,
mark size=2.6pt,
mark options={solid, draw=green!60!black, fill=green!60!black, line width=0.55pt}
]
table[
x index=0,
y index=4,
restrict x to domain=0.215:0.275,
each nth point=1
] {figures/amiri_scales/large/interface_equilibrium_large.dat};

\addplot[
red,
dash pattern=on 4pt off 2pt,
line width=1.05pt,
mark=x,
mark repeat=50,
mark size=2.8pt,
mark options={solid, draw=red, line width=0.65pt}
]
table[
x index=0,
y index=5,
restrict x to domain=0.215:0.275,
each nth point=1
] {figures/amiri_scales/large/interface_equilibrium_large.dat};

\addplot[
cyan!70!blue,
solid,
line width=1.10pt,
mark=square,
mark repeat=50,
mark size=2.2pt,
mark options={solid, draw=cyan!70!blue, fill=white, line width=0.65pt}
]
table[
x index=0,
y index=6,
restrict x to domain=0.215:0.275,
each nth point=1
] {figures/amiri_scales/large/interface_equilibrium_large.dat};

% angoli del riquadro zoom
\coordinate (insetSW) at (rel axis cs:0,0);
\coordinate (insetNW) at (rel axis cs:0,1);

\end{axis}

% linee oblique di connessione
\draw[black, line width=0.4pt] (zoomB) -- (insetNW);
\draw[black, line width=0.4pt] (zoomD) -- (insetSW);

\end{tikzpicture}
\caption{$H^*/L^*=40$.}
\label{fig:large_amiri_scales}
\end{subfigure}
\vspace{-0.25em}
\caption{Axial profiles of dimensionless temperature and concentration for the scale-separation analysis. Barred quantities denote averages over each $x^*$--$y^*$ cross-section, whereas the remaining profiles are evaluated along the axial line $x^* = 3L^*/4$ and $y^*=W^*/2$.}
\label{fig:comparison_amiri_scales}
\end{figure}
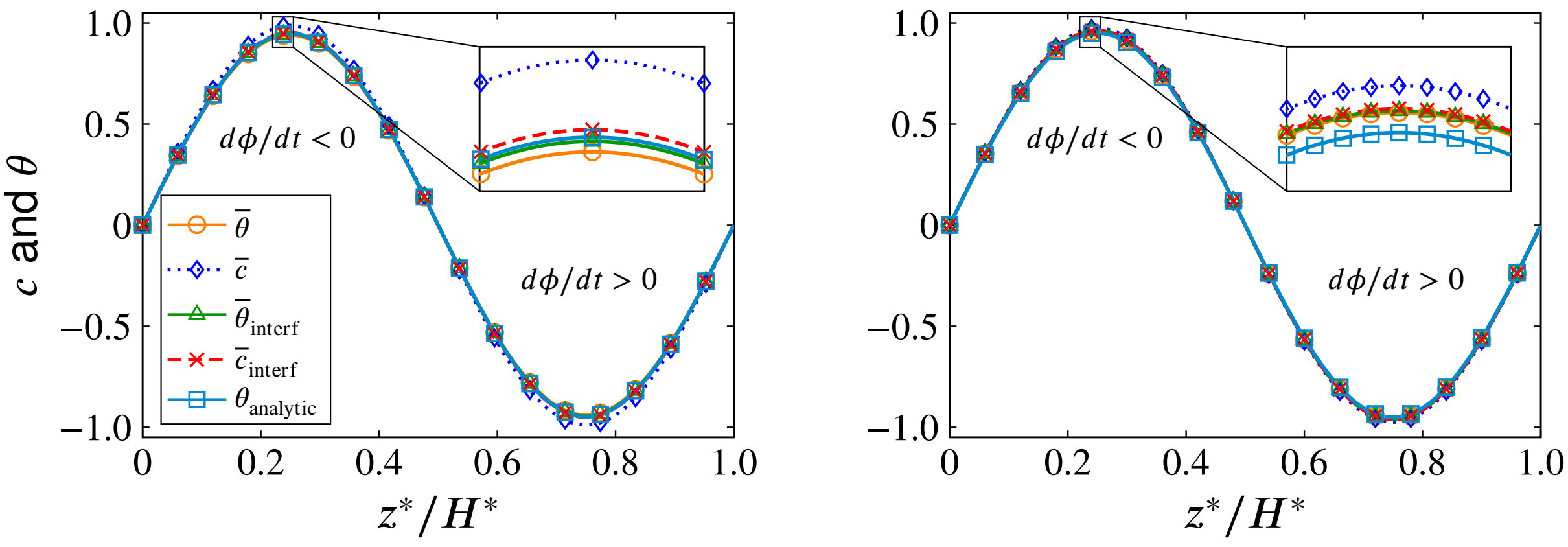
Figure~\ref{fig:comparison_amiri_scales} compares the results obtained for the two domains at an early stage of the transient, $t/\tau_{\mathrm{mushy}} \simeq 0.05$, where $\tau_{\mathrm{mushy}} = (H^*)^2 / 4\pi^2 A_{1,2}^{(\mathrm{diff})}$ is the characteristic time scale associated with the exponential decay of the perturbations. At the solid--liquid interface, the dimensionless temperature and concentration remain coupled through the liquidus relation. In contrast, the discrepancy between the cross-sectionally averaged temperature and concentration profiles decreases as the separation between transverse and axial length scales increases. This shows that finite transverse diffusion can produce a measurable departure from the one-dimensional local-equilibrium behaviour when $H^*/L^*$ is not sufficiently large. Based on this comparison, the ratio $H^*/L^* \simeq 40$, shown in Fig.~\ref{fig:large_amiri_scales}, is taken as a lower bound for adequate scale separation. All domains considered in the subsequent simulations are therefore chosen to satisfy this constraint. The same figure also shows the associated phase response, with $d\phi/dt<0$ in freezing regions and $d\phi/dt>0$ in melting regions.

\medskip
\noindent\textit{Diffusion-induced phase-change response for NH$_4$Cl}\par
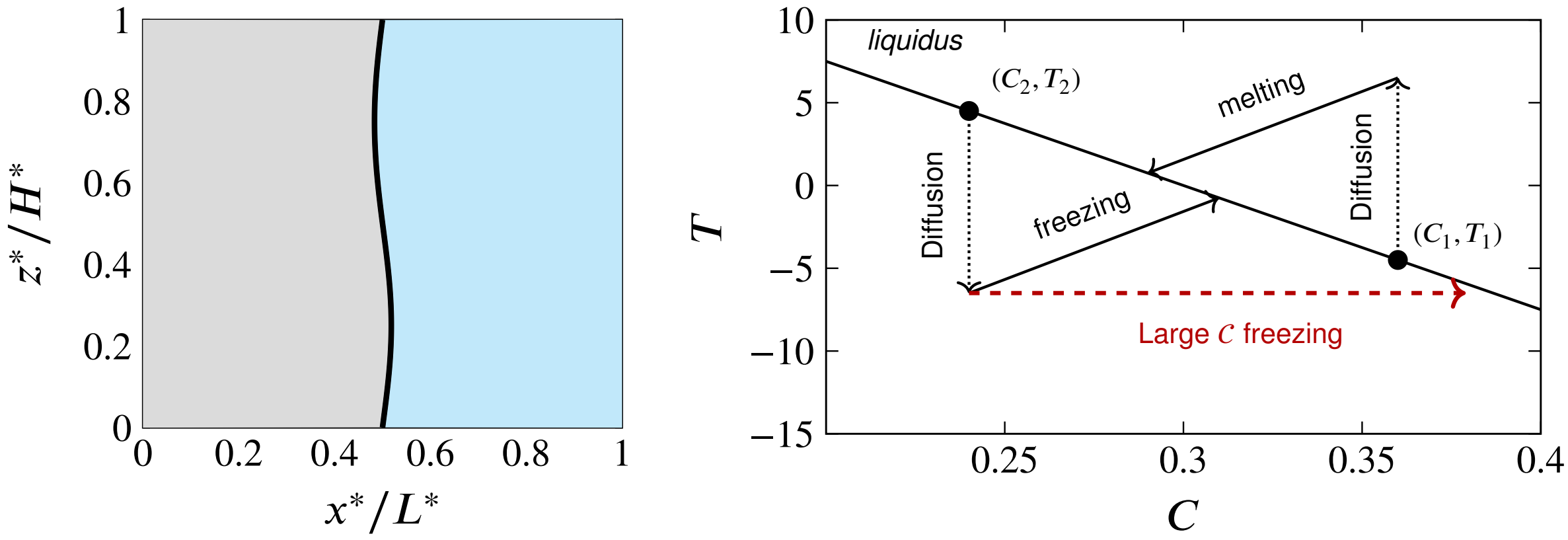
\begin{figure}[pos=htbp]
\centering
% Same vertical space reserved for the two plots
\newlength{\amiriPanelHeight}
\setlength{\amiriPanelHeight}{0.35\textwidth}
\begin{subfigure}[t]{0.31\textwidth}
\centering
\vspace{0pt}

\begin{minipage}[t][\amiriPanelHeight][t]{\linewidth}
\centering

\setlength\figurewidth{0.94\linewidth}
\setlength\figureheight{0.80\linewidth}

\begin{tikzpicture}[font=\sffamily\sansmath]

\begin{axis}[
width=\figurewidth,
height=\figureheight,
scale only axis,
xmin=0,
xmax=1,
ymin=0,
ymax=1,
xlabel={\(x^*/L^*\)},
ylabel={\(z^*/H^*\)},
xlabel style={
    font=\fontsize{13}{8}\selectfont\sffamily\sansmath,
    at={(axis description cs:0.5,-0.13)},
    anchor=north
},
ylabel style={
    font=\fontsize{13}{8}\selectfont\sffamily\sansmath,
    at={(axis description cs:-0.17,0.5)},
    anchor=south
},
grid=none,
axis line style={black},
tick align=inside,
tick pos=both,
tick label style={
    font=\fontsize{11}{8}\selectfont\sffamily\sansmath
},
label style={
    font=\fontsize{13}{8}\selectfont\sffamily\sansmath
},
legend style={
    draw=none,
    fill=none
},
clip=true,
]

% Interface curve
\addplot[
name path=interface,
black,
line width=1.4pt
]
table[
x index=1,
y index=0,
col sep=space,
comment chars={\#}
]
{figures/amiri_B_sens/amiri_interface_B_8e-3_iter_25000.dat};

% Left domain boundary
\path[name path=leftboundary]
(axis cs:0,0) -- (axis cs:0,1);

% Right domain boundary
\path[name path=rightboundary]
(axis cs:1,0) -- (axis cs:1,1);

% Solid region, left of the interface
\addplot[
draw=none,
fill=gray!28,
fill opacity=1
]
fill between[
of=leftboundary and interface
];

% Liquid region, right of the interface
\addplot[
draw=none,
fill=cyan!22,
fill opacity=1
]
fill between[
of=interface and rightboundary
];

% Redraw the interface on top
\addplot[
black,
line width=1.6pt
]
table[
x index=1,
y index=0,
col sep=space,
comment chars={\#}
]
{figures/amiri_B_sens/amiri_interface_B_8e-3_iter_25000.dat};

\end{axis}

\end{tikzpicture}

\end{minipage}

\vspace{-0.75em}
\caption{Porosity contour from LBM results.}
\label{fig:diffusive_reactive_interface}
\end{subfigure}
\hfill
\begin{subfigure}[t]{0.61\textwidth}
\centering
\vspace{0pt}

\begin{minipage}[t][\amiriPanelHeight][t]{\linewidth}
\centering

\setlength\figurewidth{0.87\linewidth}
\setlength\figureheight{0.57\linewidth}

% Requires:
% \usepackage{pgfplots}
% \pgfplotsset{compat=1.18}

\begin{tikzpicture}[font=\sffamily\sansmath]

\begin{axis}[
    width=\figurewidth,
    height=\figureheight,
    xmin=0.20, xmax=0.40,
    ymin=-15, ymax=10,
    xlabel={$C$},
    ylabel={$T$},
    axis lines=box,
    tick align=inside,
    tick pos=both,
    xtick={0.25,0.30,0.35,0.40},
    ytick={-15,-10,-5,0,5,10},
    major tick length=3.5pt,
    line width=0.8pt,
xlabel style={
    font=\fontsize{13}{8}\selectfont\sffamily\sansmath,
    at={(axis description cs:0.5,-0.13)},
    anchor=north
},
ylabel style={
    font=\fontsize{13}{8}\selectfont\sffamily\sansmath,
    at={(axis description cs:-0.13,0.5)},
    anchor=south
},
tick label style={
    font=\fontsize{11}{8}\selectfont\sffamily\sansmath
},
label style={
    font=\fontsize{13}{8}\selectfont\sffamily\sansmath
},
    clip=false
]

% =========================================================
% LIQUIDUS -- negative-slope branch only
% =========================================================

\draw[
    black,
    thick
]
(axis cs:0.20,7.5)
--
(axis cs:0.40,-7.5);

% Horizontal label for a cleaner appearance
\node[
    font=\footnotesize\itshape,
    anchor=west
] at (axis cs:0.208,8.65)
{liquidus};

% =========================================================
% EQUILIBRIUM STATES ON THE LIQUIDUS
% =========================================================

% Upper state: (C2,T2)
\addplot[
    only marks,
    mark=*,
    mark size=2.4pt,
    black
]
coordinates {(0.24,4.5)};

\node[
    font=\footnotesize,
    anchor=south west
] at (axis cs:0.243,4.8)
{$(C_2,T_2)$};

% Lower state: (C1,T1)
\addplot[
    only marks,
    mark=*,
    mark size=2.4pt,
    black
]
coordinates {(0.36,-4.5)};

\node[
    font=\footnotesize,
    anchor=north west
] at (axis cs:0.361,-1.5)
{$(C_1,T_1)$};

% =========================================================
% DIFFUSION DEPARTURES FROM THE LIQUIDUS
% =========================================================

% Diffusion from upper state
\draw[
    ->,
    black,
    densely dotted,
    thick
]
(axis cs:0.24,4.1)
--
(axis cs:0.24,-6.5);

\node[
    font=\footnotesize,
    rotate=90,
    anchor=south
] at (axis cs:0.236,-1.2)
{Diffusion};

% Diffusion from lower state
\draw[
    ->,
    black,
    densely dotted,
    thick
]
(axis cs:0.36,-4.1)
--
(axis cs:0.36,6.5);

\node[
    font=\footnotesize,
    rotate=90,
    anchor=south
] at (axis cs:0.356,1.1)
{Diffusion};

% =========================================================
% STANDARD RETURN TO THE LIQUIDUS
% =========================================================

% Freezing
\draw[
    ->,
    black,
    thick
]
(axis cs:0.24,-6.5)
--
node[
    midway,
    sloped,
    above=2pt,
    font=\footnotesize
]
{freezing}
(axis cs:0.31,-0.75);

% Melting
\draw[
    ->,
    black,
    thick
]
(axis cs:0.36,6.5)
--
node[
    midway,
    sloped,
    above=2pt,
    font=\footnotesize
]
{melting}
(axis cs:0.29,0.75);

% =========================================================
% LARGE-CONCENTRATION-RATIO FREEZING
% =========================================================

\draw[
    ->,
    red!70!black,
    very thick,
    dashed
]
(axis cs:0.24,-6.5)
--
(axis cs:0.379,-6.5);

\node[
    font=\footnotesize,
    text=red!70!black,
    anchor=north,
    align=center
] at (axis cs:0.316,-7.55)
{Large $\mathcal{C}$ freezing};

\end{axis}

\end{tikzpicture}
\end{minipage}
\vspace{-0.75em}
\caption{Schematic representation of the diffusion-induced departure from and restoration to the liquidus. Temperature and concentration values are illustrative and taken from~\cite{butler11}.}
\label{fig:butler_largeC_diffusion}
\end{subfigure}
\vspace{-0.25em}
\caption{Diffusion-induced phase-change response in the $y^*=W^*/2$ section of the ice domain at $t/\tau_{\mathrm{mushy}} \simeq 0.2$.}
\label{fig:diffusion-brine-reaction}
\end{figure}

\smallskip
\noindent
The simulation shown in Fig.~\ref{fig:large_amiri_scales} is further evolved in time to examine how the LBM framework captures the temporal phase-change response of the mushy layer. In this case, the interface dynamics is computed using the link-based formulation of the interfacial rate, Eq.~\eqref{eq:rate_link}. In the simulated diffusive regime, heat diffuses faster than solute. Thermal perturbations are therefore smoothed before the concentration field can adjust appreciably. At a warm and dilute peak, thermal diffusion removes heat from the region while the local concentration remains comparatively low. The system consequently departs from the liquidus following the left-side diffusion path in Fig.~\ref{fig:butler_largeC_diffusion}. Solidification therefore occurs near the temperature maximum, i.e. at low $z^*/H^*$, as shown in Fig.~\ref{fig:diffusive_reactive_interface}. Conversely, at a cold and concentrated trough, thermal diffusion supplies heat faster than solute can diffuse out. The state is then displaced along the opposite diffusion path, shown on the right-hand side of Fig.~\ref{fig:butler_largeC_diffusion}, leading to melting at high $z^*/H^*$. This behaviour is analogous to the diffusion-induced phase change described by Butler \emph{et al.}~\cite{butler11}. In the present configuration, however, no external thermal invasion is imposed and the local curvature of the sinusoidal perturbation produces an analogous departure from liquidus equilibrium.
The Modified NH$_4$Cl configuration considered here corresponds to the large-concentration-ratio regime reported in Table~\ref{tab:high_diff_parameters}. As discussed in Refs.~\cite{butler11,amiri24}, melting and solidification provide a mechanism by which departures from thermodynamic equilibrium are compensated through simultaneous changes in temperature and composition. The relative efficiency with which phase change modifies the two fields can be inferred directly from Eqs.~\eqref{eq:temperature_dimensionless} and~\eqref{eq:concentration_dimensionless}. In the energy equation, Eq.~\eqref{eq:temperature_dimensionless}, the phase-change contribution is proportional to $S\,\partial\phi/\partial t^*$, whereas in the solute conservation equation, Eq.~\eqref{eq:concentration_dimensionless}, the corresponding contribution is proportional to $\rho_s'(\mathcal{C}-c)\,\partial\phi/\partial t^*$. In the present test, where the concentration ratio is large, a relatively small amount of melting or solidification can therefore produce a much larger adjustment of the liquid concentration than of the temperature. This mechanism is illustrated by the red dashed arrow in Fig.~\ref{fig:butler_largeC_diffusion}.

Consistently, the effective diffusivity predicted by the analytical solution is $A_{1,2}^{(\mathrm{diff})}\simeq0.392$, which is substantially closer to the effective thermal diffusivity of the corresponding passive porous layer, $\alpha_{\mathrm{eff}}\simeq0.481$, than to the solutal diffusivity, $D\simeq0.033$. This indicates that, in this regime, phase change acts mainly by adapting the concentration field to the diffusive evolution of the temperature field.

\medskip
\noindent\textit{LBM sensitivity to interfacial kinetic relaxation}\par
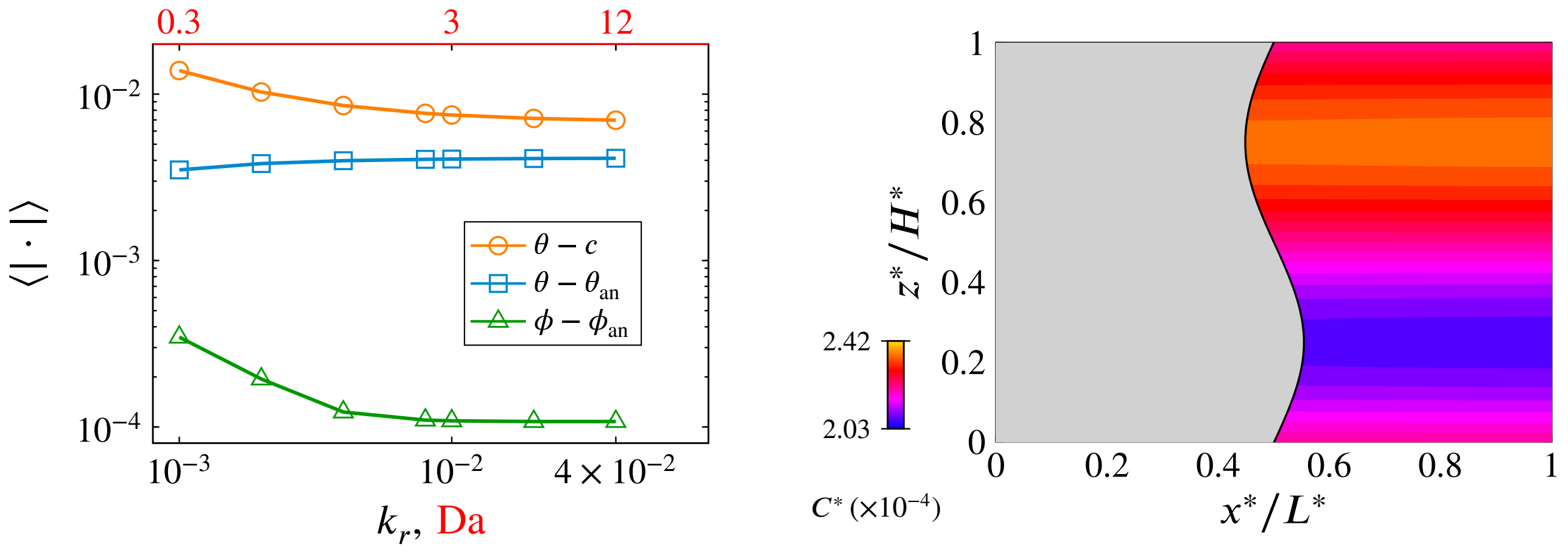
\begin{figure}[pos=htbp]
\centering

\newlength{\sensPanelHeight}
\setlength{\sensPanelHeight}{0.33\textwidth}

\begin{subfigure}[t]{0.47\textwidth}
\centering
\begin{minipage}[t][\sensPanelHeight][t]{\linewidth}
\centering
\setlength\figurewidth{0.78\linewidth}
\setlength\figureheight{0.56\linewidth}
\begin{tikzpicture}[font=\sffamily\sansmath]

% =========================================================
% Bottom axis: kinetic velocity coefficient B_Da used in the simulations
% Top axis: corresponding kinetic Damkohler number
%
% Da = (L^*/D) B_Da
%
% Present configuration:
% L^* = 10
% D   = (0.6-0.5)/3 = 1/30
%
% Therefore:
% Da = 300 B_Da
% =========================================================

\pgfmathsetmacro{\DaFactor}{300}

% =========================================================
% Main axis: B_Da
% =========================================================
\begin{axis}[
name=mainaxis,
width=\figurewidth,
height=\figureheight,
scale only axis,
xmode=log,
ymode=log,
xmin=8.0e-4,
xmax=8.75e-2,
ymin=8.0e-5,
ymax=2.0e-2,
xlabel={\(k_r,\,\textcolor{red}{\mathrm{Da}}\)},
ylabel={\(\langle |\cdot| \rangle\)},
grid=none,
axis line style={black},
tick align=inside,
tick pos=both,
xtick={1e-3,1e-2,4e-2},
xticklabels={
{$10^{-3}$},
{$10^{-2}$},
{$4\times10^{-2}$}
},
minor x tick num=8,
ytick={1e-4,1e-3,1e-2},
yticklabels={
{$10^{-4}$},
{$10^{-3}$},
{$10^{-2}$}
},
minor y tick num=8,
legend style={
font=\fontsize{9.5}{8}\selectfont\sffamily\sansmath,
anchor=center,
at={(0.72,0.4)},
draw=black,
fill=white,
inner sep=1.5pt,
row sep=0pt
},
legend cell align=left,
tick label style={
font=\fontsize{11}{8}\selectfont\sffamily\sansmath
},
label style={
font=\fontsize{13}{8}\selectfont\sffamily\sansmath
},
xlabel style={
font=\fontsize{13}{8}\selectfont\sffamily\sansmath,
at={(axis description cs:0.5,-0.13)},
anchor=north
},
ylabel style={
font=\fontsize{13}{8}\selectfont\sffamily\sansmath,
at={(axis description cs:-0.17,0.5)},
anchor=south
},
clip=true
]

% =========================================================
% Mean |theta-c|
% =========================================================
\addplot[
orange,
solid,
mark=o,
mark size=2.7pt,
line width=1.1pt,
mark options={
solid,
draw=orange,
fill=white,
line width=0.65pt
}
]
table[
x index=0,
y index=2,
col sep=space,
comment chars={\#}
]
{figures/amiri_B_sens/B_sensitivity_overlay_summary_thetaAn.dat};

\addlegendentry{\(\theta-c\)}

% =========================================================
% Mean |theta-theta_an|
% =========================================================
\addplot[
cyan!70!blue,
solid,
mark=square,
mark size=2.6pt,
line width=1.1pt,
mark options={
solid,
draw=cyan!70!blue,
fill=white,
line width=0.65pt
}
]
table[
x index=0,
y index=4,
col sep=space,
comment chars={\#}
]
{figures/amiri_B_sens/B_sensitivity_overlay_summary_thetaAn.dat};

\addlegendentry{\(\theta-\theta_{\mathrm{an}}\)}

% =========================================================
% Mean |phi_num-phi_an|
% =========================================================
\addplot[
green!60!black,
solid,
mark=triangle,
mark size=3.4pt,
line width=1.1pt,
mark options={
solid,
draw=green!60!black,
fill=white,
line width=0.55pt
}
]
table[
x index=0,
y index=6,
col sep=space,
comment chars={\#}
]
{figures/amiri_B_sens/B_sensitivity_overlay_summary_thetaAn.dat};

\addlegendentry{\(\phi-\phi_{\mathrm{an}}\)}

\end{axis}

% =========================================================
% Secondary top axis: Da = 300 B_Da
% =========================================================
\begin{axis}[
at={(mainaxis.south west)},
anchor=south west,
width=\figurewidth,
height=\figureheight,
scale only axis,
xmode=log,
xmin=0.24,
xmax=26.25,
ymin=8.0e-5,
ymax=2.0e-2,
axis x line*=top,
axis y line=none,
axis line style={red},
xtick={0.3,3,12},
xticklabels={
{$0.3$},
{$3$},
{$12$}
},
minor x tick num=8,
tick align=inside,
xticklabel style={
    /pgf/number format/fixed,
    /pgf/number format/precision=1,
    red,
    font=\fontsize{11}{8}\selectfont\sffamily\sansmath
},
clip=false
]
\end{axis}

\end{tikzpicture}
\vfill
\end{minipage}
\vspace{-1.2em}
\caption{Mean errors as a function of $k_r$ and $\mathrm{Da}$.}
\label{fig:amiri_B_sensitivity_errors}
\end{subfigure}
\hfill
\begin{subfigure}[t]{0.47\textwidth}
\centering
\begin{minipage}[t][\sensPanelHeight][t]{\linewidth}
\centering
\setlength\figurewidth{0.78\linewidth}
\setlength\figureheight{0.56\linewidth}
\raisebox{0.4em}{%
\begin{tikzpicture}[font=\sffamily\sansmath]
% ============================================================
% Custom concentration colormap
% ============================================================
\pgfplotsset{
    colormap={dario}{
        rgb255=(0,0,255)
        rgb255=(255,0,255)
        rgb255=(255,0,0)
        rgb255=(255,220,0)
    }
}
% ============================================================
% MAIN AXIS: concentration contour + moving interface
% ============================================================
\begin{axis}[
name=fieldaxis,
width=\figurewidth,
height=\figureheight,
scale only axis,
xmin=0,
xmax=1,
ymin=0,
ymax=1,
xlabel={$x^*/L^*$},
ylabel={$z^*/H^*$},
axis line style={black},
tick align=inside,
tick pos=both,
grid=none,
clip=true,
scaled x ticks=false,
scaled y ticks=false,
xtick={0,0.2,0.4,0.6,0.8,1},
ytick={0,0.2,0.4,0.6,0.8,1},
tick label style={
    font=\fontsize{11}{13}\selectfont\sffamily\sansmath
},
label style={
    font=\fontsize{13}{15}\selectfont\sffamily\sansmath
},
xlabel style={yshift=1pt},
ylabel style={yshift=-2pt},
]
% ============================================================
% RASTER CONCENTRATION FIELD
% The black solid--liquid interface is already contained
% in the exported PNG.
% ============================================================
\addplot graphics[
xmin=0,
xmax=1,
ymin=0,
ymax=1
]
{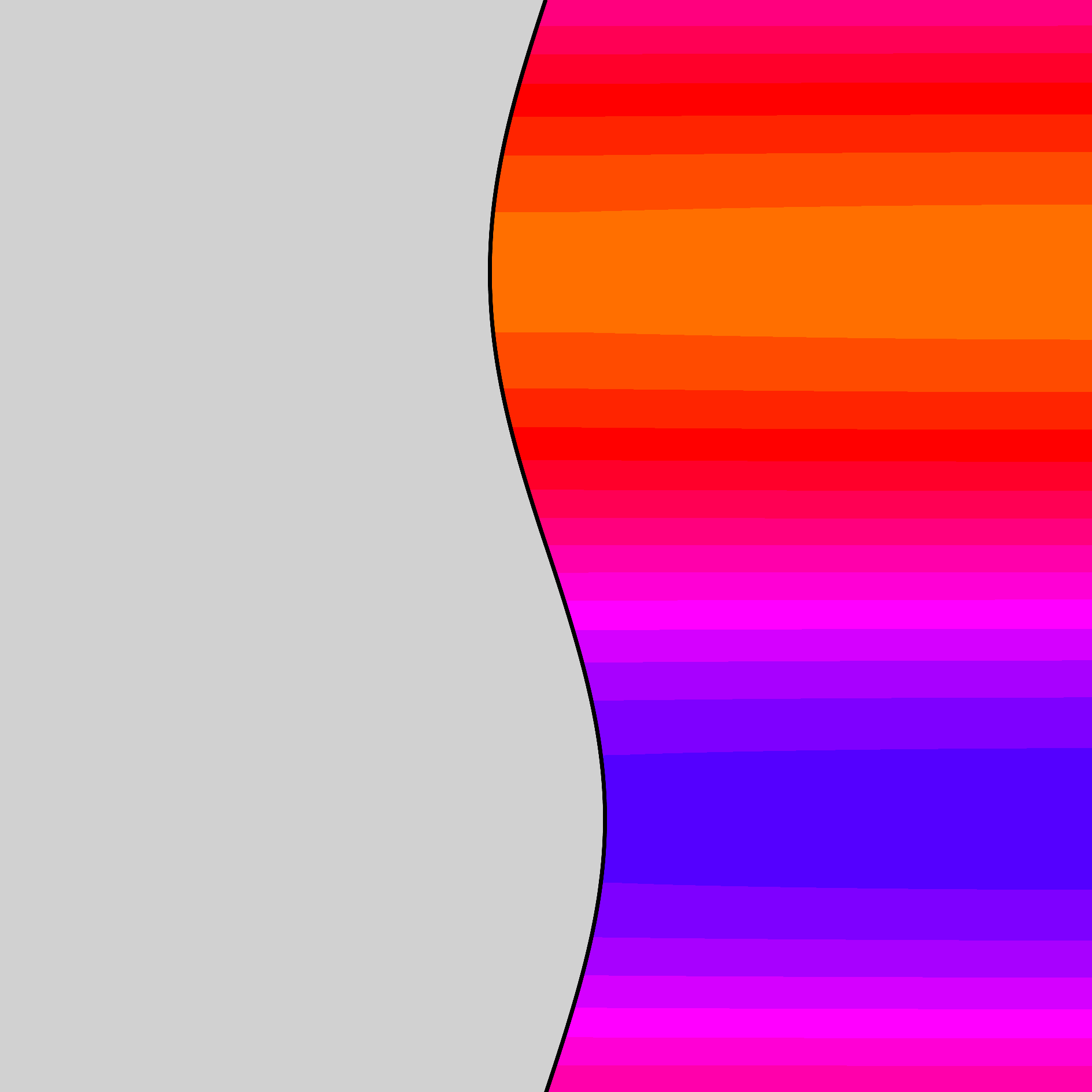};
\end{axis}
% ============================================================
% COMPACT COLORBAR - LOWER LEFT
%
% Physical range:
% C = 2.026e-4 ... 2.415e-4
%
% The colorbar itself is normalized from 0 to 1 only for
% graphical purposes. Physical values are written explicitly.
% ============================================================
\begin{axis}[
at={(fieldaxis.south west)},
anchor=south east,
xshift=-30pt,
yshift=4pt,
width=3.5pt,
height=0.22\figureheight,
scale only axis,
hide axis,
colormap name=dario,
point meta min=0,
point meta max=1,
colorbar,
colorbar style={
    width=5pt,
    height=0.22\figureheight,
    at={(0,0)},
    anchor=south west,
    % only minimum and maximum
    ytick={0,1},
    yticklabels={$2.03$,$2.42$},
    % put numbers on the LEFT of the bar
    yticklabel pos=left,
    tick align=outside,
    tick label style={
        font=\fontsize{8.5}{9}\selectfont\sffamily\sansmath
    },
    % one compact title instead of C + scientific exponent
    title={$C^*\,(\times10^{-4})$},
    title style={
        font=\fontsize{9}{10}\selectfont\sffamily\sansmath,
        xshift=-6pt,
        yshift=-65pt
    }
}
]
% Dummy values only to construct the colorbar
\addplot[
scatter,
only marks,
mark size=0pt,
scatter src=explicit
]
coordinates {
    (0,0) [0]
    (0,1) [1]
};
\end{axis}
\end{tikzpicture}%
}
\vfill
\end{minipage}
\vspace{-1.2em}
\caption{Representation of interface and $C^*$ contour, $k_r=\mathcal{O}(10^{-2})$.}
\label{fig:amiri_interface_B_selected}
\end{subfigure}

\vspace{-0.2em}
\caption{Sensitivity of the coupled LBM solution to the kinetic coefficient at
$t/\tau_{\mathrm{mushy}}\simeq0.61$ and $y^*=W^*/2$.}
\label{fig:amiri_B_sensitivity}
\end{figure}

\smallskip
\noindent
A sensitivity analysis with respect to the kinetic coefficient $k_r$ in Eq.~\eqref{eq:rate_link} is performed to assess the influence of the interfacial kinetic time scale on the LBM solution and to identify a suitable kinetic regime for the mushy-layer simulations. The scale-separated lattice $20\times800\times20$ is simulated, with half of the domain occupied by solid and half by liquid. The corresponding dimensionless number is $\mathrm{Da} = L^* \, k_r / A_1$, where $L^* = 10$ is the initial pore diameter. The modified high-diffusivity NH$_4$Cl properties reported in the third column of Table~\ref{tab:high_diff_parameters} are adopted. For each value of $k_r$, three quantities are monitored along a sampled axial profile at a $x^*$ within the liquid region and at the mid-plane $y^*=W^*/2$. In Fig.~\ref{fig:amiri_B_sensitivity_errors}, $|\theta-c|$ represents the departure from local thermodynamic equilibrium, $|\theta-\theta_{\mathrm{an}}|$ is the $\theta$ error with respect to the analytical solution and $|\phi_{\mathrm{num}}-\phi_{\mathrm{an}}|$ denotes the porosity discrepancy with respect to the analytical prediction. Note that the porosity profile is obtained by averaging the local liquid fraction across the $x^*$-direction, as $\phi(z^*)=N_x^{-1}\sum_{j}\alpha_j(x_j^*,z^*)$. The symbol $\langle|\cdot|\rangle$ in Fig.~\ref{fig:amiri_B_sensitivity_errors} denotes the spatial mean along the sampled axial profile. 
Fig.~\ref{fig:amiri_B_sensitivity_errors} shows that, for small $k_r$, the solution remains sensitive to the kinetic coefficient, as the interface response is slow compared with the diffusive transport time scale. Increasing $k_r$ accelerates relaxation towards liquidus equilibrium and progressively reduces all three error measures. The errors eventually approach a plateau for $k_r=\mathcal{O}(10^{-2})$, or equivalently $\mathrm{Da}=\mathcal{O}(10^{0})$, indicating that the interface kinetics are sufficiently fast for further increases in $k_r$ to produce only limited improvement. For larger values of $k_r$, not shown here, the interface response becomes stiff and numerical oscillations progressively develop. The range $\mathrm{Da}=\mathcal{O}(10^{0})$ is therefore adopted as a suitable kinetic regime, providing sufficiently rapid relaxation while remaining below the onset of numerical instability. Fig.~\ref{fig:amiri_interface_B_selected} provides a qualitative view of the coupled response at the selected kinetic coefficient. The concentration field follows the imposed axial disturbance while remaining smooth across the transverse liquid extent, with no pronounced boundary-layer accumulation near the moving interface. This supports the effectively one-dimensional solutal behaviour obtained under the chosen scale separation. The interface deformation is likewise consistent with the diffusion-induced phase-change pattern, with freezing and melting occurring in the expected regions.

\medskip
\noindent\textit{Reference NH$_4$Cl configuration}\par
\smallskip
\noindent
The reference NH$_4$Cl configuration is then considered through the original dimensionless parameters reported in the NH$_4$Cl column of Tab.~\ref{tab:high_diff_parameters}. This case is used to assess the coupled LBM framework under the most representative NH$_4$Cl conditions considered here, without reducing the computational cost through modified material parameters. The interface dynamics is computed using the link-based implementation of the kinetic relation, Eq.~\eqref{eq:rate_link}.

For the reference NH$_4$Cl configuration, the diffusion eigenvalue is
$A_{1,2}^{(\mathrm{NH_4Cl})}\simeq273.75$, which is substantially larger than the value obtained for the modified NH$_4$Cl configuration,
$A_{1,2}^{(\mathrm{mod})}\simeq11.77$. Since the characteristic mushy-layer time scale is $\tau_{\mathrm{mushy}}
= H^{*2} / 4\pi^2 A_{1,2}^{(\mathrm{diff})}$,
a larger value of $A_{1,2}^{(\mathrm{diff})}$ implies a faster axial evolution of the mushy-layer disturbance. To retain the same degree of scale separation established in the previous tests, the ratio between the transverse solutal diffusion time and the mushy-layer time scale is kept approximately constant:
\begin{equation}
\tau_{\perp} / \tau_{\mathrm{mushy}}=
(L^{*2}/D) \, / \, \left[
     {H^{*2}/\left(4\pi^2 A_{1,2}^{(\mathrm{diff})}\right)}\right]
=
(4\pi^2 A_{1,2}^{(\mathrm{diff})}L^{*2}) \, / \,
     {D H^{*2}}
\simeq \mathrm{const},
\label{eq:scale_separation_ratio}
\end{equation}
where $\tau_{\perp}\sim L^{*2}/D$ is the transverse solutal diffusion time. Using the same transverse pore size and solutal diffusivity in the reference and modified NH$_4$Cl simulations, $L^* = 10$ and $A_1 = 0.033$ respectively , Eq.~\eqref{eq:scale_separation_ratio} reduces to
\begin{equation}
H^*_{\mathrm{NH_4Cl}} \, / \, H^*_{\mathrm{mod}}
=
\left(
A_{1,2}^{(\mathrm{NH_4Cl})}
\, / \,
A_{1,2}^{(\mathrm{mod})}
\right)^{1/2}
\simeq 4.8 .
\label{eq:NH4Cl_domain_scaling}
\end{equation}
The domain of the reference NH$_4$Cl simulation is therefore extended by a factor of approximately $5$ along the $z$-direction, leading to a $20\times4000\times20$ lattice domain. The kinetic coefficient $k_r$ is chosen according to the regime identified in the sensitivity analysis above, corresponding to $\mathrm{Da}=\mathcal{O}(10^0)$. The resulting numerical profiles are compared with the analytical predictions for temperature, concentration, and porosity at $t/\tau_{\mathrm{mushy}}\approx0.5$.
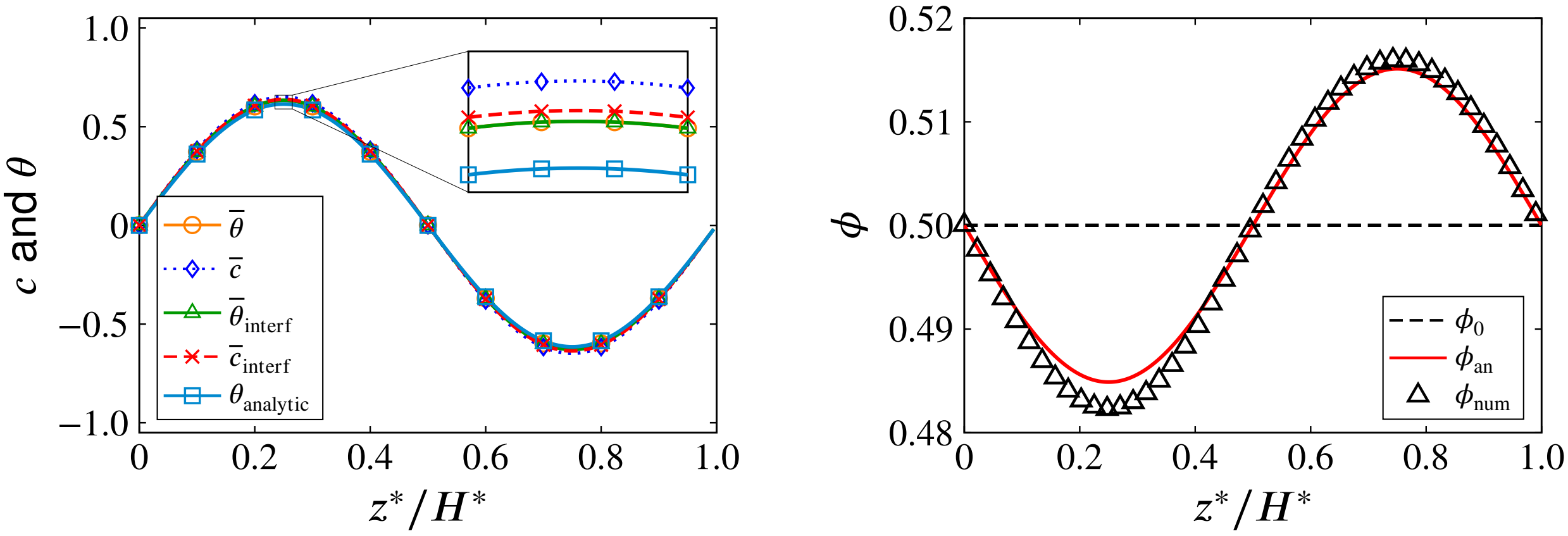
\begin{figure}[pos=htbp]
\centering
\begin{subfigure}[t]{0.47\textwidth}
\centering
\setlength\figurewidth{0.78\linewidth}
\setlength\figureheight{0.56\linewidth}
\begin{tikzpicture}[font=\sffamily\sansmath]
\def\plotstride{20}
\def\zoomstride{5}
\begin{axis}[
width=\figurewidth,
height=\figureheight,
scale only axis,
xmin=0,
xmax=1,
ymin=-1.05,
ymax=1.05,
xlabel={$z^*/H^*$},
ylabel={$c$ and $\theta$},
axis line style={black},
tick align=inside,
tick pos=both,
xtick={0,0.2,0.4,0.6,0.8,1.0},
ytick={-1.0,-0.5,0,0.5,1.0},
xticklabels={$0$,$0.2$,$0.4$,$0.6$,$0.8$,$1.0$},
yticklabels={$-1.0$,$-0.5$,$0$,$0.5$,$1.0$},
tick label style={
    font=\fontsize{11}{8}\selectfont\sffamily\sansmath
},
label style={
    font=\fontsize{13}{8}\selectfont\sffamily\sansmath
},
legend style={
    font=\fontsize{8.7}{8}\selectfont\sffamily\sansmath,
    anchor=south west,
    at={(0.03,0.03)},
    draw=black,
    fill=white,
    inner sep=1.5pt,
    row sep=0pt
},
legend cell align=left,
clip=true,
]

% theta all dump 20000
\addplot[
orange,
solid,
line width=1.05pt,
mark=o,
mark repeat=20,
mark size=2.4pt,
mark options={solid, draw=orange, fill=white, line width=0.65pt}
]
table[x index=0,y index=2,each nth point=\plotstride] {figures/amiri_plots/exact/interface_equilibrium_dump_800000_tikz.dat};
\addlegendentry{$\overline{\theta}$};

% c liquid mean dump 18000
\addplot[
blue,
dotted,
line width=1.00pt,
mark=diamond,
mark repeat=20,
mark size=2.4pt,
mark options={solid, draw=blue, fill=white, line width=0.65pt}
]
table[x index=0,y index=3,each nth point=\plotstride] {figures/amiri_plots/exact/interface_equilibrium_dump_800000_tikz.dat};
\addlegendentry{$\overline{c}$};

% theta interface dump 18000
\addplot[
green!60!black,
solid,
line width=1.05pt,
mark=triangle,
mark repeat=20,
mark size=2.6pt,
mark options={solid, draw=green!60!black, fill=green!60!black, line width=0.55pt}
]
table[x index=0,y index=4,each nth point=\plotstride] {figures/amiri_plots/exact/interface_equilibrium_dump_800000_tikz.dat};
\addlegendentry{$\overline{\theta}_{\mathrm{interf}}$};

% c interface dump 18000
\addplot[
red,
dash pattern=on 4pt off 2pt,
line width=1.05pt,
mark=x,
mark repeat=20,
mark size=2.8pt,
mark options={solid, draw=red, line width=0.65pt}
]
table[x index=0,y index=5,each nth point=\plotstride] {figures/amiri_plots/exact/interface_equilibrium_dump_800000_tikz.dat};
\addlegendentry{$\overline{c}_{\mathrm{interf}}$};

% theta analytic modo 2
\addplot[
cyan!70!blue,
solid,
line width=1.10pt,
mark=square,
mark repeat=20,
mark size=2.2pt,
mark options={solid, draw=cyan!70!blue, fill=white, line width=0.65pt}
]
table[x index=0,y index=6,each nth point=\plotstride] {figures/amiri_plots/exact/interface_equilibrium_dump_800000_tikz.dat};
\addlegendentry{$\theta_{\mathrm{analytic}}$};

% quadratino sul picco
\coordinate (zoomA) at (axis cs:0.235,0.59);
\coordinate (zoomB) at (axis cs:0.265,0.66);
\coordinate (zoomC) at (axis cs:0.235,0.66);
\coordinate (zoomD) at (axis cs:0.265,0.59);

\draw[black, line width=0.2pt]
(axis cs:0.235,0.59) rectangle (axis cs:0.265,0.66);

\end{axis}

% -------------------------------------------------------------------------
% Zoom inset sul picco, senza labels né numeri sugli assi
% -------------------------------------------------------------------------
\begin{axis}[
    at={(rel axis cs:0.95,0.92)},
    anchor=north east,
    width=0.38\figurewidth,
    height=0.34\figureheight,
    scale only axis,
xmin=0.235,
xmax=0.265,
ymin=0.605,
ymax=0.663,
    grid=major,
    axis background/.style={fill=white},
    axis line style={black},
    tick align=inside,
    tick pos=both,
    xtick=\empty,
    ytick=\empty,
    xlabel={},
    ylabel={},
    clip=true,
]

\addplot[
orange,
solid,
line width=1.05pt,
mark=o,
mark repeat=8,
mark size=2.4pt,
mark options={solid, draw=orange, fill=white, line width=0.65pt}
]
table[
x index=0,
y index=2,
restrict x to domain=0.235:0.265,
each nth point=\zoomstride
] {figures/amiri_plots/exact/interface_equilibrium_dump_800000_tikz.dat};

\addplot[
blue,
dotted,
line width=1.00pt,
mark=diamond,
mark repeat=8,
mark size=2.4pt,
mark options={solid, draw=blue, fill=white, line width=0.65pt}
]
table[
x index=0,
y index=3,
restrict x to domain=0.235:0.265,
each nth point=\zoomstride
] {figures/amiri_plots/exact/interface_equilibrium_dump_800000_tikz.dat};
\addplot[
green!60!black,
solid,
line width=1.05pt,
mark=triangle,
mark repeat=8,
mark size=2.6pt,
mark options={solid, draw=green!60!black, fill=green!60!black, line width=0.55pt}
]
table[
x index=0,
y index=4,
restrict x to domain=0.235:0.265,
each nth point=\zoomstride
] {figures/amiri_plots/exact/interface_equilibrium_dump_800000_tikz.dat};

\addplot[
red,
dash pattern=on 4pt off 2pt,
line width=1.05pt,
mark=x,
mark repeat=8,
mark size=2.8pt,
mark options={solid, draw=red, line width=0.65pt}
]
table[
x index=0,
y index=5,
restrict x to domain=0.235:0.265,
each nth point=\zoomstride
] {figures/amiri_plots/exact/interface_equilibrium_dump_800000_tikz.dat};

\addplot[
cyan!70!blue,
solid,
line width=1.10pt,
mark=square,
mark repeat=8,
mark size=2.2pt,
mark options={solid, draw=cyan!70!blue, fill=white, line width=0.65pt}
]
table[
x index=0,
y index=6,
restrict x to domain=0.235:0.265,
each nth point=\zoomstride
] {figures/amiri_plots/exact/interface_equilibrium_dump_800000_tikz.dat};

% angoli del riquadro zoom
\coordinate (insetSW) at (rel axis cs:0,0);
\coordinate (insetNW) at (rel axis cs:0,1);

\end{axis}

% linee oblique di connessione
\draw[black, line width=0.1pt] (zoomB) -- (insetNW);
\draw[black, line width=0.1pt] (zoomD) -- (insetSW);

\end{tikzpicture}
\vspace{-1.25em}
\caption{Dimensionless temperature and concentration profiles.}
\label{fig:exact_amiri_temperature_concentration}
\end{subfigure}
\hfill
\begin{subfigure}[t]{0.47\textwidth}
\centering
\setlength\figurewidth{0.78\linewidth}
\setlength\figureheight{0.56\linewidth}
\begin{tikzpicture}[font=\sffamily\sansmath]

\begin{axis}[
width=\figurewidth,
height=\figureheight,
scale only axis,
xmin=0,
xmax=1,
ymin=0.48,
ymax=0.52,
xlabel={$z^*/H^*$},
ylabel={$\phi$},
axis line style={black},
tick align=inside,
tick pos=both,
xtick={0,0.2,0.4,0.6,0.8,1.0},
ytick={0.48,0.49,0.50,0.51,0.52},
xticklabels={$0$,$0.2$,$0.4$,$0.6$,$0.8$,$1.0$},
yticklabels={$0.48$,$0.49$,$0.50$,$0.51$,$0.52$},
tick label style={
    font=\fontsize{11}{8}\selectfont\sffamily\sansmath
},
label style={
    font=\fontsize{13}{8}\selectfont\sffamily\sansmath
},
legend style={
    font=\fontsize{8.7}{8}\selectfont\sffamily\sansmath,
    anchor=north east,
    at={(0.97,0.33)},
    draw=black,
    fill=white,
    inner sep=1.5pt,
    row sep=0pt
},
legend cell align=left,
clip=true,
]

% Initial porosity
\addplot[
black,
dash pattern=on 4pt off 2pt,
line width=1.05pt
]
coordinates {
(0,0.5)
(1,0.5)
};
\addlegendentry{$\phi_0$};

% Analytical porosity
\addplot[
red,
solid,
line width=1.10pt
]
table[
x index=0,
y index=2,
col sep=space,
comment chars={\#}
]
{figures/amiri_plots/exact/porosity_profiles_dump_800000_tikz.dat};
\addlegendentry{$\phi_{\mathrm{an}}$};

% Numerical porosity
\addplot[
black,
only marks,
mark=triangle,
mark repeat=90,
mark size=3.4pt,
mark options={solid}
]
table[
x index=0,
y index=1,
col sep=space,
comment chars={\#}
]
{figures/amiri_plots/exact/porosity_profiles_dump_800000_tikz.dat};
\addlegendentry{$\phi_{\mathrm{num}}$};

\end{axis}

\end{tikzpicture}
\vspace{-1.25em}
\caption{Numerical and analytical porosity profiles.}
\label{fig:modified_amiri_porosity}
\end{subfigure}
\vspace{-0.25em}
\caption{Comparison between the numerical and analytical solutions, real NH$_4$Cl, $t/\tau_{\mathrm{mushy}}\approx0.5$.}
\label{fig:exact_amiri_profiles}
\end{figure}
Fig.~\ref{fig:exact_amiri_profiles} shows that the simulation remains in good agreement with the ideal mushy-layer prediction for the original NH$_4$Cl configuration. The temperature profile closely follows the analytical solution, and the interfacial temperature and concentration remain close to liquidus equilibrium, $\theta_{\mathrm{interf}}\simeq c_{\mathrm{interf}}$. The porosity perturbation is also small, consistently with the large-concentration-ratio regime of NH$_4$Cl. A weak asymmetry is nevertheless visible in the numerical porosity profile between the freezing and melting portions of the disturbance. This feature is absent, or strongly reduced, in the analytical prediction because Eq.~\eqref{eq:phi_eq2} is obtained under the linearizing assumptions $\mathcal{C}\gg\theta$ and $\phi\simeq\phi_0$. In the full numerical model, the coefficient controlling the porosity response retains its local dependence on temperature and porosity. 

\medskip
\noindent\textit{Low-concentration-ratio thermosolutal response}\par
\smallskip
\noindent
A final test is performed using the Modified NH$_4$Cl (2) configuration reported in Table~\ref{tab:high_diff_parameters}. This configuration retains the high-diffusivity NH$_4$Cl parameters, but reduces the concentration ratio from $\mathcal{C}=11.42$ to $\mathcal{C}=1$. This choice deliberately moves the system outside the large-concentration-ratio regime used to derive the analytical solution, in order to assess the coupled LBM response under a more sea-ice-relevant thermosolutal condition. In the present nondimensionalization,  the initial liquidus equilibrium can be written as
$C^*=|\Delta C^*|(\mathcal{C}-\theta)$. Thus, $\mathcal{C}$ controls the relative amplitude of the concentration variation induced by the thermal perturbation. For the large-concentration-ratio case, $\mathcal{C}=11.42$, the imposed range $-1\leq\theta\leq1$ gives $10.42|\Delta C^*|\leq C^*\leq12.42|\Delta C^*|$, so that the concentration field remains nearly uniform relative to its reference value. In contrast, for $\mathcal{C}=1$, the same thermal perturbation gives $0\leq C^*\leq2|\Delta C^*|$: the warm maximum, $\theta=1$, corresponds to a solute-free region, whereas the cold minimum corresponds to the largest solute concentration $2|\Delta C^*|$, as visible in Fig.~\ref{fig:amiri_Cbiglow_profiles}. The liquidus relation therefore maps the initially sinusoidal thermal perturbation into a strongly non-uniform concentration field.

The asymmetry observed in both the profiles of Fig.~\ref{fig:amiri_Cbiglow_profiles} and the interface morphology of Fig.~\ref{fig:interface_amiri_Cbiglow} can be understood from the factor $\mathcal{C}-\theta$ appearing in the analytical model. As shown in Eqs.~\eqref{eq:theta_analytical_diff} and~\eqref{eq:phi_eq2}, this term enters both the effective transport coefficient for $\theta$ and the coefficient governing the $\phi$ response. For $\mathcal{C}=\mathcal{O}(1)$, it can no longer be treated as constant: near the warm maximum, $\mathcal{C}-\theta$ is small, whereas near the cold minimum it is larger. As a result, an initially symmetric perturbation does not produce a symmetric evolution of temperature, concentration, and phase change. For the porosity response specifically, Eq.~\eqref{eq:phi_eq2} shows that, for a given temperature curvature, the magnitude of the local evolution scales as
$
\partial \phi/\partial t^* \propto 1/(\mathcal{C}-\theta).
$
Therefore, the phase-change response is larger in the freezing region, leading to a more pronounced advance of the solid phase and a smaller melting displacement.
\par
\vspace{0.1em}
\begin{figure}[pos=htbp]
\centering
\vspace*{0.5em}
\begin{subfigure}[t]{0.45\textwidth}
    \centering
    \vspace{0pt}
    \begin{minipage}[t][0.86\linewidth][b]{\linewidth}
        \centering
        \setlength{\figurewidth}{0.82\linewidth}
        \setlength{\figureheight}{0.58\linewidth}
        \begin{tikzpicture}[font=\sffamily\sansmath]

\begin{axis}[
width=\figurewidth,
height=\figureheight,
scale only axis,
xmin=0,
xmax=1,
ymin=0.990,
ymax=1.011,
xlabel={\(z^*/H^*\)},
ylabel={\(T^*\)},
grid=none,
axis line style={black},
tick align=inside,
tick pos=both,
xtick={0,0.2,0.4,0.6,0.8,1.0},
xticklabels={
{$0$},
{$0.2$},
{$0.4$},
{$0.6$},
{$0.8$},
{$1.0$}
},
ytick={0.990,0.995,1.000,1.005,1.010},
yticklabels={
{$0.990$},
{$0.995$},
{$1.000$},
{$1.005$},
{$1.010$}
},
legend style={
    font=\fontsize{9.5}{8}\selectfont\sffamily\sansmath,
    anchor=north east,
    at={(0.99,0.99)},
    draw=black,
    fill=white,
    inner sep=1.5pt,
    row sep=0pt
},
legend cell align=left,
tick label style={
    font=\fontsize{11}{8}\selectfont\sffamily\sansmath
},
label style={
    font=\fontsize{13}{8}\selectfont\sffamily\sansmath
},
xlabel style={
    font=\fontsize{13}{8}\selectfont\sffamily\sansmath,
    at={(axis description cs:0.5,-0.13)},
    anchor=north
},
ylabel style={
    font=\fontsize{13}{8}\selectfont\sffamily\sansmath,
    at={(axis description cs:-0.20,0.5)},
    anchor=south
},
clip=true,
]

% Data layout:
% column 0 = axial node index z
% column 1 = dimensionless temperature T*
%
% The abscissa is normalized internally as z/H, with H = 800.
% Dimensionless time: t* = iteration * D / H^2,
% with D = 0.03333 and H = 800.

% ============================================================
% iteration = 10000, t* = 5.2078125e-4
% ============================================================
\addplot[
orange,
solid,
mark=o,
mark repeat=40,
mark phase=0,
mark size=2.7pt,
line width=1.1pt,
mark options={
    solid,
    draw=orange,
    fill=white,
    line width=0.65pt
}
]
table[
x expr=\thisrowno{0}/800,
y index=1,
col sep=space,
comment chars={\#}
]
{figures/amiri_Cbiglow/tem_col17_10000.dat};
\addlegendentry{\(t^*=0.0005\)}

% ============================================================
% iteration = 20000, t* = 1.0415625e-3
% ============================================================
\addplot[
cyan!70!blue,
solid,
mark=square,
mark repeat=40,
mark phase=13,
mark size=2.6pt,
line width=1.1pt,
mark options={
    solid,
    draw=cyan!70!blue,
    fill=white,
    line width=0.65pt
}
]
table[
x expr=\thisrowno{0}/800,
y index=1,
col sep=space,
comment chars={\#}
]
{figures/amiri_Cbiglow/tem_col17_20000.dat};
\addlegendentry{\(t^*=0.001\)}

% ============================================================
% iteration = 60000, t* = 3.1246875e-3
% ============================================================
\addplot[
green!60!black,
solid,
mark=triangle,
mark repeat=40,
mark phase=26,
mark size=3.4pt,
line width=1.1pt,
mark options={
    solid,
    draw=green!60!black,
    fill=white,
    line width=0.55pt
}
]
table[
x expr=\thisrowno{0}/800,
y index=1,
col sep=space,
comment chars={\#}
]
{figures/amiri_Cbiglow/tem_col17_60000.dat};
\addlegendentry{\(t^*=0.0031\)}

\end{axis}

\end{tikzpicture}
        \par
    \end{minipage}
    \vspace{-1em}
    \caption{Temperature profiles at different times.}
    \label{fig:temp_amiri_Cbiglow}
\end{subfigure}
\hfill
\begin{subfigure}[t]{0.45\textwidth}
    \centering
    \vspace{0pt}
    \begin{minipage}[t][0.86\linewidth][b]{\linewidth}
        \centering
        \setlength{\figurewidth}{0.82\linewidth}
        \setlength{\figureheight}{0.58\linewidth}
        \begin{tikzpicture}[font=\sffamily\sansmath]

\begin{axis}[
width=\figurewidth,
height=\figureheight,
scale only axis,
xmin=0,
xmax=1,
ymin=0,
ymax=4.0e-5,
xlabel={\(z^*/H^*\)},
ylabel={\(C^*\)},
grid=none,
axis line style={black},
tick align=inside,
tick pos=both,
xtick={0,0.2,0.4,0.6,0.8,1.0},
xticklabels={
{$0$},
{$0.2$},
{$0.4$},
{$0.6$},
{$0.8$},
{$1.0$}
},
ytick={0,1e-5,2e-5,3e-5,4e-5},
yticklabels={
    {$0$},
    {$1$},
    {$2$},
    {$3$},
    {$4$}
},
scaled y ticks=false,
extra description/.code={
    \node[anchor=south west]
    at ([yshift=2pt]axis description cs:0,1)
    {$\times 10^{-5}$};
},
legend style={
    font=\fontsize{9.5}{8}\selectfont\sffamily\sansmath,
    anchor=north east,
    at={(0.99,0.35)},
    draw=black,
    fill=white,
    inner sep=1.5pt,
    row sep=0pt
},
legend cell align=left,
tick label style={
    font=\fontsize{11}{8}\selectfont\sffamily\sansmath
},
label style={
    font=\fontsize{13}{8}\selectfont\sffamily\sansmath
},
xlabel style={
    font=\fontsize{13}{8}\selectfont\sffamily\sansmath,
    at={(axis description cs:0.5,-0.13)},
    anchor=north
},
ylabel style={
    font=\fontsize{13}{8}\selectfont\sffamily\sansmath,
    at={(axis description cs:-0.1,0.5)},
    anchor=south
},
clip=true,
]

% Data layout:
% column 0 = axial node index z
% column 1 = dimensionless concentration c*
%
% The abscissa is normalized internally as z/H, with H = 800.
% Dimensionless time: t* = iteration * D / H^2,
% with D = 0.03333 and H = 800.

% ============================================================
% iteration = 10000, t* = 5.2078125e-4
% ============================================================
\addplot[
orange,
solid,
mark=o,
mark repeat=40,
mark phase=0,
mark size=2.7pt,
line width=1.1pt,
mark options={
    solid,
    draw=orange,
    fill=white,
    line width=0.65pt
}
]
table[
x expr=\thisrowno{0}/800,
y index=1,
col sep=space,
comment chars={\#}
]
{figures/amiri_Cbiglow/con_col17_10000.dat};
\addlegendentry{\(t^*=0.0005\)}

% ============================================================
% iteration = 20000, t* = 1.0415625e-3
% ============================================================
\addplot[
cyan!70!blue,
solid,
mark=square,
mark repeat=40,
mark phase=13,
mark size=2.6pt,
line width=1.1pt,
mark options={
    solid,
    draw=cyan!70!blue,
    fill=white,
    line width=0.65pt
}
]
table[
x expr=\thisrowno{0}/800,
y index=1,
col sep=space,
comment chars={\#}
]
{figures/amiri_Cbiglow/con_col17_20000.dat};
\addlegendentry{\(t^*=0.001\)}

% ============================================================
% iteration = 60000, t* = 3.1246875e-3
% ============================================================
\addplot[
green!60!black,
solid,
mark=triangle,
mark repeat=40,
mark phase=26,
mark size=3.4pt,
line width=1.1pt,
mark options={
    solid,
    draw=green!60!black,
    fill=white,
    line width=0.55pt
}
]
table[
x expr=\thisrowno{0}/800,
y index=1,
col sep=space,
comment chars={\#}
]
{figures/amiri_Cbiglow/con_col17_60000.dat};
\addlegendentry{\(t^*=0.0031\)}

\end{axis}

\end{tikzpicture}
        \par
    \end{minipage}
    \vspace{-1em}
    \caption{Concentration profiles at different times.}
    \label{fig:con_amiri_Cbiglow}
\end{subfigure}
\vspace{-0.20em}
\caption{Comparison of the lattice-unit temperature and concentration profiles
for $\mathcal{C}=1$.}
\label{fig:amiri_Cbiglow_profiles}
\end{figure}
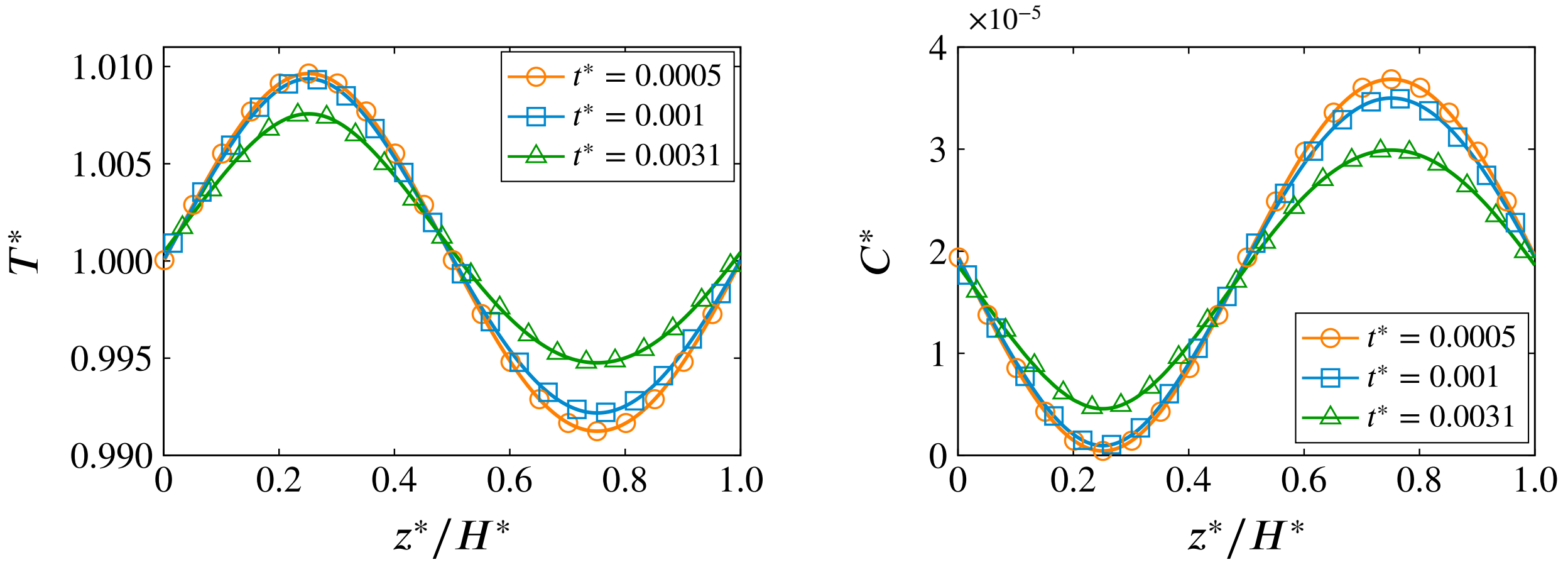
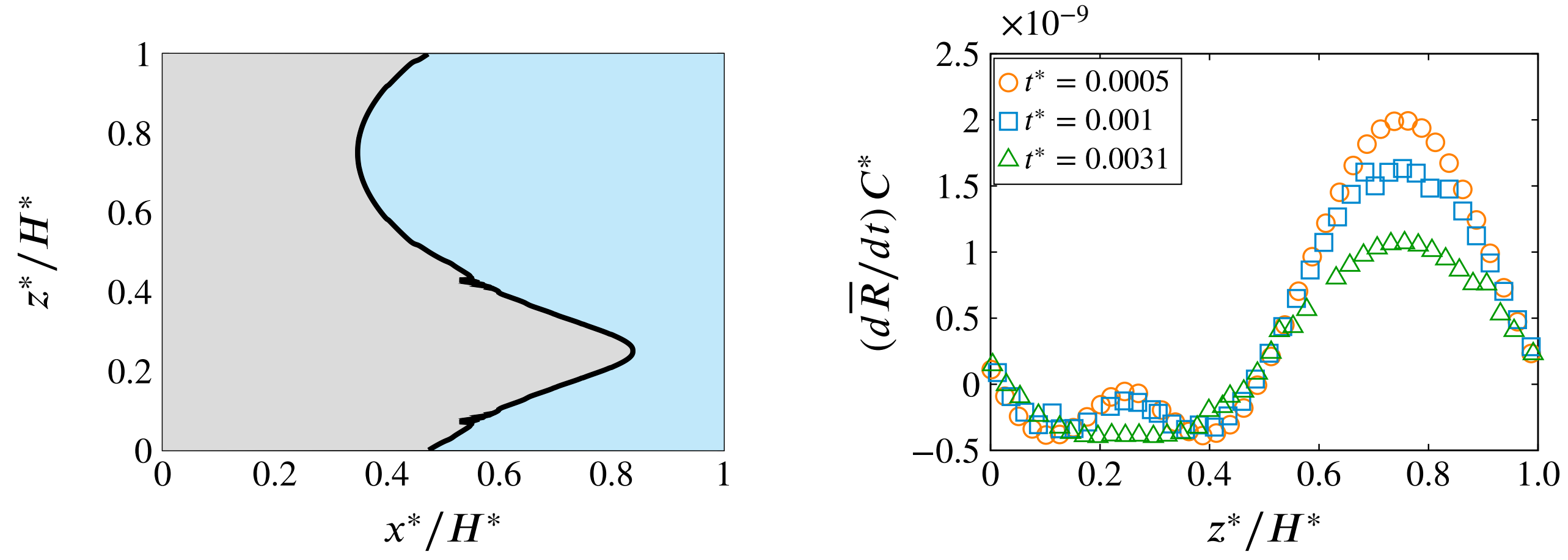
\begin{figure}[pos=htbp]
\centering
\begin{subfigure}[t]{0.45\textwidth}
    \centering
    \setlength\figurewidth{0.82\textwidth}
    \setlength\figureheight{0.58\textwidth}
    \begin{tikzpicture}[font=\sffamily\sansmath]

\begin{axis}[
width=\figurewidth,
height=\figureheight,
scale only axis,
xmin=0,
xmax=1,
ymin=0,
ymax=1,
xlabel={\(x^*/H^*\)},
ylabel={\(z^*/H^*\)},
xlabel style={
    font=\fontsize{13}{8}\selectfont\sffamily\sansmath,
    at={(axis description cs:0.5,-0.13)},
    anchor=north
},
ylabel style={
    font=\fontsize{13}{8}\selectfont\sffamily\sansmath,
    at={(axis description cs:-0.17,0.5)},
    anchor=south
},
grid=none,
axis line style={black},
tick align=inside,
tick pos=both,
tick label style={
    font=\fontsize{11}{8}\selectfont\sffamily\sansmath
},
label style={
    font=\fontsize{13}{8}\selectfont\sffamily\sansmath
},
legend style={
    draw=none,
    fill=none
},
clip=true,
]

% Interface curve
\addplot[
name path=interface,
black,
line width=1.4pt
]
table[
x index=1,
y index=0,
col sep=space,
comment chars={\#}
]
{figures/amiri_Cbiglow/interface_60000.dat};

% Left domain boundary
\path[name path=leftboundary]
(axis cs:0,0) -- (axis cs:0,1);

% Right domain boundary
\path[name path=rightboundary]
(axis cs:1,0) -- (axis cs:1,1);

% Solid region, left of the interface
\addplot[
draw=none,
fill=gray!28,
fill opacity=1
]
fill between[
of=leftboundary and interface
];

% Liquid region, right of the interface
\addplot[
draw=none,
fill=cyan!22,
fill opacity=1
]
fill between[
of=interface and rightboundary
];

% Redraw the interface on top
\addplot[
black,
line width=1.6pt
]
table[
x index=1,
y index=0,
col sep=space,
comment chars={\#}
]
{figures/amiri_Cbiglow/interface_60000.dat};

\end{axis}

\end{tikzpicture}
\vspace{-1.25em}
    \caption{Interface contour at \(t^*=0.0031\).}
    \label{fig:interface_amiri_Cbiglow}
\end{subfigure}
\hfill
\begin{subfigure}[t]{0.45\textwidth}
    \centering
    \setlength\figurewidth{0.80\textwidth}
    \setlength\figureheight{0.58\textwidth}
    \begin{tikzpicture}[font=\sffamily\sansmath]

\begin{axis}[
width=\figurewidth,
height=\figureheight,
scale only axis,
xmin=0,
xmax=1,
ymin=-5e-10,
ymax=2.5e-9,
xlabel={\(z^*/H^*\)},
ylabel={\(\displaystyle (d \overline{R} /dt) \, C^*\)},
grid=none,
axis line style={black},
tick align=inside,
tick pos=both,
xtick={0,0.2,0.4,0.6,0.8,1.0},
xticklabels={
{$0$},
{$0.2$},
{$0.4$},
{$0.6$},
{$0.8$},
{$1.0$}
},
ytick={-5e-10,0,5e-10,1e-9,1.5e-9,2e-9,2.5e-9},
yticklabels={
{$-0.5$},
{$0$},
{$0.5$},
{$1$},
{$1.5$},
{$2$},
{$2.5$}
},
scaled y ticks=false,
extra description/.code={
    \node[
        anchor=south west,
        font=\fontsize{11}{8}\selectfont\sffamily\sansmath
    ]
    at (axis description cs:0,1.02)
    {$\times 10^{-9}$};
},
legend style={
    font=\fontsize{9.5}{8}\selectfont\sffamily\sansmath,
    anchor=north east,
    at={(0.35,0.99)},
    draw=black,
    fill=white,
    inner sep=1.5pt,
    row sep=0pt
},
legend cell align=left,
tick label style={
    font=\fontsize{11}{8}\selectfont\sffamily\sansmath
},
label style={
    font=\fontsize{13}{8}\selectfont\sffamily\sansmath
},
xlabel style={
    font=\fontsize{13}{8}\selectfont\sffamily\sansmath,
    at={(axis description cs:0.5,-0.13)},
    anchor=north
},
ylabel style={
    font=\fontsize{13}{8}\selectfont\sffamily\sansmath,
    at={(axis description cs:-0.15,0.5)},
    anchor=south
},
clip=true
]

% Data layout:
% column 0 = axial node index z
% column 1 = interface radius R
% column 2 = (dR/dt) C*(r)
%
% The abscissa is normalized as z/H, with H = 800.
% Dimensionless time:
% t* = iteration * D / H^2,
% with D = 0.03333 and H = 800.

% ============================================================
% iteration = 10000, t* = 0.00052
% ============================================================
\addplot[
orange,
only marks,
mark=o,
mark repeat=20,
mark phase=0,
mark size=2.7pt,
line width=1.1pt,
mark options={
    solid,
    draw=orange,
    fill=white,
    line width=0.65pt
}
]
table[
x expr=\thisrowno{0}/800,
y index=2,
col sep=space,
comment chars={\#}
]
{figures/amiri_Cbiglow/conwet_interface_10000.dat};

\addlegendentry{\(t^*=0.0005\)}

% ============================================================
% iteration = 20000, t* = 0.001
% ============================================================
\addplot[
cyan!70!blue,
only marks,
mark=square,
mark repeat=20,
mark phase=13,
mark size=2.6pt,
line width=1.1pt,
mark options={
    solid,
    draw=cyan!70!blue,
    fill=white,
    line width=0.65pt
}
]
table[
x expr=\thisrowno{0}/800,
y index=2,
col sep=space,
comment chars={\#}
]
{figures/amiri_Cbiglow/conwet_interface_20000.dat};

\addlegendentry{\(t^*=0.001\)}

% ============================================================
% iteration = 60000, t* = 0.0031
% ============================================================
\addplot[
green!60!black,
only marks,
mark=triangle,
mark repeat=20,
mark phase=26,
mark size=3.4pt,
line width=1.1pt,
mark options={
    solid,
    draw=green!60!black,
    fill=white,
    line width=0.55pt
}
]
table[
x expr=\thisrowno{0}/800,
y index=2,
col sep=space,
comment chars={\#}
]
{figures/amiri_Cbiglow/conwet_interface_60000.dat};

\addlegendentry{\(t^*=0.0031\)}

\end{axis}

\end{tikzpicture}
\vspace{-1.25em}
    \caption{Reactive Robin interfacial flux at different dimensionless times.}
    \label{fig:conwet_interface_amiri_Cbiglow}
\end{subfigure}
\vspace{-0.25em}
\caption{Interface morphology and corresponding reactive Robin contribution for the low-concentration-ratio configuration, $\mathcal{C}=1$. The results are obtained using the link-based reaction formulation, and $\mathrm{d}\overline{R}/\mathrm{d}t$ denotes the mean of the link-based reaction rates $A_2$.}
\label{fig:interface_robin_flux_amiri_Cbiglow}
\end{figure}
Fig.~\ref{fig:conwet_interface_amiri_Cbiglow} shows the temporal evolution of the reactive Robin contribution at the moving interface. As heat diffuses away from the warm peak at $z^*/H^*\simeq0.25$, freezing is triggered in a region that is initially almost solute-free. A strong local freezing rate, $A_2$, is therefore generated to reject sufficient solute into the adjacent liquid, increase the local concentration, and drive the system back towards the liquidus condition. This produces a spike in the reactive interfacial contribution, which generates the localized advance of the solid phase visible at $z^* / H^* \simeq 0.25$ in Fig.~\ref{fig:interface_amiri_Cbiglow}. As the temperature and concentration fields progressively adjust, the solute content in the freezing region increases and the interfacial contribution smooths out towards a more stable distribution.

Overall, this low-concentration-ratio case shows that reducing $\mathcal{C}$ to $\mathcal{O}(1)$ does not simply modify the amplitude of the phase-change response. Instead, it makes the freezing and melting dynamics intrinsically asymmetric and can generate strongly localized interface evolution, as observed here for $\mathcal{C}=1$ near the initially solute-poor warm peak.
\section{Conclusions}
\label{sec:conclusions}
In this work, a pore-scale lattice Boltzmann framework has been developed to resolve the coupled evolution of heat transport, solute redistribution, and solid--liquid interface motion in mushy layers, with particular application to sea-ice-like systems. A central feature of the proposed formulation is its ability to couple local departures from thermodynamic equilibrium directly to solute transport at the phase boundary and to the resulting evolution of the pore geometry.

The Volume-of-Pixel (VOP) method has been extended to a directional, link-based formulation, in which the phase-boundary displacement is evaluated independently along each solid--fluid lattice link. This allows local melting and solidification rates to drive direction-dependent interface motion, capturing anisotropic geometry evolution induced by heterogeneous transport conditions. In addition, the Robin boundary treatment of Huang \emph{et al.}~\cite{huang15} has been extended to moving phase boundaries, providing a unified treatment of solute redistribution during both melting and freezing. The formulation preserves the coupling between local interface position, interface velocity, and solute balance, and has been generalized from the original two-dimensional setting to a D3Q19 lattice.

The thermal tests were used to assess the behaviour of the implemented VOP formulation under diffusion-controlled and spatially heterogeneous phase-change conditions. The solutal boundary treatment was validated against analytical Robin-boundary solutions and moving-boundary redistribution tests. The extended scheme retains convergence rates between first and second order over the enlarged admissible interface-position range, allowing the phase boundary to move continuously across the local control volume associated with each lattice node. Repeated moving-boundary simulations further show that the coupling between the Robin boundary condition and the VOP update preserves the global solute inventory to good accuracy.

The complete model was then applied to a simplified pore-scale mushy layer representing sea ice in a diffusion-dominated regime. When sufficient separation is achieved between the transverse pore scale and the axial perturbation length, the numerical temperature, concentration, and porosity fields approach the predictions of ideal mushy-layer theory. The simulations recover the transport-induced phase-change mechanism described by Butler~\cite{butler11} and later investigated at the pore scale by Amiri and Butler~\cite{amiri24}. The present framework also makes it possible to explore regimes in which the assumptions underlying the large-$\mathcal{C}$ linearization cease to hold. When the concentration ratio is reduced to $\mathcal{C}=\mathcal{O}(1)$, equal positive and negative thermal perturbations no longer generate symmetric solidification and melting responses. Instead, stronger solidification develops in the warm, solute-poor region, while melting remains weaker in the cold, solute-rich region. The present work therefore extends pore-scale mushy-layer modelling beyond fixed-geometry and equilibrium numerical descriptions by coupling local thermosolutal disequilibrium, solute redistribution, and deforming solid--liquid interfaces within a coupled LBM-based formulation.\\

\noindent\textbf{Acknowledgements.}
This work was supported by the Swedish Research Council (Vetenskapsrådet) under grant agreement no. 2024-05283. Computational resources were provided by the National Academic Infrastructure for Supercomputing in Sweden (NAISS), funded by the Swedish Research Council.\\

\noindent\textbf{Declaration of interests.}
The authors report no conflicts of interest.\\

\noindent\textbf{Declaration of generative AI and AI-assisted technologies in the manuscript preparation process.} During the preparation of this work, the authors used ChatGPT (OpenAI) to assist with language editing and improving the clarity of the manuscript. The authors reviewed and edited all AI-assisted content and take full responsibility for the content of the published article.\\
%\input{appendix}

% Numbered bibliography appropriate for JCP.
% The CAS documentation recommends model1-num-names for numbered references;
% this bundle does not include that .bst, so the official Elsevier
% elsarticle-num style is included locally to preserve the journal's numbered output.
\bibliographystyle{elsarticle-num}
\bibliography{references}

\end{document}